\documentclass[
preprint,
preprintnumbers,
 superscriptaddress,
 amsmath,amssymb,
 aps,nofootinbib
]{revtex4-1}

\usepackage{graphicx}
\usepackage{dcolumn}
\usepackage{bm}
\usepackage[utf8]{inputenc}
\usepackage{braket}
\usepackage{textcomp}
\usepackage{natbib}
\usepackage[dvipsnames]{xcolor}
\usepackage[colorlinks]{hyperref} 
\hypersetup{
      colorlinks=true,
      linkcolor=blue,
      filecolor=blue,
      citecolor = teal,      
      urlcolor=blue,
      }
\begin{document}

\count\footins = 1000
\title{First results from the INDRA-FAZIA apparatus on isospin diffusion in $^{58,64}$Ni+$^{58,64}$Ni systems at Fermi energies}
\author{C.~Ciampi}
 \email{ciampi@fi.infn.it}
\affiliation{INFN - Sezione di Firenze, 50019 Sesto Fiorentino, Italy}
\affiliation{Dipartimento di Fisica e Astronomia, Universit\`{a} di Firenze, 50019 Sesto Fiorentino, Italy}
\author{S.~Piantelli}
\affiliation{INFN - Sezione di Firenze, 50019 Sesto Fiorentino, Italy}
\author{G.~Casini}
\affiliation{INFN - Sezione di Firenze, 50019 Sesto Fiorentino, Italy}
\author{G.~Pasquali}
\affiliation{INFN - Sezione di Firenze, 50019 Sesto Fiorentino, Italy}
\affiliation{Dipartimento di Fisica e Astronomia, Universit\`{a} di Firenze, 50019 Sesto Fiorentino, Italy}
\author{J.~Quicray}
\affiliation{Normandie University, ENSICAEN, UNICAEN, CNRS/IN2P3, LPC Caen, F-14000 Caen, France}
\author{L.~Baldesi}
\affiliation{INFN - Sezione di Firenze, 50019 Sesto Fiorentino, Italy}
\author{S.~Barlini}
\affiliation{INFN - Sezione di Firenze, 50019 Sesto Fiorentino, Italy}
\affiliation{Dipartimento di Fisica e Astronomia, Universit\`{a} di Firenze, 50019 Sesto Fiorentino, Italy}
\author{B.~Borderie}
\affiliation{Université Paris-Saclay, CNRS/IN2P3, IJCLab, 91405 Orsay, France}
\author{R.~Bougault}
\affiliation{Normandie University, ENSICAEN, UNICAEN, CNRS/IN2P3, LPC Caen, F-14000 Caen, France}
\author{A.~Camaiani}
\affiliation{KU Leuven, Instituut voor Kern- en Stralingsfysica, 3001 Leuven, Belgium}
\author{A.~Chbihi}
\affiliation{Grand Accélérateur National d'Ions Lourds (GANIL), CEA/DRF–CNRS/IN2P3, Boulevard Henri Becquerel, F-14076 Caen, France}
\author{D.~Dell'Aquila}
\affiliation{Dipartimento di Chimica e Farmacia, Università degli Studi di Sassari, Sassari, Italy}
\affiliation{INFN - Laboratori Nazionali del Sud, Catania, Italy}
\author{M.~Cicerchia}
\affiliation{INFN Laboratori Nazionali di Legnaro, 35020 Legnaro, Italy}
\author{J.~A.~Dueñas}
\affiliation{Departamento de Ingeniería Eléctrica y Centro de Estudios Avanzados en Física, Matemáticas y Computación, Universidad de Huelva, 21007 Huelva, Spain}
\author{Q.~Fable}
\affiliation{Laboratoire des 2 Infinis - Toulouse (L2IT-IN2P3), Universit\'e de Toulouse, CNRS, UPS, F-31062 Toulouse Cedex 9 (France)}
\author{D.~Fabris}
\affiliation{INFN Sezione di Padova, 35131 Padova, Italy}
\author{J.~D.~Frankland}
\affiliation{Grand Accélérateur National d'Ions Lourds (GANIL), CEA/DRF–CNRS/IN2P3, Boulevard Henri Becquerel, F-14076 Caen, France}
\author{C.~Frosin}
\affiliation{INFN - Sezione di Firenze, 50019 Sesto Fiorentino, Italy}
\affiliation{Dipartimento di Fisica e Astronomia, Universit\`{a} di Firenze, 50019 Sesto Fiorentino, Italy}
\author{T.~Génard}
\affiliation{Grand Accélérateur National d'Ions Lourds (GANIL), CEA/DRF–CNRS/IN2P3, Boulevard Henri Becquerel, F-14076 Caen, France}
\author{F.~Gramegna}
\affiliation{INFN Laboratori Nazionali di Legnaro, 35020 Legnaro, Italy}
\author{D.~Gruyer}
\affiliation{Normandie University, ENSICAEN, UNICAEN, CNRS/IN2P3, LPC Caen, F-14000 Caen, France}
\author{K.~I.~Hahn}
\affiliation{Department of Science Education, Ewha Womans University, Seoul 03760, Republic of Korea}
\author{M.~Henri}
\affiliation{Grand Accélérateur National d'Ions Lourds (GANIL), CEA/DRF–CNRS/IN2P3, Boulevard Henri Becquerel, F-14076 Caen, France}
\author{B.~Hong}
\affiliation{Center for Extreme Nuclear Matters (CENuM), Korea University, Seoul 02841, Republic of Korea}
\affiliation{Department of Physics, Korea University, Seoul 02841, Republic of Korea}
\author{S.~Kim}
\affiliation{Institute for Basic Science, Daejeon 34126, Republic of Korea}
\author{T.~Kozik}
\affiliation{Marian Smoluchowski Institute of Physics Jagiellonian University, 30-348 Krakow, Poland}
\author{M.~J.~Kweon}
\affiliation{Center for Extreme Nuclear Matters (CENuM), Korea University, Seoul 02841, Republic of Korea}
\affiliation{Department of Physics, Inha University, Incheon 22212, Republic of Korea}
\author{J.~Lemarié}
\affiliation{Grand Accélérateur National d'Ions Lourds (GANIL), CEA/DRF–CNRS/IN2P3, Boulevard Henri Becquerel, F-14076 Caen, France}
\author{N.~Le~Neindre}
\affiliation{Normandie University, ENSICAEN, UNICAEN, CNRS/IN2P3, LPC Caen, F-14000 Caen, France}
\author{I.~Lombardo}
\affiliation{INFN Sezione di Catania, 95123 Catania, Italy}
\author{O.~Lopez}
\affiliation{Normandie University, ENSICAEN, UNICAEN, CNRS/IN2P3, LPC Caen, F-14000 Caen, France}
\author{T.~Marchi}
\affiliation{INFN Laboratori Nazionali di Legnaro, 35020 Legnaro, Italy}
\author{S.~H.~Nam}
\affiliation{Center for Extreme Nuclear Matters (CENuM), Korea University, Seoul 02841, Republic of Korea}
\affiliation{Department of Physics, Korea University, Seoul 02841, Republic of Korea}
\author{A.~Ordine}
\affiliation{INFN Sezione di Napoli, 80126 Napoli, Italy}
\author{P.~Ottanelli}
\affiliation{INFN - Sezione di Firenze, 50019 Sesto Fiorentino, Italy}
\affiliation{Dipartimento di Fisica e Astronomia, Universit\`{a} di Firenze, 50019 Sesto Fiorentino, Italy}
\author{J.~Park}
\affiliation{Center for Extreme Nuclear Matters (CENuM), Korea University, Seoul 02841, Republic of Korea}
\affiliation{Department of Physics, Korea University, Seoul 02841, Republic of Korea}
\author{J.~H.~Park}
\affiliation{Center for Extreme Nuclear Matters (CENuM), Korea University, Seoul 02841, Republic of Korea}
\affiliation{Department of Physics, Inha University, Incheon 22212, Republic of Korea}
\author{M.~P\^{a}rlog}
\affiliation{Normandie University, ENSICAEN, UNICAEN, CNRS/IN2P3, LPC Caen, F-14000 Caen, France}
\affiliation{``Horia Hulubei'' National Institute for R\&D in Physics and Nuclear Engineering (IFIN-HH), P.~O.~Box MG-6, Bucharest Magurele, Romania}
\author{G.~Poggi}
\affiliation{INFN - Sezione di Firenze, 50019 Sesto Fiorentino, Italy}
\affiliation{Dipartimento di Fisica e Astronomia, Universit\`{a} di Firenze, 50019 Sesto Fiorentino, Italy}
\author{A.~Rebillard-Soulié}
\affiliation{Normandie University, ENSICAEN, UNICAEN, CNRS/IN2P3, LPC Caen, F-14000 Caen, France}
\author{A.~A.~Stefanini}
\affiliation{INFN - Sezione di Firenze, 50019 Sesto Fiorentino, Italy}
\affiliation{Dipartimento di Fisica e Astronomia, Universit\`{a} di Firenze, 50019 Sesto Fiorentino, Italy}
\author{S.~Upadhyaya}
\affiliation{Marian Smoluchowski Institute of Physics Jagiellonian University, 30-348 Krakow, Poland}
\author{S.~Valdré}
\affiliation{INFN - Sezione di Firenze, 50019 Sesto Fiorentino, Italy}
\author{G.~Verde}
\affiliation{INFN Sezione di Catania, 95123 Catania, Italy}
\affiliation{Laboratoire des 2 Infinis - Toulouse (L2IT-IN2P3), Universit\'e de Toulouse, CNRS, UPS, F-31062 Toulouse Cedex 9 (France)}
\author{E.~Vient}
\affiliation{Normandie University, ENSICAEN, UNICAEN, CNRS/IN2P3, LPC Caen, F-14000 Caen, France}
\author{M.~Vigilante}
\affiliation{INFN Sezione di Napoli, 80126 Napoli, Italy}
\affiliation{Dipartimento di Fisica, Università di Napoli, 80126 Napoli, Italy}

\collaboration{INDRA-FAZIA collaboration}

\begin{abstract}
 An investigation of the isospin equilibration process in the reactions $^{58,64}$Ni+$^{58,64}$Ni at two bombarding energies in the Fermi regime ($32\,$MeV/nucleon and $52\,$MeV/nucleon) is presented. Data have been acquired during the first experimental campaign of the coupled INDRA-FAZIA apparatus in GANIL. Selecting from peripheral to semi-central collisions, both the neutron content of the quasiprojectile residue and that of the light ejectiles coming from the quasiprojectile evaporation have been used as probes of the dynamical process of isospin diffusion between projectile and target for the asymmetric systems. The isospin transport ratio technique has been employed. 
 The relaxation of the initial isospin imbalance with increasing centrality has been clearly evidenced.
 The isospin equilibration appears stronger for the reactions at $32\,$MeV/nucleon, as expected due to the longer projectile-target interaction time than at $52\,$MeV/nucleon.  
 Coherent indications of isospin equilibration come from the quasiprojectile residue characteristics and from particles ascribed to the quasiprojectile decay.
\end{abstract}

\maketitle
\section{Introduction}
Heavy ion collisions in the Fermi energy regime ($20-100\,$MeV/nucleon) are a widely employed tool to collect information on the properties of nuclear matter \cite{Li1998,DiToro2010}. The theoretical description of the reactions in this transition regime between the low and the high energies requires both the mean field contributions and the nucleon-nucleon collisions to be taken into account. 
A variety of outputs can be observed, mostly depending on the centrality of the reaction \cite{Lukasik1997}.
In particular, in semi-peripheral and peripheral reactions, a binary exit channel, characterized by the production of two heavy fragments, namely the quasiprojectile (QP) and the quasitarget (QT), is generally observed.
Together with the two heavy fragments, lighter ejectiles such as Light Charged Particles (LCPs, $Z=1,2$) and Intermediate Mass Fragments (IMFs, in this work we refer to $Z=3,4$) are also produced. The latter are mostly emitted at midvelocity, i.e. at a typical velocity between those of the QP and the QT, supposedly before, during or after the rupture in two main bodies of the deformed transient system \cite{Lionti2005}. 

Many efforts have been dedicated to the study of the isotopic composition of the products of heavy ion reactions, looking for details of the microscopic mechanisms not entirely washed out by the statistical evaporation.
Such studies assume that nucleon exchange processes taking place during the interaction phase of the reaction between the entrance channel nuclei are at the origin of the kinetic energy dissipation into internal degrees of freedom and of the evolution towards their equilibration. Among these latter, a special focus has been put on the isospin transport phenomena \cite{Baran2005}. 
Two main experimental observations in the field of isospin dynamics have been widely recognized by the scientific community: the process of isospin equilibration between reacting nuclei having different initial neutron-to-proton ratio $N/Z$ \cite{Hong2002,Tsang2004,Liu2007,Galichet2009,Lombardo2010,Barlini2013, Keksis2010, May2018,Fable2018, Chbihi2020, Piantelli2021, Camaiani2021}, and the neutron enrichment of the midvelocity emissions \cite{Larochelle2000,Milazzo2002,Lionti2005,Theriault2006,Lombardo2010,Barlini2013,Fable2018,Chbihi2020, Piantelli2021}. 
An interpretation of these two phenomena can be found in the framework of the Nuclear Equation of State (NEoS).
The isospin equilibration between asymmetric projectile and target can be ascribed to the \emph{isospin diffusion} mechanism, which is driven by an isospin gradient and depends on the contact time between the two colliding nuclei \cite{Tsang2004}, which in turn varies as a function of the bombarding energy and of the impact parameter of the collision \cite{Colonna2006}. For a long enough interaction time, a full isospin equilibration, i.e., a homogeneous distribution of the isospin content among the reaction products, could be expected.
On the other hand, the neutron enrichment at midvelocity has been associated with the \emph{isospin drift} (or migration), an effect driven by the presence of a nuclear density gradient, leading to a net neutron flux towards low-density regions \cite{DiToro2010}. A neutron enrichment can be therefore expected for the diluted intermediate neck region, formed in the separation of the projectile-target system, and hence for the midvelocity species there originated. The neutron enrichment of the IMFs coming from the neck can, however, be partially counterbalanced by the isospin fractionation mechanism, which tends to produce neutron rich gas and neutron poor fragments \cite{Muller1995,Ono2003,Ono2004}.
The two aforementioned processes of drift and diffusion act simultaneously during the projectile-target interaction, with the exception of symmetric reactions, in which the diffusion contribution is absent and only the isospin drift takes place. 
On the other hand, the drift process, being started by density variations, is basically absent or strongly reduced for too low beam energies and/or too small systems.
The strength of the isospin transport processes depends on the symmetry energy term $E_{sym}(\rho)$ of the NEoS \cite{Baran2005}. Therefore such effects are studied to gather information on the NEoS parametrization by comparing the experimental observations with the predictions of theoretical models. In particular, in the framework of transport models, the isospin diffusion has been interpreted as related to the value of $E_{sym}$, and the isospin drift to the first derivative of $E_{sym}$ with respect to the nuclear density \cite{Colonna2020}.

In order to perform this kind of studies, it is mandatory to access the information on the chemical composition of fragments and particles produced in the reactions. Therefore, experimental apparatuses capable of detecting and isotopically identifying the ejectiles over the largest possible range in $Z$ and energy are required.
In this paper, we focus on the experimental observation of isospin diffusion in semi-peripheral and peripheral reactions by analyzing the isotopic composition of the QP. The isospin content of the QP can be probed either by inspecting the characteristics of its deexcitation emissions \cite{Lombardo2010, Keksis2010,Piantelli2021} or by the direct isotopic identification of its remnant \cite{Piantelli2021,Camaiani2021}.
Common experimental apparatuses generally access only QP evaporation or, less frequently, only the isotopic composition of the QP residue.
The former can be obtained by means of arrays with large angular coverage, providing a good reconstruction of the global event, often with serious limitations in terms of isotopic discrimination beyond $Z=6-8$.
On the other hand, the direct investigation of the isospin of the QP remnant can be performed, e.g., by using mass spectrometers \cite{Souliotis2011, Fable2018}, that provide a precise mass identification for a wide range of heavy fragment isotopes, albeit for only one product per event collected within their angular acceptance.
The recently coupled INDRA-FAZIA apparatus \cite{Lopez2018} combines the excellent isotopic identification performance of FAZIA, exploited for the QP phase space, and the large angular coverage of INDRA, in order to collect the most comprehensive information on the event\footnote{Some first examples of experiments in which the isospin diffusion phenomenon has been evidenced both on the QP remnant and on the QP light emissions can be found in Ref.~\cite{Piantelli2017}, for reactions at energies below $20\,$MeV/nucleon, and in Ref.~\cite{Piantelli2021}, though with a limited angular coverage and only for a tail of the QP charge distribution. Moreover, among the rare setups providing both large angular coverage and isotopic identification also for heavy fragments we recall the NIMROD array \cite{Wuenschel2009}.}.
A description of the apparatus can be found in Sec.~\ref{sec:experiment}. 

In this paper, we present the first results obtained from the E789 experiment in GANIL, which is the first one exploiting the INDRA-FAZIA setup: the four reactions  $^{58,64}$Ni+$^{58,64}$Ni at two bombarding energies ($32\,$MeV/nucleon and $52\,$MeV/nucleon) have been investigated. We compare the products of the two asymmetric reactions with those of both the neutron rich ($^{64}$Ni+$^{64}$Ni) and the neutron poorer ($^{58}$Ni+$^{58}$Ni) symmetric systems, thus gathering information on the isospin equilibration. Moreover, the use of two different incident energies allows to test the evolution of the isospin transport with variations of the dynamical conditions, namely the reaction timescales, the nuclear densities, and the possible deformations and angular momenta.
Taking advantage of such a complete set of data, we can study the isospin equilibration between asymmetric projectile and target by exploiting the isospin transport ratio method proposed by \citeauthor{Rami2000}~\cite{Rami2000}. Let $A$ and $B$ be two nuclides (in our case, $A=^{64}$Ni and $B=^{58}$Ni), and $AA,AB,BA,BB$ the possible combinations of colliding systems; the isospin transport ratio is defined as:
\begin{equation}
 \label{eq:imbratio}
 R(X_{i}) = \frac{2X_i-X_{AA}-X_{BB}}{X_{AA}-X_{BB}}
\end{equation}
where $i=AA,AB,BA,BB$ and $X$ is an observable which is somehow sensitive to the effect of isospin diffusion (for example, the $\langle N/Z \rangle$ of the projectile-like fragment); by construction, for the two symmetric systems $i=AA,BB$, $R(X_{i})$ takes value $+1$ and $-1$ respectively. Since data for both asymmetric reactions $i=AB,BA$ are available, two different ``branches'' of the isospin transport ratio can be built, using symmetric reactions as references. The limit of fully non-equilibrated condition corresponds to $R(X_i)=\pm1$, 
while, as the projectile of the $AB$ reaction is the target of the $BA$ reaction, in the case of isospin equilibration between $A$ and $B$ in these two reactions (which are really just the same in the center of mass frame) we should have $X_{AB}=X_{BA}$ and therefore $R(X_{AB})=R(X_{BA})$.
If the experimental conditions are similar for the four reactions, we can expect that by exploiting eq.~\eqref{eq:imbratio} the systematic uncertainties related to the apparatus are strongly reduced.   
In general, the isospin transport ratio technique allows to bypass any perturbation introducing a linear transformation on the considered isospin observable $X$ \cite{Camaiani2020}.
The ratio $R$ is expected to be largely unaffected by the statistical deexcitation of the products \cite{Tsang2009,Mallik2022}, provided that it is similar for all systems; in Ref.~\cite{Camaiani2020} slight distortions of the isospin transport ratio due to statistical evaporation have been evidenced only for low excitation energies of the primary fragments. 
This technique has been widely adopted in the past \cite{May2018,Tsang2004,Tsang2009,Liu2007,Sun2010,Camaiani2021}, and it is expected to help constrain the density dependence of the symmetry energy of the NEoS by enhancing the differences due to the assumption of different parametrizations \cite{Napolitani2010}. 
In this paper, the isospin equilibration observables are presented as a function of an order parameter whose correlation to the reaction centrality was tested by means of simulations based on the AMD (antisymmetrized molecular dynamics \cite{Ono1992}) model coupled to GEMINI++ \cite{Charity2010} as afterburner (see Sec.~\ref{ssec:centrality}), similarly to Ref.~\cite{Camaiani2021}. The isospin transport ratio is evaluated here using two different probes, namely the isospin content of the detected QP remnant (Sec.~\ref{ssec:QPremnant}) and the chemical composition of the QP decay particles (Sec.~\ref{ssec:QPejectiles}).
To our knowledge, this is one of the first works where the experimental investigation of the isospin equilibration is performed using a complete set of observables for both the QP residue and the evaporation particles for four relevant reactions and as a function of the reaction centrality.

\section{The experiment}\label{sec:experiment}
\begin{figure}[]
\centering
\includegraphics[width=0.7\columnwidth]{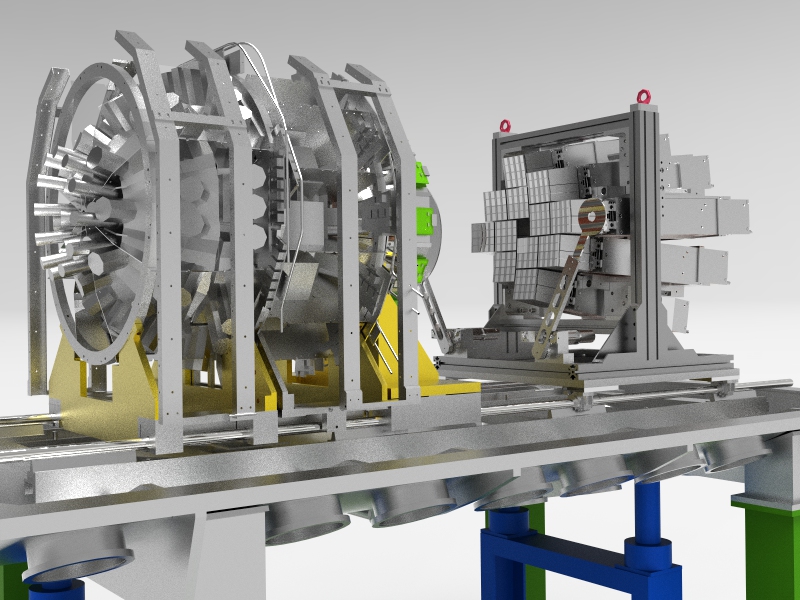}
\caption{Rendering of the INDRA-FAZIA mechanical coupling, with INDRA on the left side and FAZIA on the right side. The target holder, not visible in the picture, is located inside the INDRA apparatus.}
\label{fig:INDRA_FAZIA-render}
\end{figure}

In this section, we briefly describe the main characteristics of the two detection arrays composing the INDRA-FAZIA apparatus, located in the D5 experimental hall at GANIL.
The two devices have been coupled together as shown in Fig.\ref{fig:INDRA_FAZIA-render} in order to exploit both the large angular coverage of INDRA and the optimal $(Z,A)$ identification provided by FAZIA. 

\begin{figure}
\includegraphics[width=0.48\columnwidth]{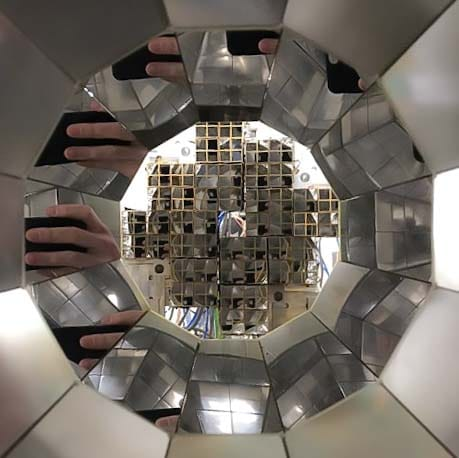}
\includegraphics[width=0.49\columnwidth]{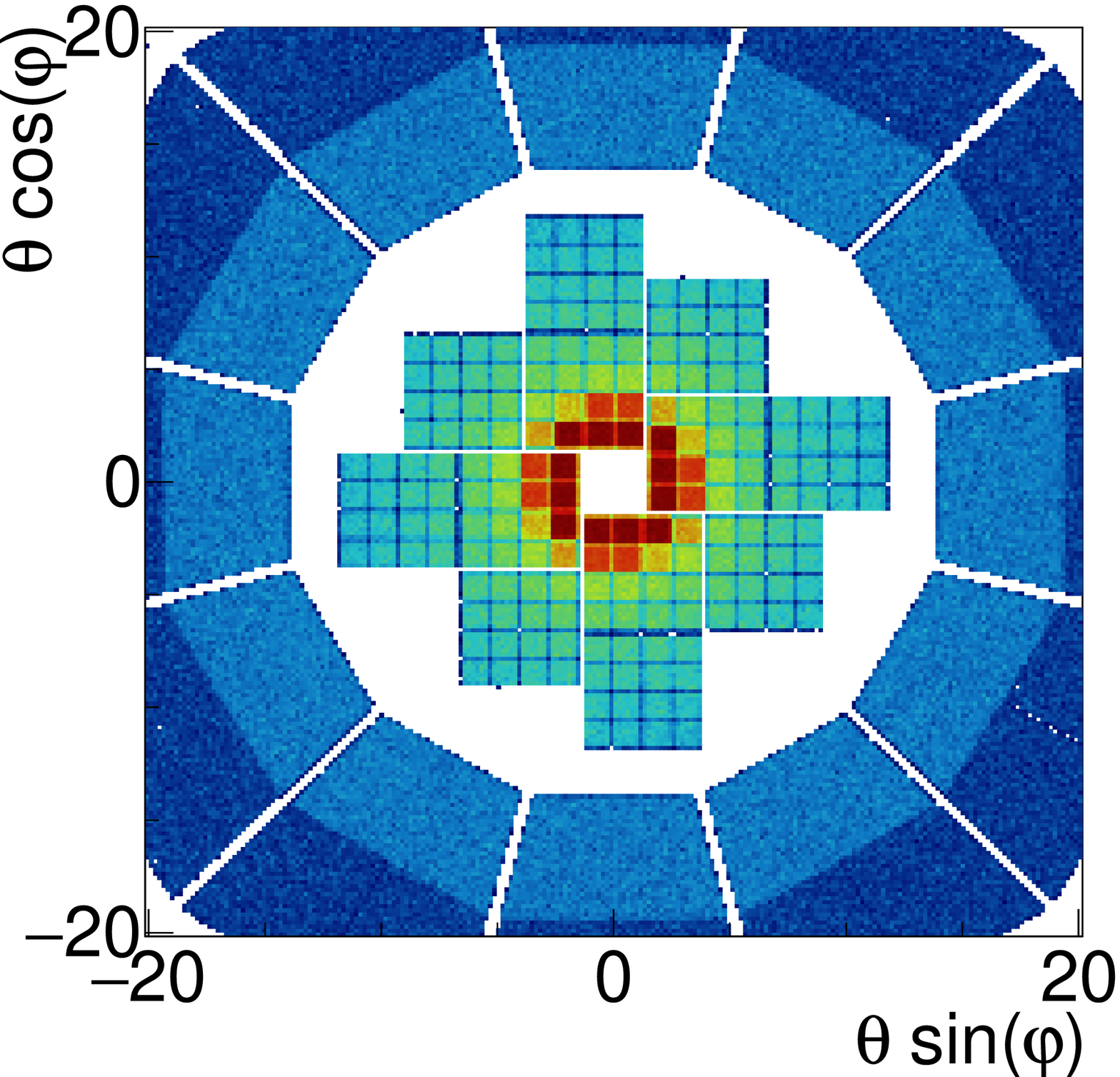}
\caption{Forward view from the target (left) and polar plot (right) of the INDRA-FAZIA apparatus, showing the 12 FAZIA blocks in a wall configuration. The polar plot is obtained by reporting the polar coordinates $(\theta \sin \varphi,\theta\cos\varphi)$ of all the detected particles, where $\theta$ and $\varphi$ are the polar and azimuthal angles in the laboratory reference frame, respectively.}
\label{fig:rosetta}
\end{figure}

FAZIA \cite{Bougault2014,Valdre2019} is a multi-detector array that represents the state of the art of nuclear fragment identification in the Fermi energy domain. The basic module of FAZIA is the \textit{block}, consisting of 16 three-stage $2\times2\,$cm$^{2}$ $\Delta$E-E telescopes. The first two layers are highly homogeneously doped Si detectors, $300\,\mu$m (Si1) and $500\,\mu$m (Si2) thick, respectively, and the third layer is a $10\,$cm thick CsI(Tl) scintillator read out by a photodiode. Each FAZIA block is equipped with the read-out electronics for all of its telescopes \cite{Valdre2019}, including preamplifiers, sampling ADCs, and FPGAs dedicated to the online digital treatment of the signals. Thanks to the digital signal processing implemented on the FPGAs, most of the relevant information is extracted in real-time from the detector signals. The ion identification is obtained by applying the $\Delta$E-E method and the Pulse Shape Analysis (the latter for Si1 and CsI(Tl) detectors). During the R\&D phase, the $\Delta$E-E charge identification capability has been successfully tested up to $Z\sim54$ \cite{Carboni2012}; isotopic discrimination has been achieved up to $Z\sim25$ with the $\Delta$E-E technique \cite{Carboni2012} and up to $Z\sim20$ with PSA in silicon \cite{Pastore2017}, as demonstrated in the first FAZIA experiments \cite{Barlini2013,Piantelli2020,Piantelli2021,Camaiani2021}.
For the present experiment, 12 FAZIA blocks are placed in a wall configuration at a distance of one meter from the target, covering the forward polar angles ($1.4$°$<\theta< 12.6$°), as shown in Fig.~\ref{fig:rosetta}, to exploit the performances of FAZIA for the identification of QP-like fragments.

The polar angles between 14° and 176° ($\sim$80\% of the $4\pi$ solid angle) are covered by 12 INDRA detection rings (i.e., rings 6 to 17 of the original INDRA configuration \cite{Pouthas1995,Pouthas1996}), with cylindrical symmetry with respect to the beam axis. Each detection ring is divided into several independent modules, providing a high granularity that allows for the reconstruction of high multiplicity events. The modules have different configurations depending on the ring \cite{Pouthas1995}; in the present experiment, the ionization chambers of INDRA have not been used. For the particles collected in INDRA rings 6 to 9, featuring a Si detector and a CsI, both $\Delta$E-E and CsI PSA techniques could be applied: isotopic identification is achieved up to $Z\sim4-7$. For INDRA rings 10 to 17, only the CsI information was available, and therefore the CsI PSA method was used. Moreover, INDRA CsI detectors belonging to rings 10 to 17 (corresponding to $\theta>45$°) could not be calibrated, and hence serve mainly as multiplicity counters, still providing isotopic identification for the collected LCPs.

The E789 experiment has been carried out at GANIL using $^{58}$Ni and $^{64}$Ni beams at the two bombarding energies of $32\,$MeV/nucleon and $52\,$MeV/nucleon, provided by the GANIL cyclotrons,
impinging on a target of $^{58}$Ni ($^{64}$Ni) with thickness $0.3\,$mg/cm$^2$ ($0.4\,$mg/cm$^2$).
The trigger condition for FAZIA was $M\geq2$, while triggers $M\geq1$ could also be acquired, downscaled by a factor of 100 (the recorded trigger pattern of FAZIA for each event allows to separate the two in the offline analysis), whereas for INDRA a minimum bias $M\geq1$ trigger was used but, due to an order of magnitude difference in the dead time of INDRA and FAZIA, INDRA events were only accepted if a coincident FAZIA trigger occurred within a pre-defined time window (note that FAZIA could trigger alone, as long as INDRA was not in dead time). Each event was given a $10\,$ns-resolution timestamp using a GANIL VXI CENTRUM module \cite{CENTRUM_ref} which then permitted the online merging of coincident events in order to be written to file. In the offline analysis, the difference in the recorded timestamps of each merged INDRA and FAZIA event allows to control the background of random coincidences and select well-correlated INDRA-FAZIA coincidence events.
A total of about $30\cdot10^6$ events have been acquired for each measured reaction.

\section{Data analysis and results}
As already cited, for semi-peripheral and peripheral reactions, most of the cross section is associated with a binary exit channel. Experimentally, we can expect the detection of one heavy fragment, i.e., the QP remnant, together with lighter emissions (the QT remnant being in most cases lost due to the detection energy thresholds).
The capabilities of the INDRA-FAZIA apparatus allow for an efficient selection and an exclusive analysis of the experimental events compatible with the QP evaporative channel.

In the following analysis, only the events with a trigger pattern $M\geq2$ in FAZIA have been considered in order to exclude the elastic scattering, not relevant for our purpose. 
The charge identification of at least two particles is also required.
Moreover, a preliminary selection based on global variables has been performed. A few events related to spurious coincidences (about $1\%$ of the total statistics) have been excluded by discarding events not satisfying charge and momentum conservation laws.
Finally, since we are interested in the events in which a large fraction of the reaction products has been detected, the most incomplete events have been discarded by imposing a total detected charge $Z_{tot}\geq12$ (considering all charge identified particles both in INDRA and FAZIA).
The events satisfying these preliminary conditions are about 46\% (57\%) of the total statistics for the reactions at $32\,$MeV/nucleon ($52\,$MeV/nucleon):
the analysis described in the following is carried out on this dataset. 
\begin{figure}[]
  \centering
  \includegraphics[width=0.4\columnwidth]{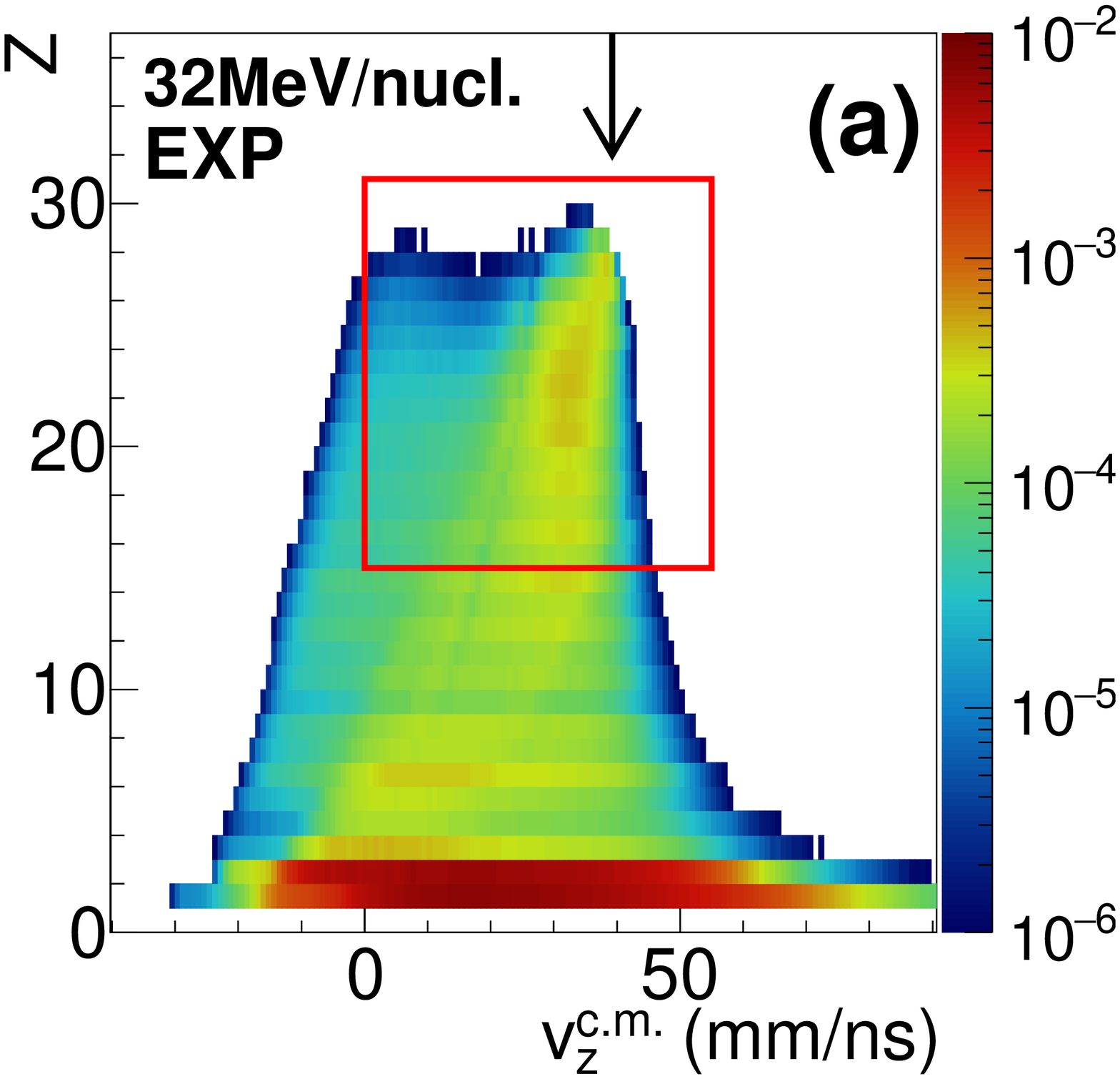}
  ~~~~
  \includegraphics[width=0.4\columnwidth]{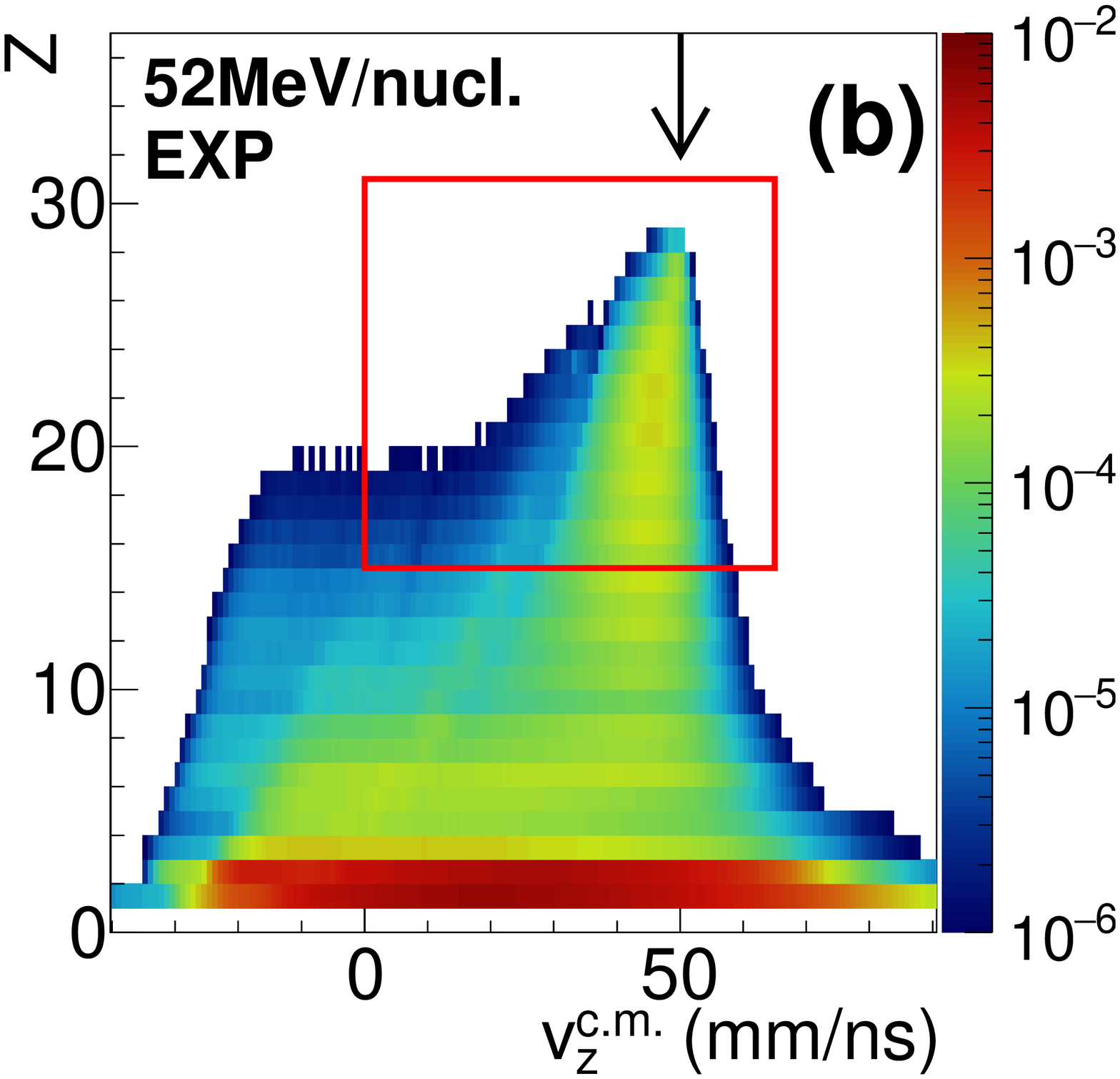}
  \caption{Experimental $Z$ vs $v_z^{c.m.}$ correlations for the fragments in the events selected by the preliminary conditions for the reactions $^{58}$Ni+$^{58}$Ni at $32\,$MeV/nucleon (a) and at $52\,$MeV/nucleon (b). The plots are normalized to their integral. The black arrows indicate the original projectile velocity for each reaction.}
  \label{fig:Z_vz_correlations}
\end{figure} 
Fig.~\ref{fig:Z_vz_correlations} shows the experimental $Z$ vs $v_z^{c.m.}$ correlations for all the identified and calibrated fragments in the events thus selected in the case of the reactions $^{58}$Ni+$^{58}$Ni at $32\,$MeV/nucleon (a) and at $52\,$MeV/nucleon (b), taken as examples. 
The velocity of each fragment is calculated from its measured kinetic energy exploiting the mass identification provided by the setup; when only the charge identification is available, a mass number based on the Evaporation Attractor Line (EAL) prediction \cite{Charity1998} is assumed. The area associated to the QP phase space is evident in each correlation, focusing towards the original projectile velocity $v_{beam}^{c.m.}$, indicated in each plot with a black arrow. However, more exclusive event selection conditions are introduced in the following.

 \subsection{Event selection: QP remnant and decay particles}
 The selection criterion for the QP evaporation channel requires the presence of only one heavy fragment, accompanied only by LCPs and, at most, IMFs ($Z=3,4$). In order to be recognized as QP remnant, the fragment must be forward emitted in the center of mass (c.m.) reference frame. Moreover, a lower limit for its charge range is imposed: the QP remnant is required to have $Z>14$, corresponding to a maximum projectile charge loss $\Delta Z=13$, i.e., about 46\% of the original projectile charge. This choice is slightly lower but nevertheless consistent with what can be found in the literature (see, e.g., Refs.~\cite{Camaiani2021,Piantelli2021}); however, the results presented in this paper are not substantially modified by setting a different lower limit on the QP remnant charge, within a reasonable interval (we tested a variation of $\pm3$ charge units with respect to the chosen value). No condition is imposed a priori on which array detects the QP remnant; however, due to the kinematics of the reactions, the QP remnant is collected by FAZIA in all cases for both beam energies.
 With the selected condition, around 40\% (30\%) of the events surviving the preliminary conditions are assigned to the QP evaporation channel selection for the four reactions at $32\,$MeV/nucleon ($52\,$MeV/nucleon).

 \begin{figure*}
  \centering
  \includegraphics[width=0.23\textwidth]{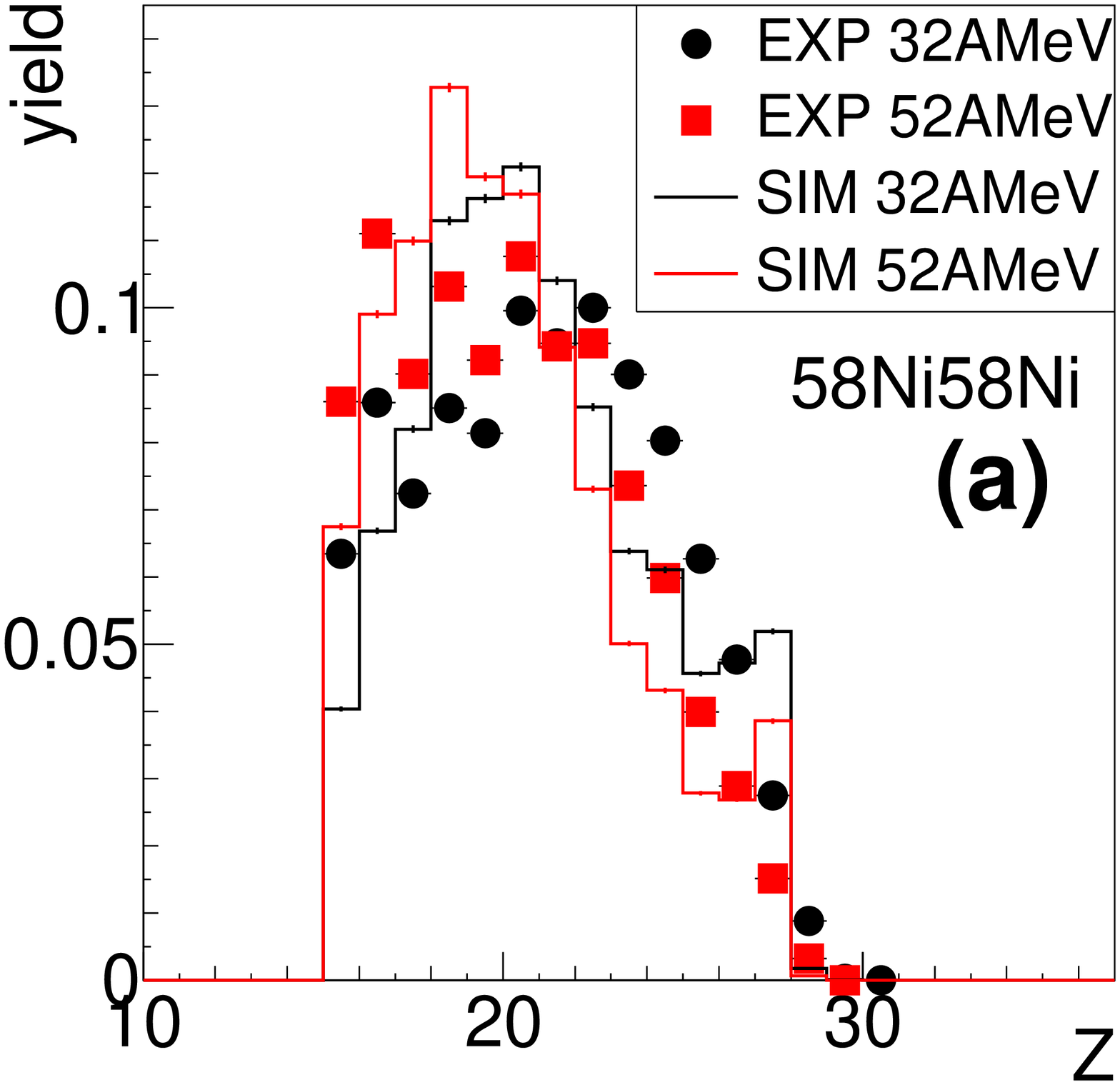}~~~
  \includegraphics[width=0.23\textwidth]{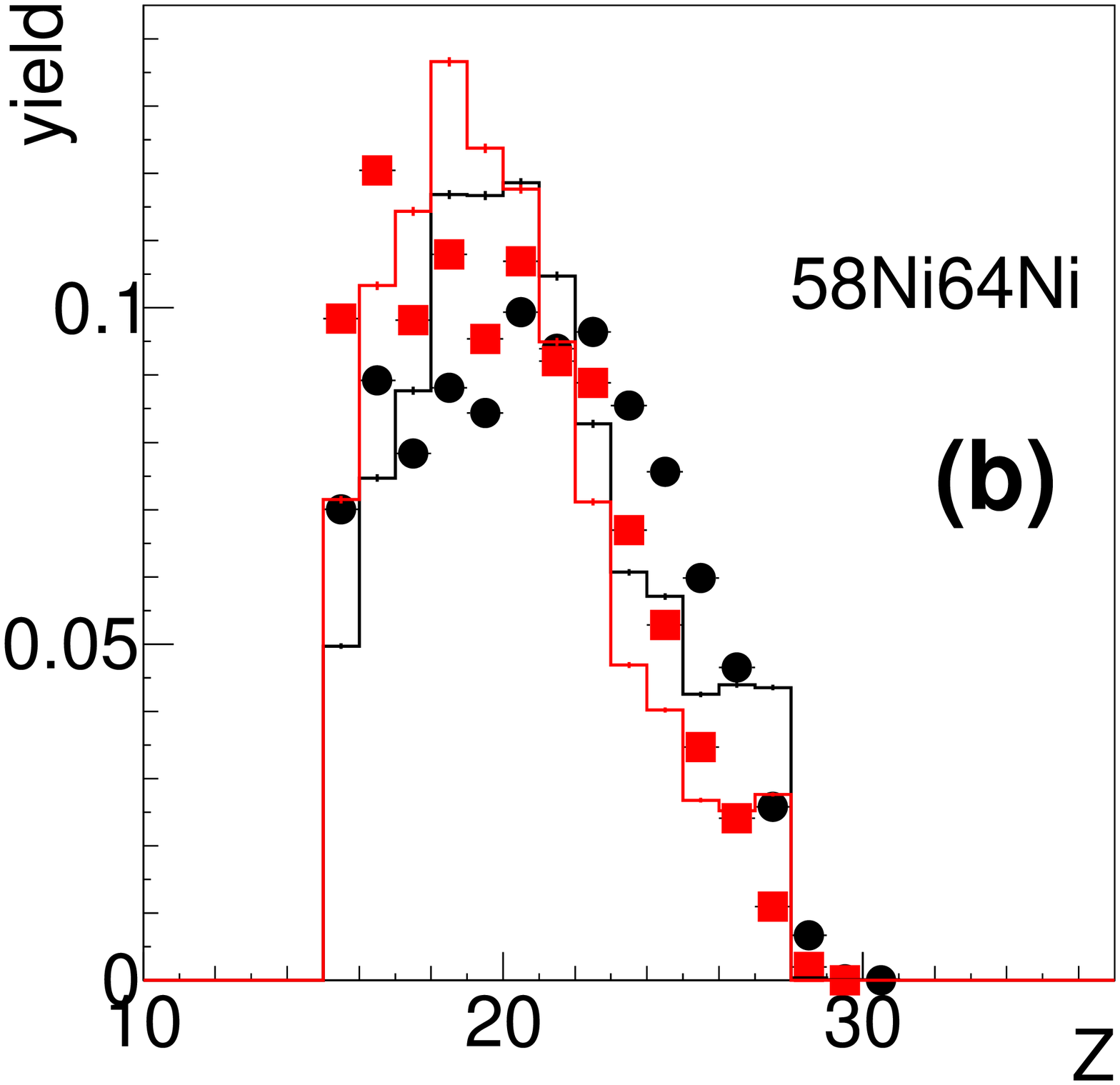}~~~
  \includegraphics[width=0.23\textwidth]{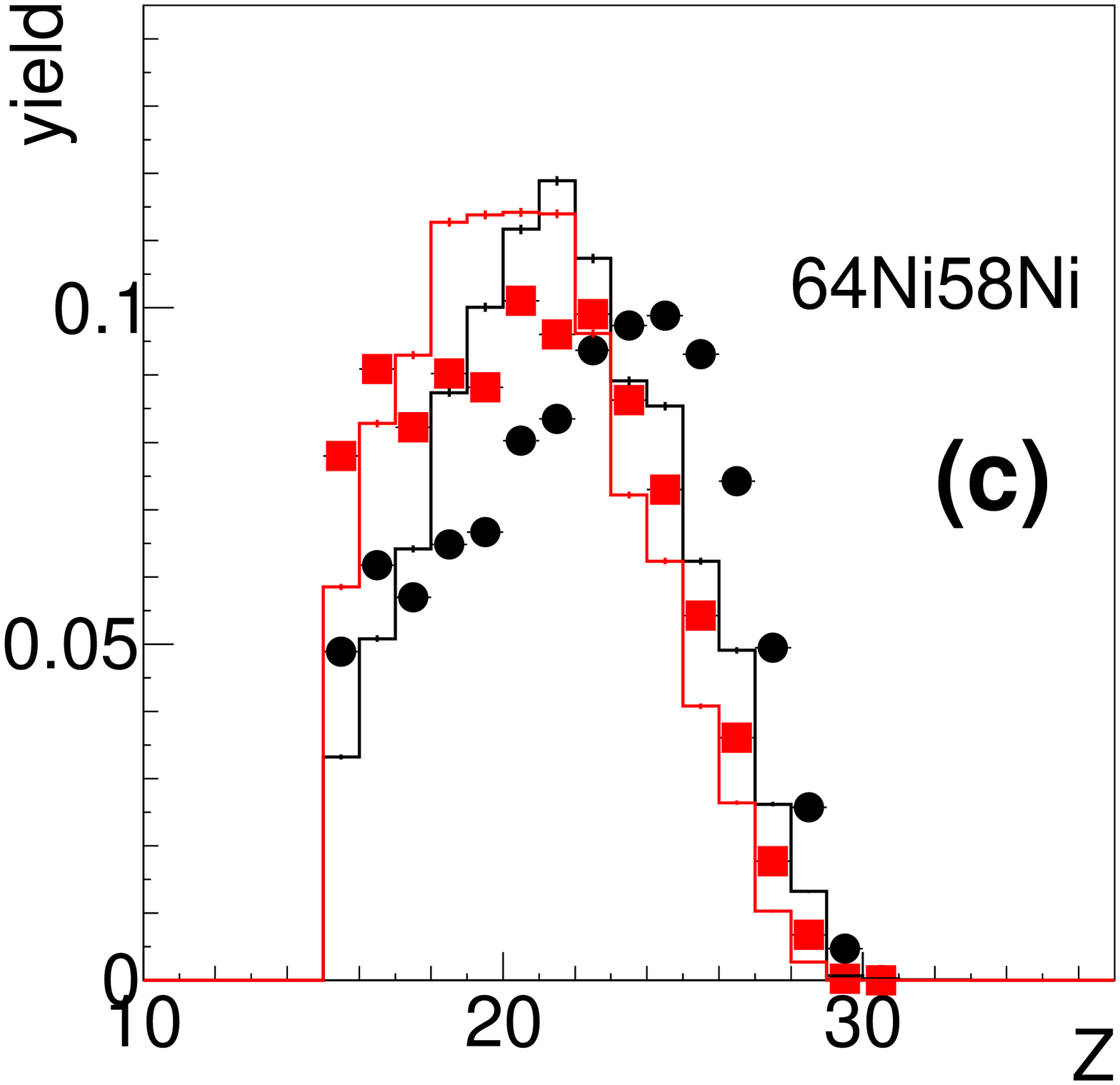}~~~
  \includegraphics[width=0.23\textwidth]{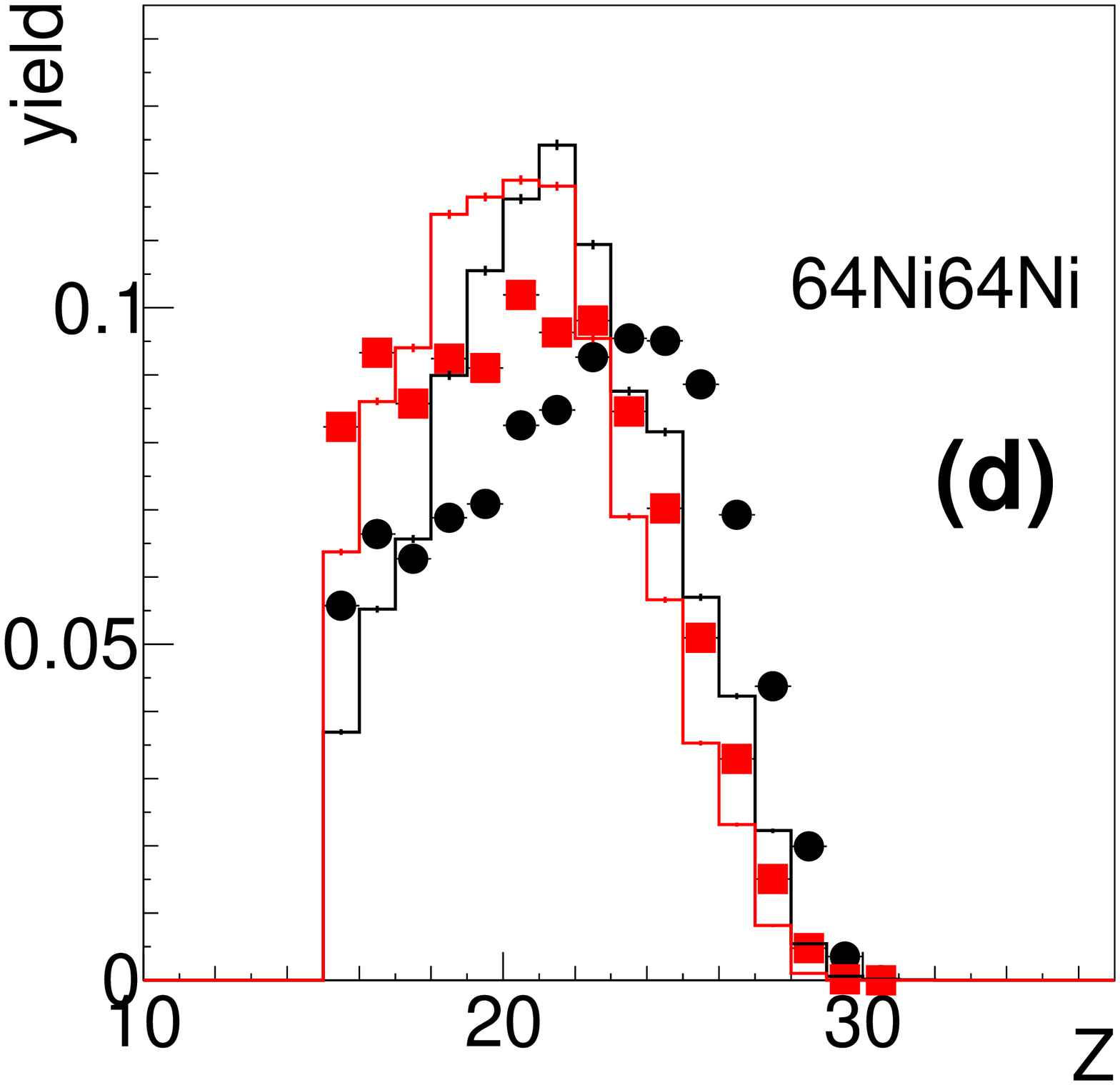}\\
  \vspace*{0.5cm}
  \includegraphics[width=0.23\textwidth]{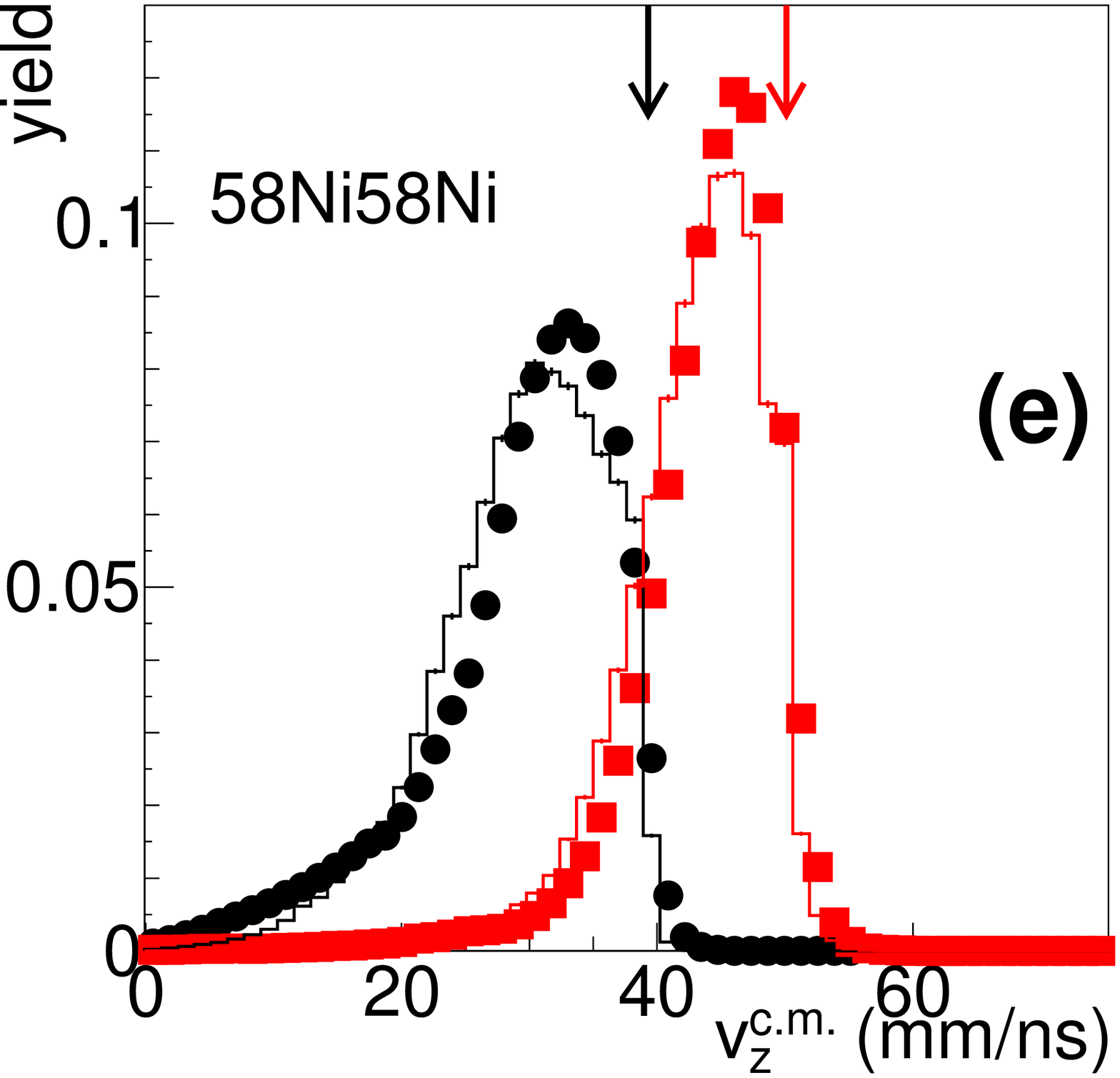}~~~
  \includegraphics[width=0.23\textwidth]{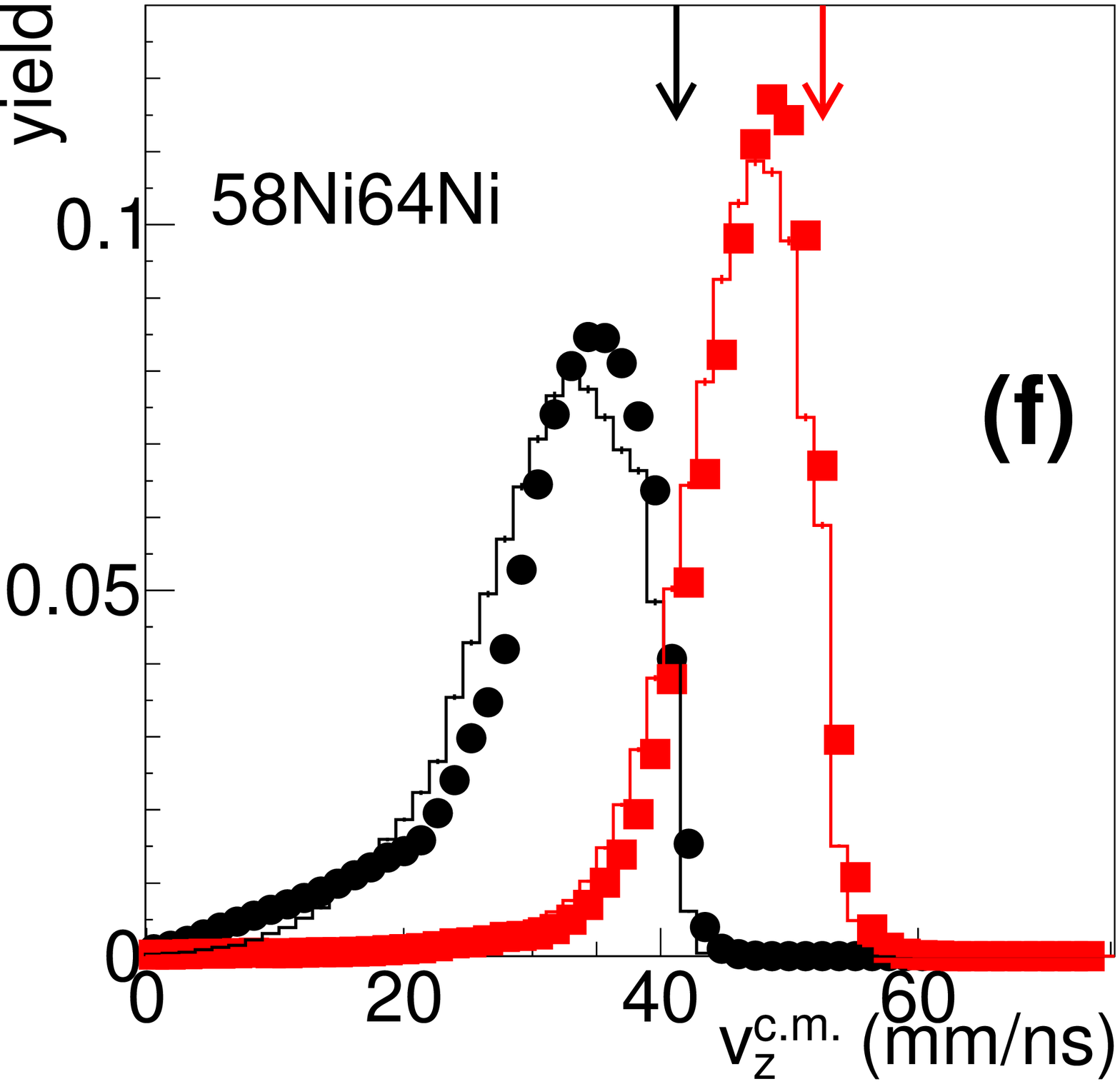}~~~
  \includegraphics[width=0.23\textwidth]{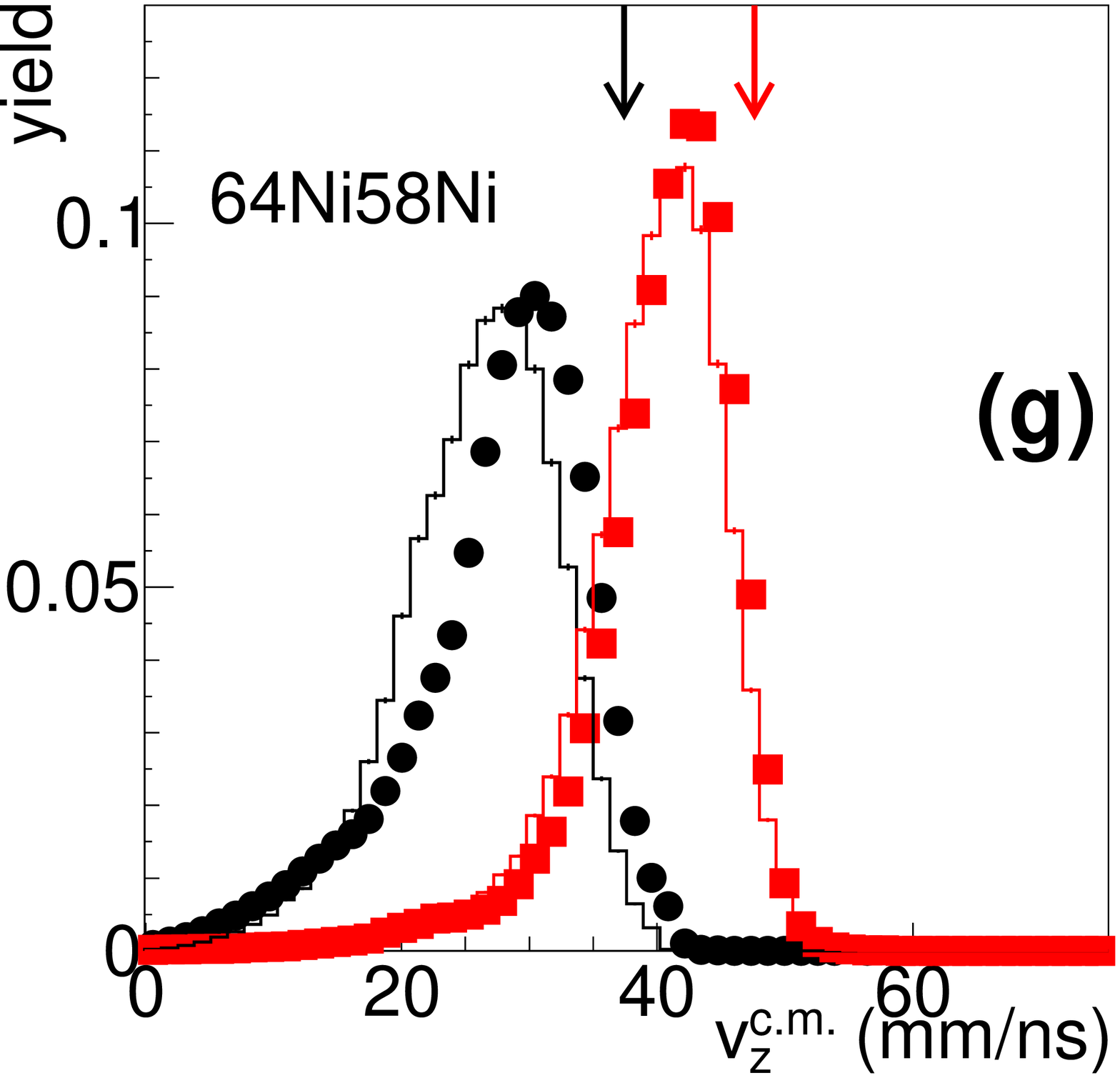}~~~
  \includegraphics[width=0.23\textwidth]{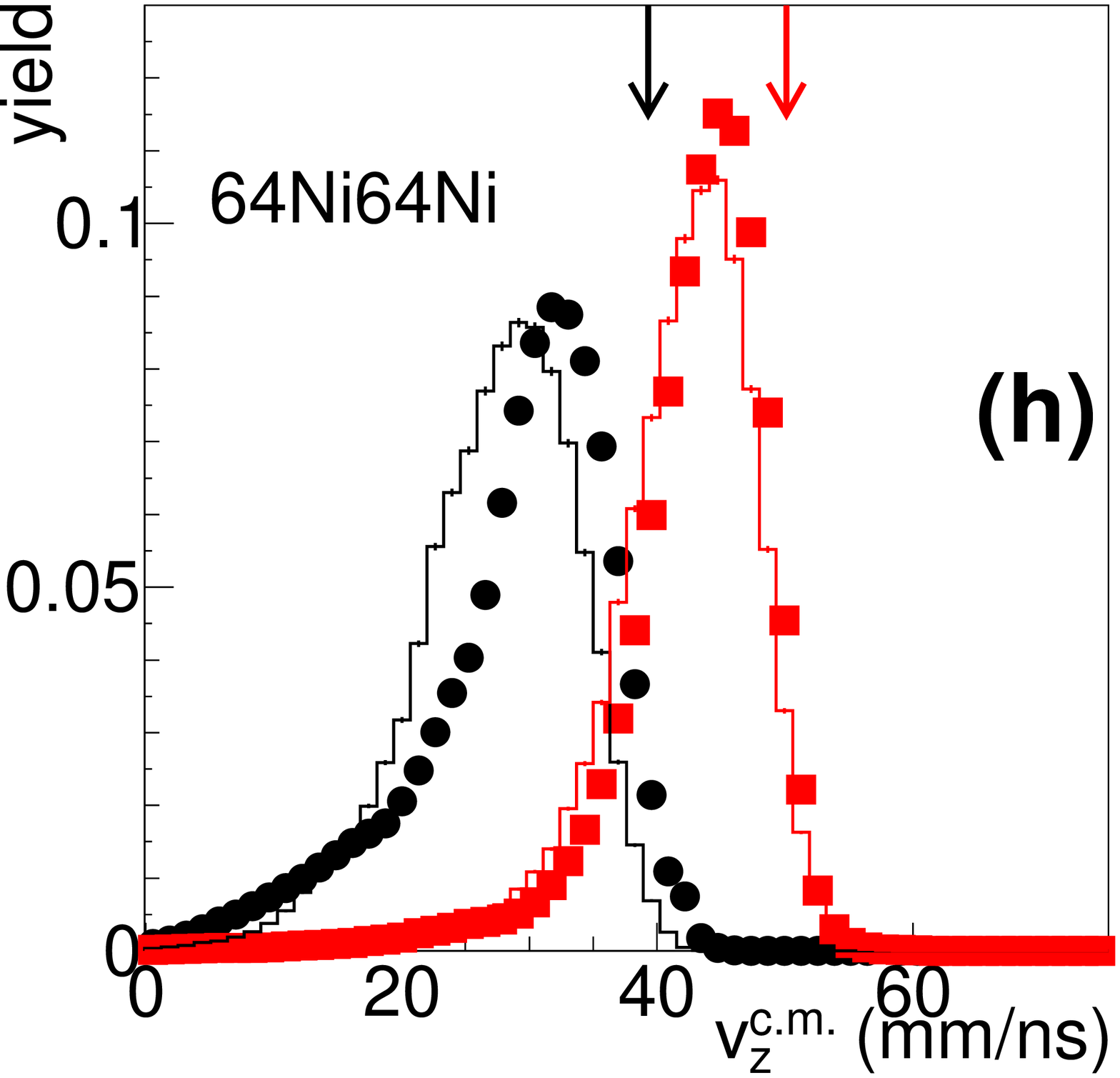}
  \caption{Charge distributions (top) and velocity distributions (bottom) of the QP remnant for all the studied reactions. Plots (a),(e) refer to the system $^{58}$Ni+$^{58}$Ni, plots (b),(f) to $^{58}$Ni+$^{64}$Ni, plots (c),(g) to $^{64}$Ni+$^{58}$Ni, and plots (d),(h) to $^{64}$Ni+$^{64}$Ni. The reactions at $32\,$MeV/nucleon ($52\,$MeV/nucleon) are plotted in black (red). Experimental data are plotted with full markers, while AMD+GEMINI++ predictions are drawn with a line. The same legend is valid for all the plots. The plots are normalized to their integral. In plots (e)-(h), the black (red) arrows indicate the original projectile velocity $v_{beam}^{c.m.}$ for each reaction at $32\,$MeV/nucleon ($52\,$MeV/nucleon).}
  \label{fig:Z_vz_distributions} 
 \end{figure*}
 In the $Z$ vs $v_z^{c.m.}$ correlations of Fig.~\ref{fig:Z_vz_correlations}, the area associated with the QP remnants selected with the cuts described above is evidenced with a red rectangle.
 It can be noticed that within this area, the two plots present a different population of the $v_z^{c.m.}\sim0\,$mm/ns region: for the reaction at $32\,$MeV/nucleon, some heavy fragments are produced in a process that is compatible with an incomplete-fusion, a possible outcome of central collisions, while for the reaction at $52\,$MeV/nucleon this region is only populated by lighter fragments.
 However, since these events correspond to more central collisions, they will be efficiently discarded by our following analysis as a function of the reaction centrality.
 
 In Fig.~\ref{fig:Z_vz_distributions} the distributions obtained for the charge $Z$ (top row) and the velocity component along the beam axis $v_z^{c.m.}$ (bottom row) of the QP remnant are shown for all the reactions. Experimental data are plotted with full markers. 
 The main features are compatible with the class of events we aim to select, with both charge and velocity distributions starting from the original beam values and extending towards lower values, as expected for increasingly dissipative collisions.
 Moreover, by comparing the charge histograms associated with the different systems in Fig.~\ref{fig:Z_vz_distributions}(a)-(d), it can also be observed that the QP remnant produced in the reactions induced by the neutron poorer $^{58}$Ni projectile generally preserves a lower nuclear charge $Z$ than that from the $^{64}$Ni-induced reactions; a similar observation can be found in Ref.~\cite{Camaiani2021} for Ca+Ca reactions.
 This can be interpreted as a result of the higher proton content of the emissions from the excited primary QP produced by the $^{58}$Ni projectile with respect to $^{64}$Ni.
 A charge distribution shifted towards slightly lower $Z$ values is generally found for all the reactions at $52\,$MeV/nucleon with respect to the same systems at $32\,$MeV/nucleon, most likely as a consequence of a stronger dynamical emission and/or a higher primary fragment excitation energy achievable at the higher bombarding energy.
 The charge histograms also reveal the presence of the well-known odd-even staggering phenomenon \cite{Lombardo2011,Piantelli2013}, with a higher production yield for the even-$Z$ QP remnants with respect to the neighbouring odd-$Z$ ones, particularly noticeable for low $Z$ values. As already recognized in Refs.~\cite{Lombardo2011, Piantelli2013}, also in the present case the effect is more evident in the neutron deficient system.
 
 On each plot of Fig.~\ref{fig:Z_vz_distributions} also the distributions obtained according to the AMD+GEMINI++ predictions are shown. This simulated dataset has been generated by producing, for each reaction, about 20000 primary events with the AMD code \cite{Ono1992} in the version described in Ref.~\cite{Piantelli2019},
 using the asy-stiff parametrization of the NEoS, corresponding to a slope parameter of the symmetry energy\footnote{The symmetry energy term of the NEoS can be expanded around the saturation density $\rho_0$ as: $\frac{E_{sym}}{A}(\rho)=S_0+L\cdot\frac{\rho-\rho_0}{3\rho_0}+O\Bigl[\bigl(\frac{\rho-\rho_0}{3\rho_0}\bigr)^2\Bigr]$. The value of the slope parameter $L$ determines whether the NEoS parametrization is asy-stiff or asy-soft.} $L=108\,$MeV, with the standard value of the symmetry energy at saturation density $S_0=32\,$MeV; for simplicity's sake, the asy-soft predictions are not shown since they are comparable to the asy-stiff ones at this level of investigation. The impact parameter in the AMD simulations follows a triangular distribution up to $b=11.6\,$fm, which is greater than the grazing impact parameter of all the studied reactions. The calculations have been stopped at $500\,$fm/$c$ \cite{Piantelli2019}. Then, the statistical code GEMINI++ \cite{Charity2010} has been applied as an afterburner: for each primary event, 100 secondary events have been produced to increase the statistics. The secondary events have then been filtered according to the apparatus acceptance (in terms of angular coverage, energy and identification thresholds, and working status of the telescopes) in order to directly compare them to the experimental data. 
 The same event selection criteria have been applied to both experimental and simulated data: the resulting charge and velocity distributions for the QP remnant are those drawn as continuous lines in the plots of Fig.~\ref{fig:Z_vz_distributions}. A reasonable overall agreement between the model predictions and the experimental results can be appreciated, generally slightly better for the reactions at $52\,$MeV/nucleon.
 However, some minor differences between the model predictions and the experimental observables can be noticed in the $Z$ distributions: in addition to the experimental odd-even staggering not completely reproduced by the simulations, the QP remnants in the experimental events tend to be heavier than those in the simulated events.
 A similar consideration can be done on the velocity distributions, with the simulations predicting slightly slower QP remnants. This discrepancy has already been highlighted in the literature for other systems simulated with AMD+GEMINI++ \cite{Piantelli2021,Camaiani2021}. 
 
 \subsection{Centrality estimation}\label{ssec:centrality}
 \begin{figure}[]
  \centering
  \includegraphics[width=0.4\columnwidth]{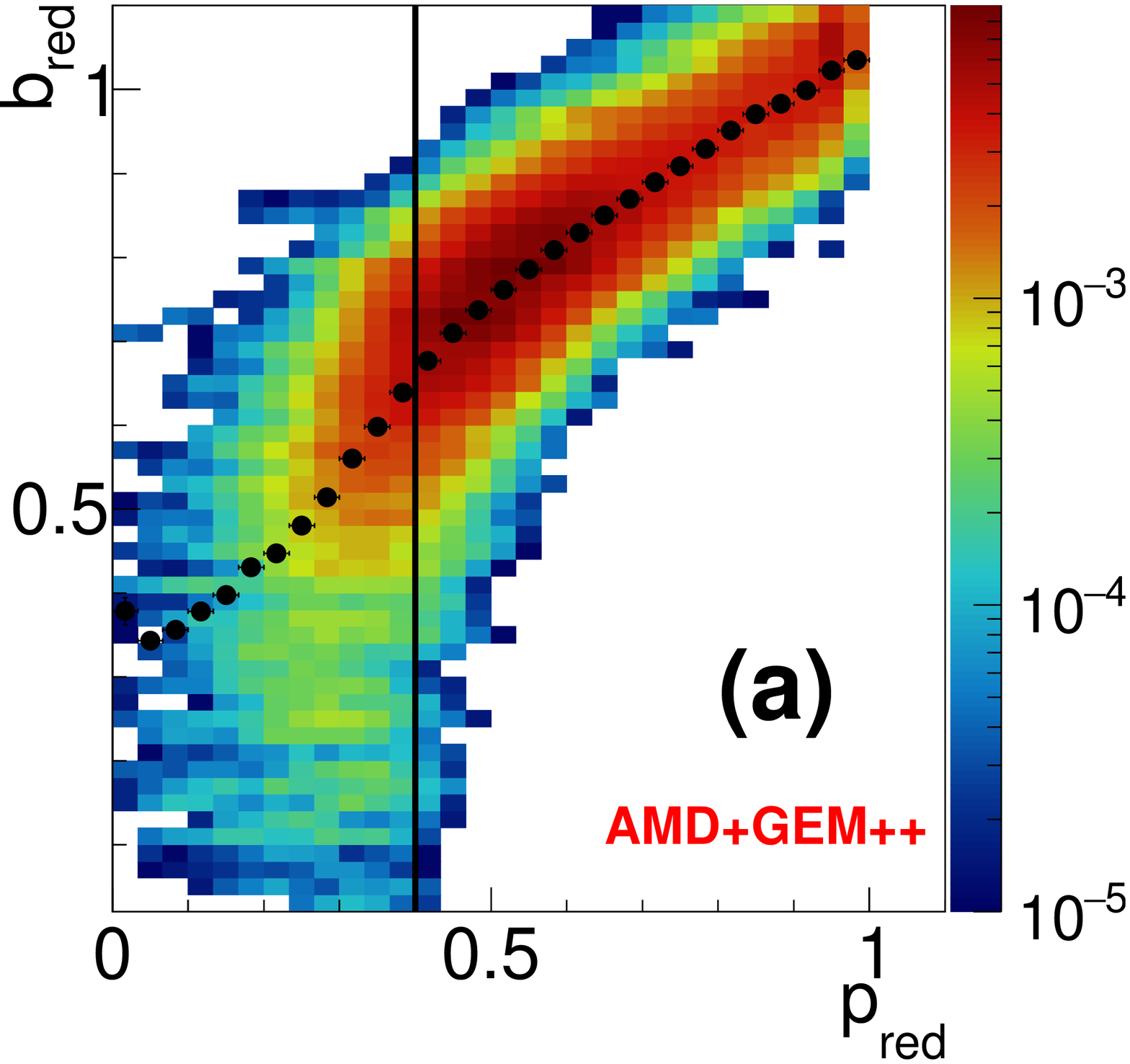}
  ~~~~
  \includegraphics[width=0.4\columnwidth]{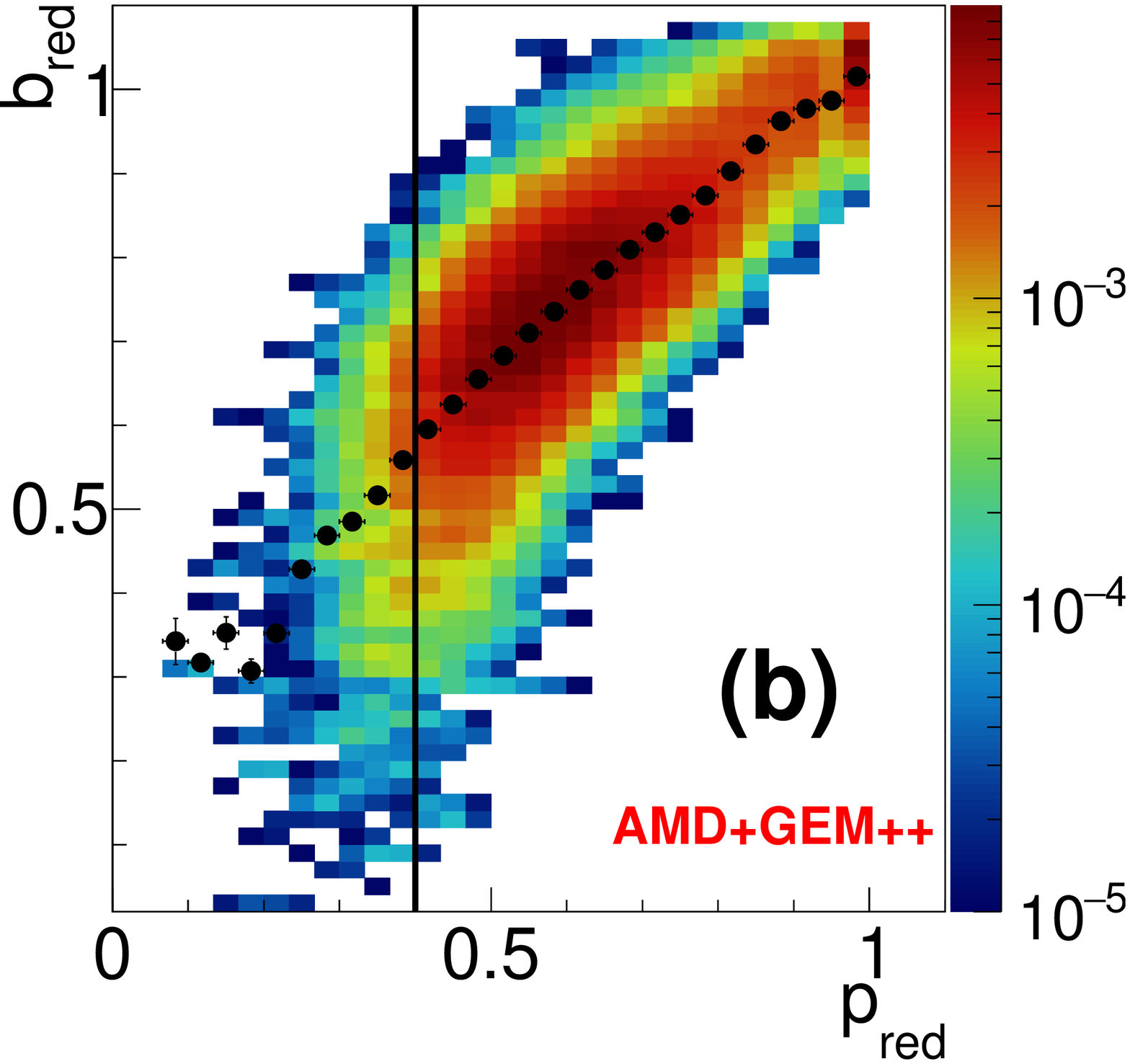}
  \caption{Correlation between $b_{red}$ and $p_{red}$ of the QP remnant in the AMD+GEMINI++ simulated events belonging to the QP evaporation channel for the reactions $^{58}$Ni+$^{58}$Ni at $32\,$MeV/nucleon (a) and at $52\,$MeV/nucleon (b). The plots are normalized to their integral. Black markers indicate the average $b_{red}$ for each $p_{red}$ bin. Vertical lines indicate the minimum $p_{red}$ value (0.4) above which the correlations are considered reliable in the following analysis.}
  \label{fig:bred_vs_pred_correlations}
 \end{figure}   
 \begin{figure*}[]
  \centering
  \includegraphics[width=0.3\textwidth]{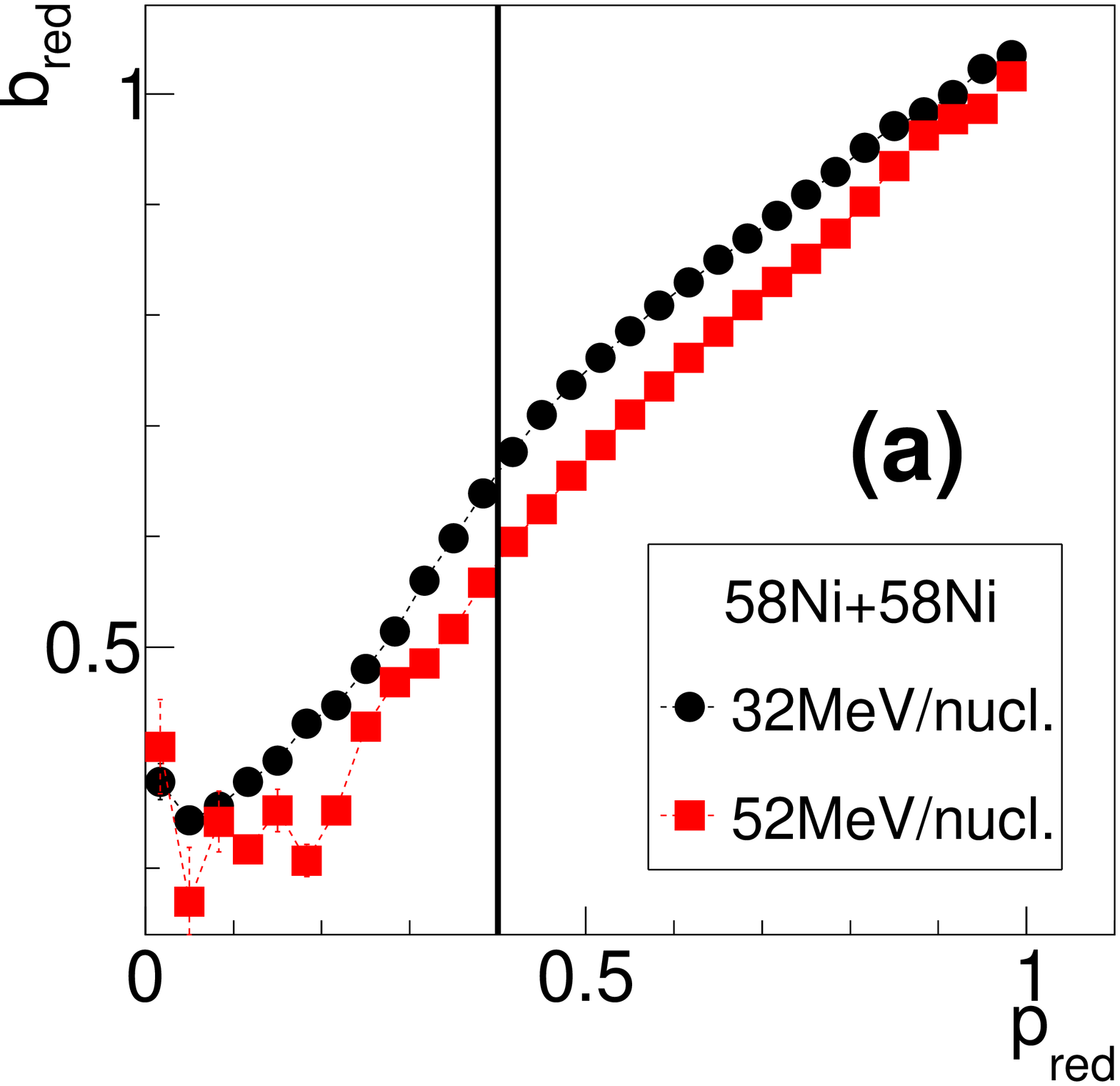}~~~
  \includegraphics[width=0.3\textwidth]{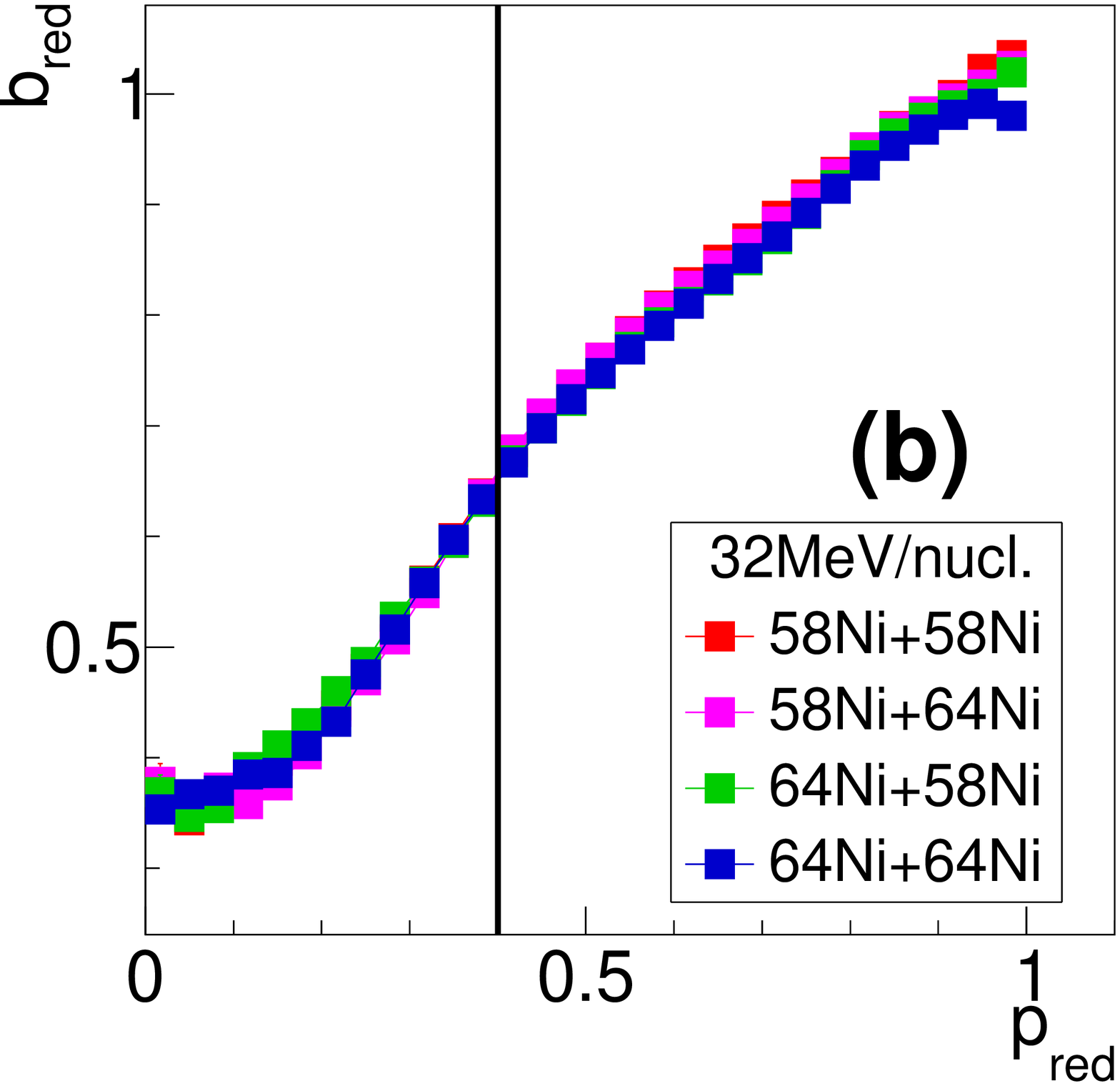}~~~
  \includegraphics[width=0.3\textwidth]{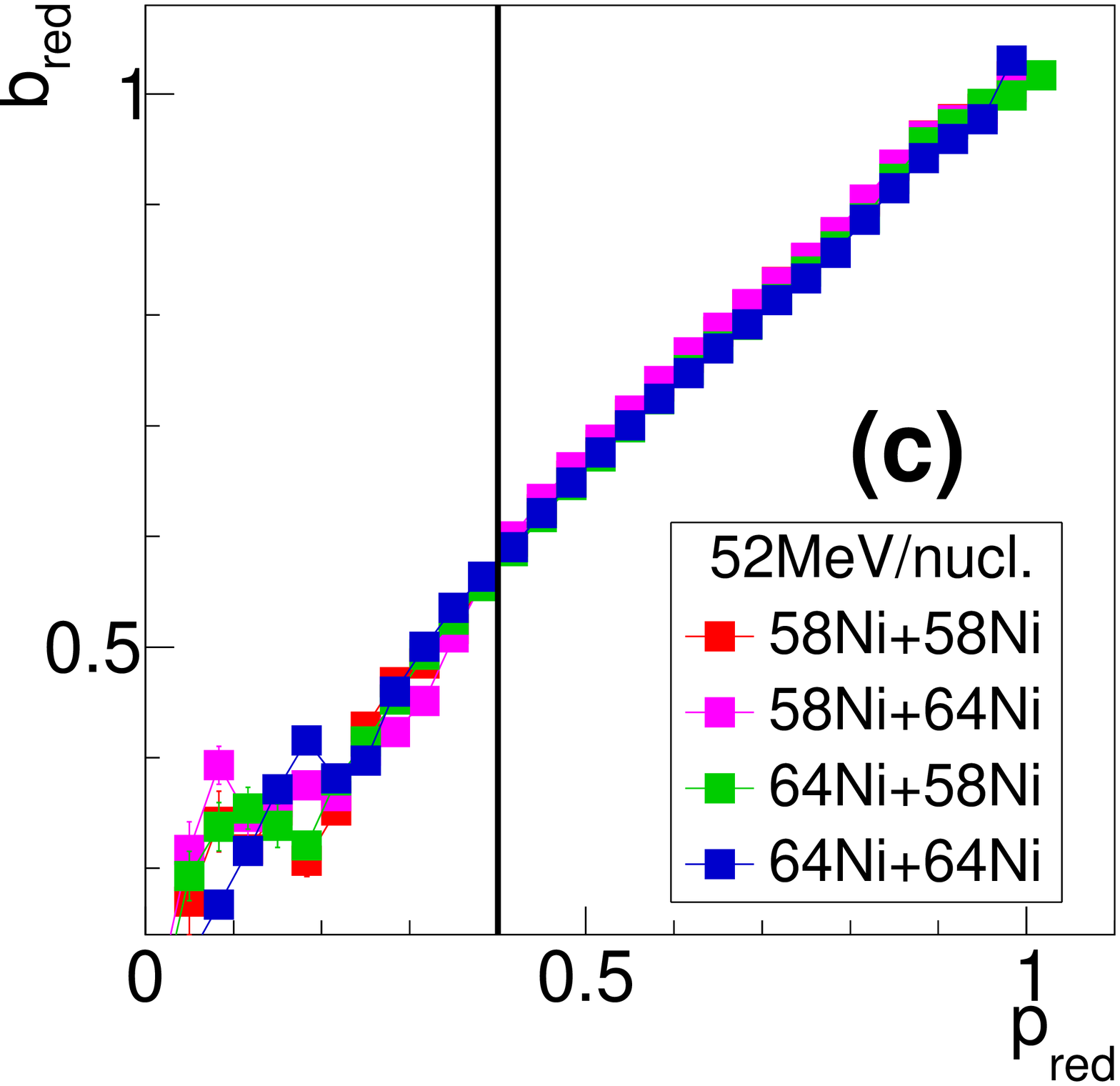}
  \caption{Comparison between the AMD+GEMINI++ predictions of the average $b_{red}$ as a function of $p_{red}$ for different energies (a) and for different systems at $32\,$MeV/nucleon (b) and $52\,$MeV/nucleon (c). Statistical errors are smaller than the marker size.} 
  \label{fig:bred_vs_pred_comparisons}
 \end{figure*}
 Our aim is to follow the isospin equilibration between projectile and target in asymmetric reactions as a function of the collision centrality. Since the impact parameter is not measurable, an experimentally observable order parameter is needed to follow the isospin equilibration as a function of the collision centrality.
 The order parameter employed in this work is the reduced QP momentum \cite{Camaiani2021}:
 \begin{equation}
  p_{red}=\Biggl(\frac{p_z^{QP}}{p_{beam}}\Biggr)_{c.m.}
 \end{equation}
 where $p_z^{QP}$ is the component along the beam axis of the momentum of the fragment identified as QP remnant and $p_{beam}$ is the original projectile momentum, both evaluated in the c.m.~reference frame.
 More peripheral collisions are in fact expected to be associated with lower kinetic energy dissipation and smaller scattering angle of the QP, and therefore to a $p_{red}$ value closer to 1.
 In Ref.~\cite{Camaiani2021}, the correlation of $p_{red}$ with centrality has also been directly tested on the experimental data by inspecting the kinetic energy spectra of LCPs attributed to the QP decay for different $p_{red}$ bins.
 Encouraged by the agreement between the experimental data and the model predictions just observed on the charge and velocity characteristics of the QP remnant, we also exploit the AMD+GEMINI++ simulations to test the behaviour of $p_{red}$ with centrality; the aim is to check whether a different behaviour is found for the two beam energies and to verify the system independence for our projectile-target combinations, which are heavier than those studied in Ref.~\cite{Camaiani2021}.

 Figure~\ref{fig:bred_vs_pred_correlations} shows the correlation between $b_{red}$ (the reduced impact parameter, defined as $b/b_{gr}$, where $b_{gr}$ represents the grazing impact parameter) and $p_{red}$ obtained according to the AMD+GEMINI++ predictions for the QP evaporation events for the reactions $^{58}$Ni+$^{58}$Ni at $32\,$MeV/nucleon (a) and $52\,$MeV/nucleon (b), as examples. Very similar correlations are found for the other reactions. The average $b_{red}$ as a function of $p_{red}$ is also plotted as black symbols. A clear correlation between the two variables is visible in both plots, particularly well defined for more peripheral reactions. 
 Judging from the correlations of Fig.~\ref{fig:bred_vs_pred_correlations}, the $p_{red}$ sorting parameter can be considered reliable (i.e., well correlated with the impact parameter) at least for values $\gtrsim0.4$ (vertical line in the plots, corresponding on average to $b_{red}\gtrsim0.6$). Therefore, $p_{red}$ is well suited for studying semiperipheral and peripheral collisions.
 The $b_{red}$ vs $p_{red}$ correlations obtained for the two bombarding energies are qualitatively similar, except for the lack of statistics for the most dissipative collisions (low $p_{red}$) at $52\,$MeV/nucleon: in fact, for such collisions, there is a higher probability of a multifragmentation-like outcome, with a consequent exclusion from the QP evaporative channel gate, due to the $Z>14$ condition on the QP remnant.
 
 In order to compare the $p_{red}$ behaviour among the different reactions in a more quantitative way, in Fig.~\ref{fig:bred_vs_pred_comparisons}(a) we show the superposition of the average $b_{red}$ as a function of $p_{red}$ obtained according to the AMD+GEMINI++ predictions for the reactions $^{58}$Ni+$^{58}$Ni at the two beam energies: only a relatively small difference can be observed between the two plots.  
 However, the behaviour at a given energy is almost independent of the system, as can be seen in Fig.~\ref{fig:bred_vs_pred_comparisons}(b) (Fig.~\ref{fig:bred_vs_pred_comparisons}(c)) for the reaction at $32\,$MeV/nucleon ($52\,$MeV/nucleon). The fact that each $p_{red}$ bin corresponds to a comparable average impact parameter value for the four reactions at the same energy justifies our use of the isospin transport ratio technique as expressed by eq.~\eqref{eq:imbratio}. 
 Due to its good sorting capability and the similar behaviour obtained for all the inspected reactions, the $p_{red}$ observable will be adopted as an order parameter to highlight the evolution of the isospin equilibration between projectile and target with centrality.
 
 \subsection{Isospin characteristics of the QP remnant}\label{ssec:QPremnant} 
 \begin{figure}[]
  \centering
  \includegraphics[width=0.49\columnwidth]{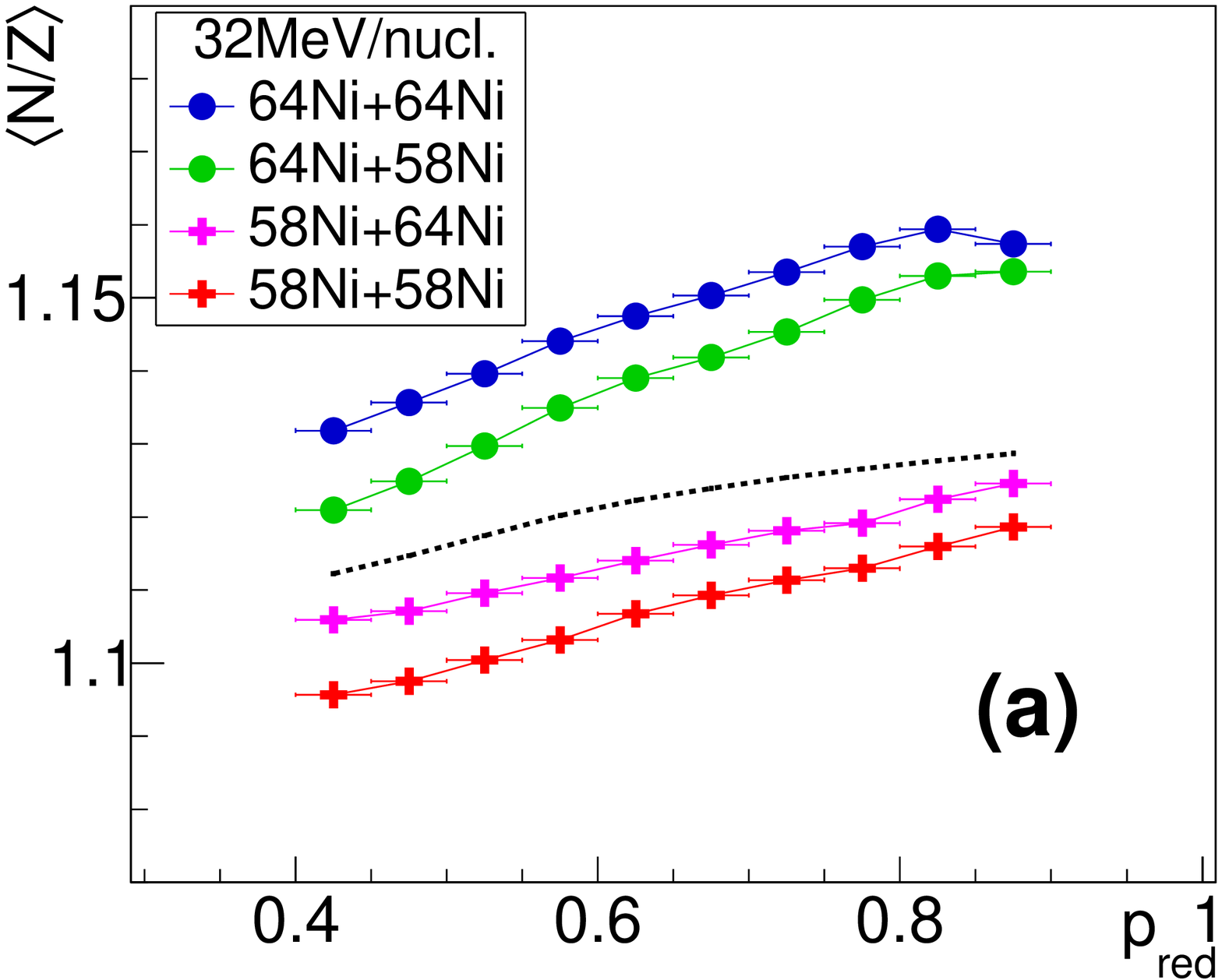}
  \includegraphics[width=0.49\columnwidth]{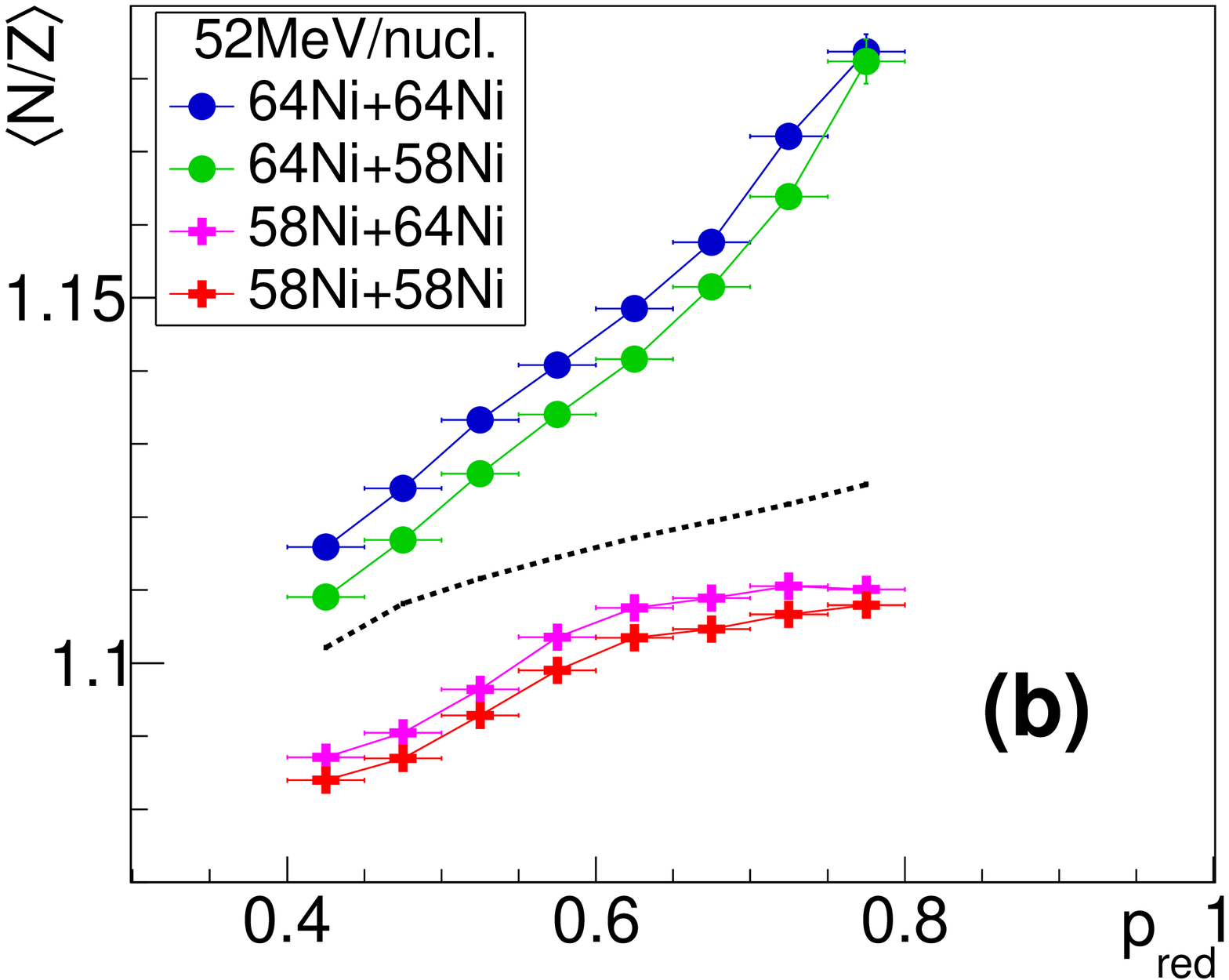} 
  \caption{Experimental neutron to proton ratio $\langle N/Z\rangle$ of the QP evaporation residue as a function of $p_{red}$ for the four reactions at $32\,$MeV/nucleon (a) and at $52\,$MeV/nucleon (b). Statistical errors are plotted on the $y$-axis (generally smaller than the marker size), while the horizontal error bars are set equal to the $p_{red}$ bin width. The black dotted line represents the $\langle N/Z\rangle$ value of the EAL calculated according to the parametrization of Ref.~\cite{Charity1998} assuming the $\langle Z\rangle$ obtained for each $p_{red}$ bin considering all the four systems.}
  \label{fig:NsuZ_pred_QP}
 \end{figure} 
 In this subsection, we exploit the average neutron to proton ratio $\langle N/Z\rangle$ of the QP remnant as isospin related observable to investigate the isospin equilibration between projectile and target as a function of $p_{red}$ \cite{Camaiani2021}.
 For the isospin analysis, the isotopic identification of the QP remnant is required.
 About 55\% (40\%) of the QP evaporation events previously selected  fulfill this requirement for the reactions at $32\,$MeV/nucleon ($52\,$MeV/nucleon).
 The experimental results are reported in Fig.~\ref{fig:NsuZ_pred_QP} for the four reactions at $32\,$MeV/nucleon (a) and at $52\,$MeV/nucleon (b). In the latter case, the four $\langle N/Z\rangle$ plots extend only up to lower $p_{red}$ values than for the reactions at $32\,$MeV/nucleon: at the higher beam energy, more QP residues are stopped in FAZIA CsI detectors and are therefore identified by exploiting the Si2-CsI $\Delta$E-E correlation. This technique features a worse mass identification with respect to the Si1-Si2 correlation (predominantly used at the lower beam energy) and hence it does not allow for the isotopic identification of heavy QP residues, related to high $p_{red}$ values.
 However, at both energies, the QP remnants produced in the reactions induced by $^{64}$Ni (full circles) feature a $\langle N/Z\rangle$ sensibly lower than that of the original projectile (1.29) and, in the asymmetric reaction, also lower than the average value of the whole system (1.18). For the reactions induced by $^{58}$Ni (full crosses), the $\langle N/Z\rangle$ of the QP residue is generally slightly larger than that of the projectile (1.07), albeit always lower than that obtained with the neutron rich projectile. Moreover, a declining trend of the isospin content with decreasing $p_{red}$, i.e., increasing centrality, can be generally noticed, more evident for the reactions induced by the neutron rich projectile. 
 These features, quite similar to those reported in Fig.~5 of Ref.~\cite{Camaiani2021}, can be mainly ascribed to the statistical emission. In fact, for increasing collision centrality, more energy is dissipated in the internal degrees of freedom, resulting in the production of more excited fragments. As pointed out in Ref.~\cite{Charity1998}, the evaporation of these hot fragments decreases their atomic and mass number, and tends to move the residues towards a locus of the nuclide chart known as Evaporation Attractor Line (EAL); the behaviour of $\langle N/Z\rangle$ of the QP remnant with increasing centrality reflects that of the EAL with decreasing atomic number $Z$. As a reference, the $\langle N/Z\rangle$ value of the EAL prediction calculated on the $\langle Z\rangle$, averaged over all systems, has been drawn in Fig.~\ref{fig:NsuZ_pred_QP} (black dotted line).
 
 However, despite the effect of statistical emission, the data clearly show the phenomenon of isospin diffusion. In fact, at both energies, a clear ordering of the four systems is respected: for each $p_{red}$ value, the $\langle N/Z\rangle$ of the QP remnants produced in the asymmetric reactions are located between those of the two symmetric reactions.
 For the most peripheral collisions the isospin content of the QP residues produced in the reactions with the same projectile and different targets is quite similar, 
 but a gap between the two develops and increases with centrality, more clearly visible for the lower beam energy. This gap due to the only change in the target isospin is a net consequence of the isospin diffusion, although with a sensitivity reduced by the evaporative decay.
 Moreover, the $\langle N/Z\rangle$ values for the asymmetric systems tend to approach each other, revealing the trend toward isospin equilibrium between projectile and target, with a higher degree of equilibration achieved for more central collisions. 
 
\begin{figure}
  \centering
  \includegraphics[width=0.6\columnwidth]{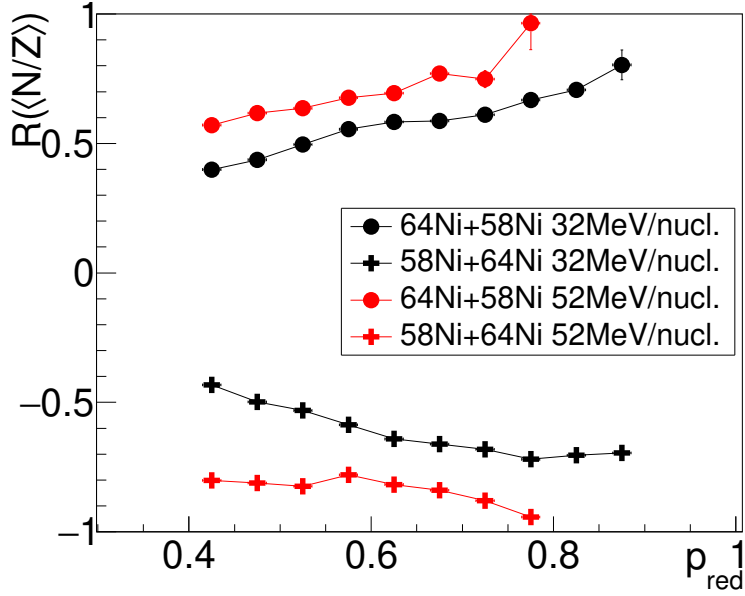}
  \caption{Isospin transport ratio calculated with the $\langle N/Z\rangle$ of the QP evaporation residue as a function of $p_{red}$ (experimental data). The results for the asymmetric reactions $^{64}$Ni+$^{58}$Ni and $^{58}$Ni+$^{64}$Ni at $32\,$MeV/nucleon ($52\,$MeV/nucleon) are plotted as black (red) symbols. The vertical error bars, often smaller than the marker size, correspond to statistical errors.}
  \label{fig:imbratio_pred_QP}
 \end{figure}

 As said, the details of the isospin equilibration can be further inspected by exploiting the isospin transport ratio technique \cite{Rami2000}. In Fig.~\ref{fig:imbratio_pred_QP}, the isospin transport ratio, as defined by eq.~\eqref{eq:imbratio}, evaluated for $X\equiv\langle N/Z\rangle$ of the QP remnant is reported as a function of $p_{red}$. The results for the reactions at $32\,$MeV/nucleon ($52\,$MeV/nucleon) are plotted as black (red) symbols. For both energies, the same symbols are associated with the different projectiles as in Fig.~\ref{fig:NsuZ_pred_QP}.
 For the most peripheral reactions, the obtained values of $R(\langle N/Z\rangle)$ are close to $\pm1$, with an absolute value always smaller than unity: in fact, since mass identification is achieved only up to $Z\sim24-25$, the events featuring the heaviest QP remnants, generally resulting from the most peripheral reactions, are not included. 
 For increasing centrality, the two $R(\langle N/Z\rangle)$ branches are driven towards each other, an evidence of evolution towards isospin equilibration.
 The full equilibration condition $R(\langle N/Z\rangle_{AB})=R(\langle N/Z\rangle_{BA})$ is not reached, but we remind that we are not investigating the most central collisions ($p_{red}<0.4$).
 The behaviour of the ratio as a function of $p_{red}$ shows a very clear and regular trend, smoother at $32\,$MeV/nucleon.
 These findings are in general agreement with observations reported in the literature at comparable energies \cite{Tsang2004,Keksis2010,May2018,Sun2010}.
 Moreover, a different degree of isospin equilibration is achieved for the reactions at the two bombarding energies, with the two $R(\langle N/Z\rangle)$ branches closer to each other at $32\,$MeV/nucleon with respect to those obtained at $52\,$MeV/nucleon. In fact, a stronger effect of isospin equilibration in the reactions at $32\,$MeV/nucleon than at $52\,$MeV/nucleon could be expected, taking into account, for instance, longer projectile-target interaction times at lower beam energy.  
 
 \begin{figure}
  \centering
  \includegraphics[width=0.6\columnwidth]{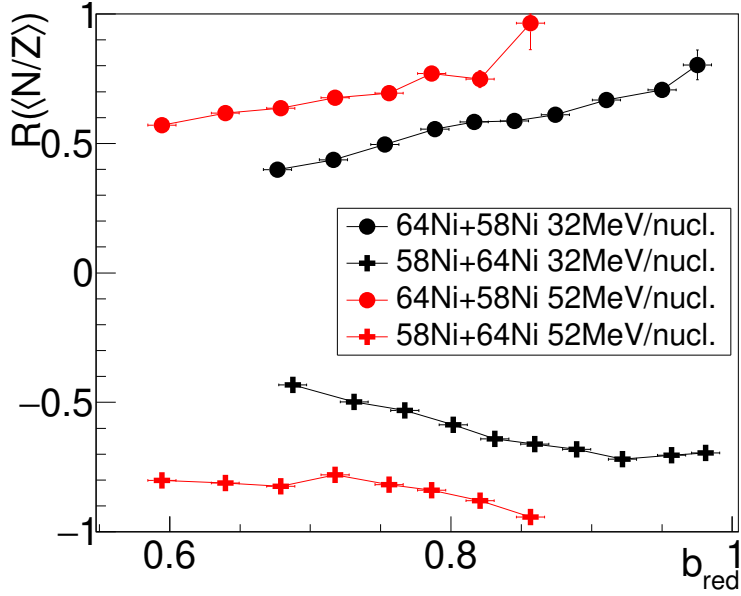}
  \caption{Isospin transport ratio calculated with the $\langle N/Z\rangle$ of the QP evaporation residue, reported as a function of $b_{red}$, obtained from the plot in Fig.~\ref{fig:imbratio_pred_QP} by rescaling $p_{red}$ into $b_{red}$ exploiting their correlation obtained within AMD+GEMINI++. The results for the asymmetric reactions $^{64}$Ni+$^{58}$Ni and $^{58}$Ni+$^{64}$Ni at $32\,$MeV/nucleon ($52\,$MeV/nucleon) are plotted as black (red) symbols. The vertical error bars, often smaller than the marker size, correspond to statistical errors.}
  \label{fig:imbratio_bred_QP}
 \end{figure}
 The plots presented in Fig.~\ref{fig:imbratio_pred_QP} are model independent since they are obtained exploiting only the information provided by the experimental data.
 However, as already evidenced, within the AMD+GEMINI++ predictions a slightly different behaviour of $p_{red}$ with the reaction centrality is found for the two bombarding energies. In order to take into account the possible role of this difference on the experimental data interpretation, in Fig.~\ref{fig:imbratio_bred_QP} we present the same isospin transport ratio plots of Fig.~\ref{fig:imbratio_pred_QP} after rescaling the $x$-axis from the $p_{red}$ variable to $b_{red}$. The rescaling has been performed by exploiting the AMD+GEMINI++ predictions on the $b_{red}$ vs $p_{red}$ correlations (see Fig.~\ref{fig:bred_vs_pred_comparisons}). In Fig.~\ref{fig:imbratio_bred_QP}, we can compare the degree of equilibration for the same (or a similar) impact parameter: also after the rescaling operation, it appears that the equilibrium condition is more closely approached in the reactions at $32\,$MeV/nucleon than at $52\,$MeV/nucleon, and such observation emerges much clearer than just as a function of $p_{red}$.
 
 So far, no distinction has been made on the basis of the particles detected in coincidence with the QP remnant, which, in a large majority of events, are only LCPs. However, we also checked whether a different behaviour is found for the QP evaporation events in which at least an IMF ($Z=3,4$) is found in coincidence with the QP remnant; the main observations are here described without showing further graphs to avoid redundancy. 
 In the subset of QP evaporation events with at least an IMF (about 10-15\% of the total events in the evaporation class), the $\langle N/Z\rangle$ of the QP remnant is slightly lower than that obtained for the more populated subclass, i.e., the one with only LCPs in coincidence with the QP. 
 A possible interpretation of this effect lies in the occurrence of some isospin drift: in fact, the accompanying IMF, which, as expected, is generally emitted at midvelocity, is likely to be generated from the neck region, showing its characteristic neutron richness and leaving behind a more neutron deficient QP remnant. As expected, the neutron enrichment of the backward emitted IMFs has been observed in this dataset \cite{Ciampi2021}. Alternatively, the reason for the lower $\langle N/Z\rangle$ for the QP remnants with coincident IMF can be that the IMF sets a bias towards slightly more excited events, for a given $p_{red}$.
 Independently of the origin, however, by exploiting the isospin transport ratio technique, the same results are obtained for both subclasses of the QP evaporation selection. This confirms the capability of the method to isolate the equilibrating action of the isospin diffusion between projectile and target, bypassing those effects acting similarly in the four reactions.
 As for the isospin drift, a more specific investigation is planned in the near future with a dedicated analysis.

 \subsection{Isospin characteristics of the QP ejectiles}\label{ssec:QPejectiles}
 The rich information collected by the INDRA-FAZIA apparatus also allows us to inspect the isospin equilibration phenomenon from a complementary point of view with respect to that presented in the previous section. In fact, also the lighter ejectiles (LCPs and IMFs) convey rich information on the isospin dynamics related to the collision event. In particular, since the isospin characteristics of the decay products of the QP are related to its isotopic composition, they can be used as tracers for the isospin diffusion between the two reaction partners. This is the approach that has been adopted in the past by various authors \cite{Lombardo2010,Galichet2009,Keksis2010}.
 
 As commonly done in the literature \cite{Piantelli2021,Theriault2006}, we study the QP decay emissions by considering the particles forward emitted with respect to the QP remnant (collected by FAZIA and by the first rings of INDRA); in fact, this selection can be expected to feature less contamination from dynamical contributions (e.g., neck emission) and from the QT statistical evaporation with respect to the backward emission.
 Various isospin sensitive observables can be built to inspect the characteristics of the light emissions, such as their $\langle N/Z\rangle$ or selected isobaric and isotopic yield ratios: we studied the behaviour of some of them, obtaining results leading to similar observations \cite{Ciampi2021}.
 For brevity, in the following, we show the results for one observable, the isospin ratio for complex particles, defined as \cite{Galichet2009}:
 \begin{equation}
  \langle N\rangle/\langle Z\rangle_{CP} =
  \sum_{i}\sum_\nu N_\nu^i / \sum_{i}\sum_\nu Z_\nu^i
 \end{equation}
 where the $\nu$ index numbers the different complex particles in the $i$-th event, the outer sum runs over all the events in the selected $p_{red}$ bin. Free protons are excluded (as well as free neutrons, which are not detected), while $d$, $t$, $^3$He, $^4$He, $^6$He, $^6$Li, $^7$Li, $^8$Li, $^9$Li, $^7$Be, $^9$Be and $^{10}$Be are taken into account as complex particles. $N_\nu^i$ ($Z_\nu^i$) is the number of neutrons (protons) bound in the $\nu$-th complex particle forward emitted with respect to the QP. 
 
 \begin{figure}[]
  \centering
  \includegraphics[width=0.49\columnwidth]{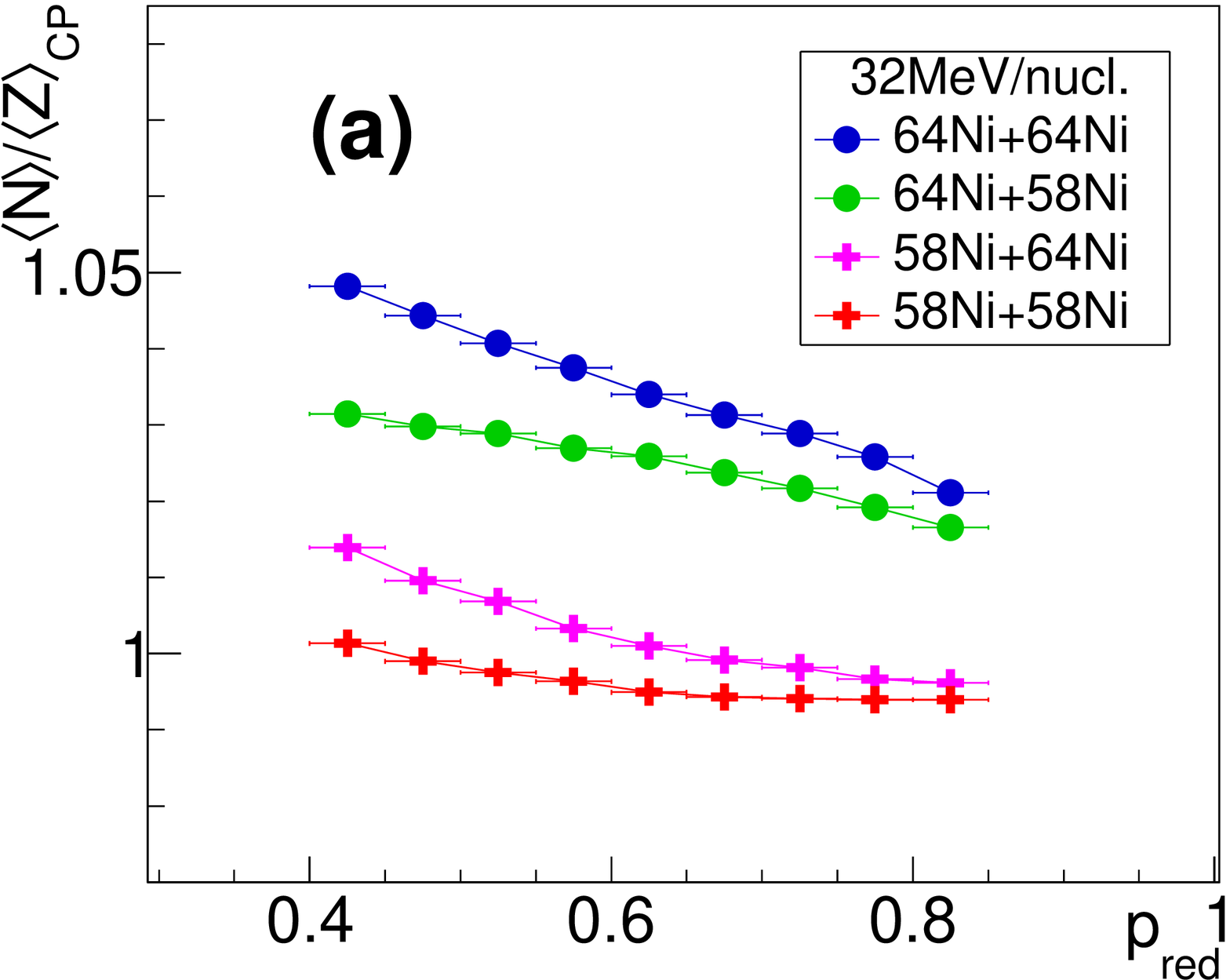}
  \includegraphics[width=0.49\columnwidth]{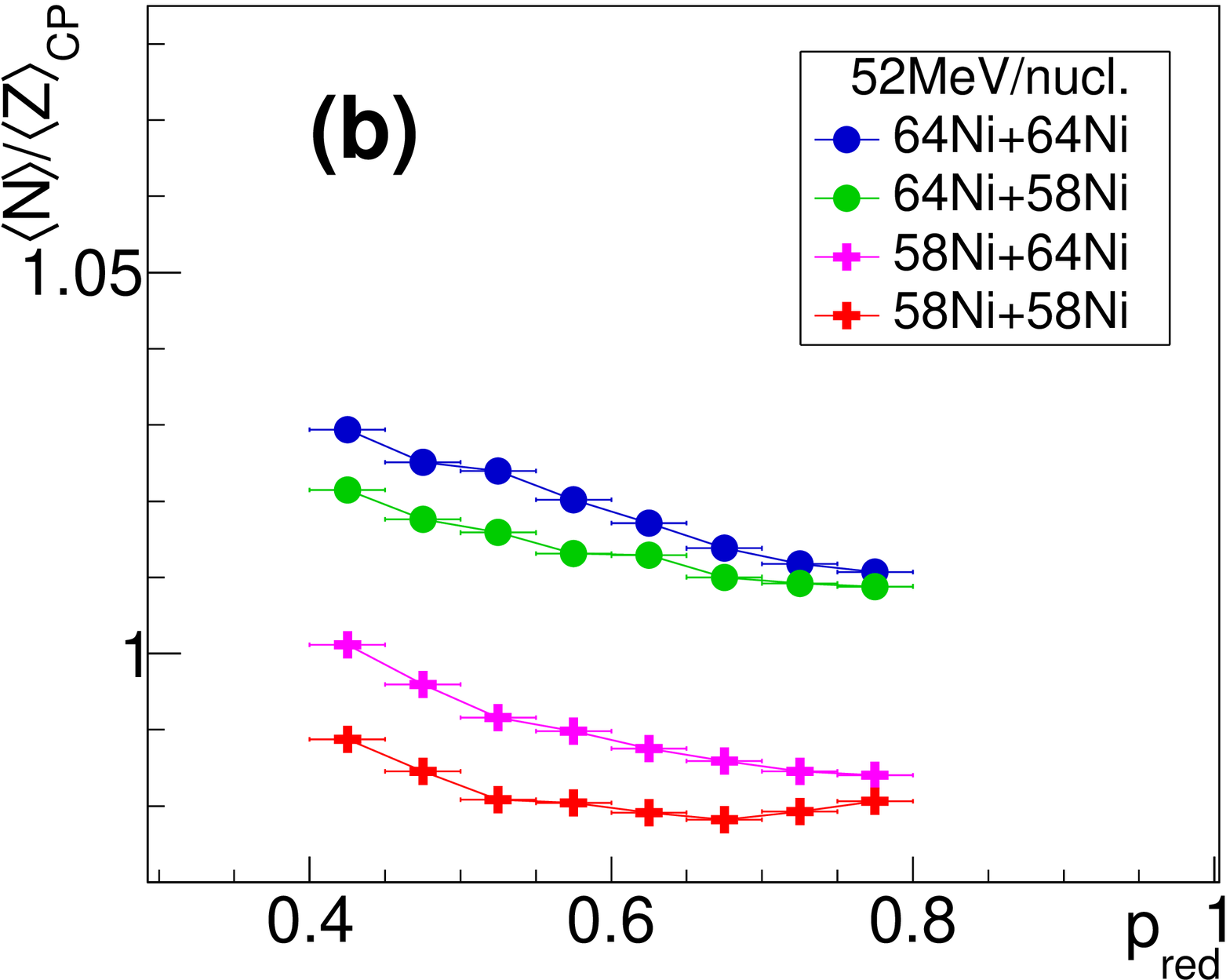}
  \caption{Experimental neutron to proton ratio $\langle N\rangle/\langle Z\rangle_{CP}$ of the complex particles (see text) forward emitted with respect to the QP remnant, as a function of $p_{red}$ for the four reactions at $32\,$MeV/nucleon (a) and at $52\,$MeV/nucleon (b). Statistical errors are plotted on the $y$-axis (generally smaller than the marker size), while the horizontal error bars are set equal to the $p_{red}$ bin width.}
  \label{fig:NsuZ_pred_Z1}
 \end{figure} 
 Figure~\ref{fig:NsuZ_pred_Z1} shows the $\langle N\rangle/\langle Z\rangle_{CP}$ of the complex particles forward emitted with respect to the QP for the four reactions at $32\,$MeV/nucleon (a) and at $52\,$MeV/nucleon (b), as a function of $p_{red}$.
 We point out that our result for the system $^{58}$Ni+$^{58}$Ni at $52\,$MeV/nucleon is quite comparable to what has been obtained for the same reaction in Ref.~\cite{Galichet2009b}.
 Complementarily to the behaviour observed for the isospin of the QP remnant, in all systems a smoothly increasing neutron content is found for forward emissions with increasing centrality. A similar behaviour is also obtained for other observables that include the contribution of free protons.
 This observation is consistent with the assumption that the trend found in the plots of Fig.~\ref{fig:NsuZ_pred_QP} is certainly affected by the role of statistical decay. 
 Furthermore, in Fig.~\ref{fig:NsuZ_pred_Z1}, the ordering of the plots for the four systems is again respected at both energies. 
 The decay products of the QP emerging from the reactions induced by the same projectile on different targets (indicated by the same marker shapes in Fig.~\ref{fig:NsuZ_pred_Z1}) feature an evidently different chemical composition, with a wider $\langle N\rangle/\langle Z\rangle_{CP}$ gap for more central collisions: such difference is due to a different composition of the primary QP, produced by the action of the isospin transport process during the contact between the two reaction partners. As a result, in Fig.~\ref{fig:NsuZ_pred_Z1}, the trend of the two asymmetric systems towards isospin equilibration for more damped collisions (lower $p_{red}$) is clear, particularly for the reactions at $32\,$MeV/nucleon.
 Such a clean evidence of isospin diffusion also suggests that the possible contribution of prompt emission within this selection of light fragments is small (as could be expected since mostly neutrons and protons, here excluded, contribute to such an emission); indeed, such a contribution of early emitted particles could wash out the effect of the isospin equilibration process.
 
\begin{figure}
  \centering
  \includegraphics[width=0.6\columnwidth]{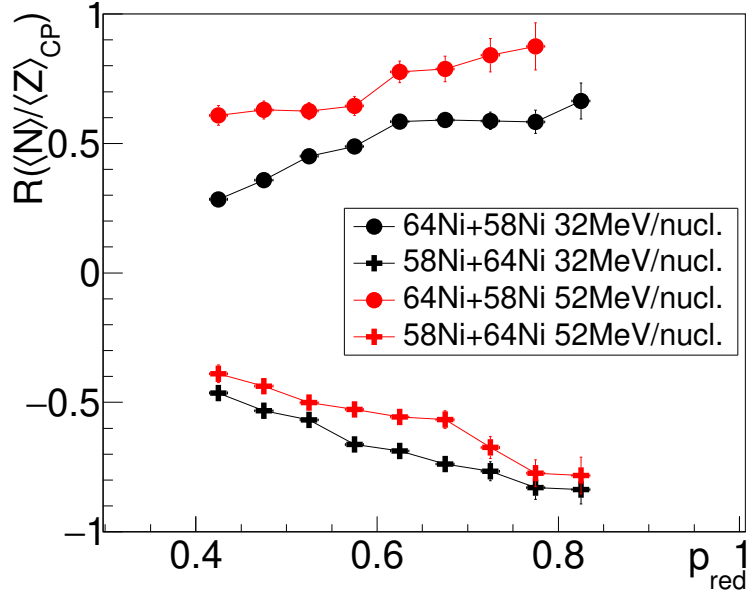}
  \caption{Isospin transport ratio calculated with the $\langle N\rangle/\langle Z\rangle_{CP}$ of the complex particles (see text) forward emitted with respect to the QP remnant, as a function of $p_{red}$ (experimental data). The results for the asymmetric reactions $^{64}$Ni+$^{58}$Ni and $^{58}$Ni+$^{64}$Ni at $32\,$MeV/nucleon ($52\,$MeV/nucleon) are plotted as black (red) symbols. The vertical error bars, often smaller than the marker size, correspond to statistical errors.}
  \label{fig:imbratio_pred_Z1} 
 \end{figure}

 The $\langle N\rangle/\langle Z\rangle_{CP}$ can also be exploited as an alternative probe to evaluate the isospin transport ratio as a function of $p_{red}$: the result is shown in Fig.~\ref{fig:imbratio_pred_Z1} for both asymmetric reactions (drawn with different markers) at both beam energies ($32\,$MeV/nucleon as black symbols, $52\,$MeV/nucleon as red symbols), and we note the strong similarity with the complementary QP remnant results in Fig.~\ref{fig:imbratio_pred_QP}. 
 The effect of isospin diffusion, driving the two systems towards equilibration for more central collisions, is evident. 
 By careful comparison of Fig.~\ref{fig:imbratio_pred_Z1} and Fig.~\ref{fig:imbratio_pred_QP}, we also notice, common features aside, a slightly different trend of the isospin transport ratio as a function of $p_{red}$.
 The fact that different isospin probes can produce different behaviours of the isospin ratio as a function of centrality has already been observed in Ref.~\cite{May2018}.
 Also by exploiting this different isospin observable, the higher degree of isospin equilibration achieved for the reactions at $32\,$MeV/nucleon is confirmed, being the two branches of $R(\langle N\rangle/\langle Z\rangle_{CP})$ closer to each other than at the higher bombarding energy: the effect is more evident for less peripheral collisions.  
 
\begin{figure}
  \centering
  \includegraphics[width=0.6\columnwidth]{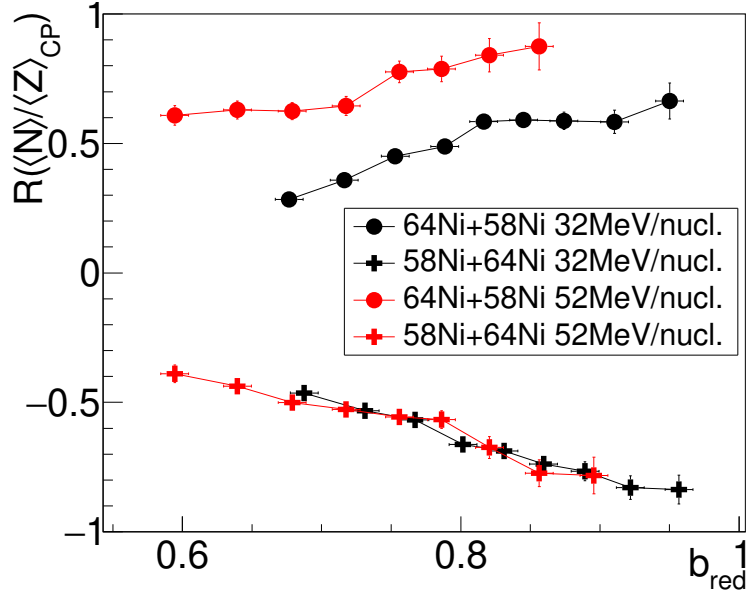}
  \caption{Isospin transport ratio calculated with the $\langle N\rangle/\langle Z\rangle_{CP}$ of the complex particles (see text) forward emitted with respect to the QP remnant, as a function of $b_{red}$, obtained from the plot of Fig.~\ref{fig:imbratio_pred_Z1} by rescaling $p_{red}$ into $b_{red}$ exploiting their correlation obtained within AMD+GEMINI++. The results for the asymmetric reactions $^{64}$Ni+$^{58}$Ni and $^{58}$Ni+$^{64}$Ni at $32\,$MeV/nucleon ($52\,$MeV/nucleon) are plotted as black (red) symbols. The vertical error bars, often smaller than the marker size, correspond to statistical errors.}
  \label{fig:imbratio_bred_Z1}
 \end{figure}

 As done for the study of the isospin content of the QP remnant in the previous paragraph, we also present the same isospin transport ratio $R(\langle N\rangle/\langle Z\rangle_{CP})$ as a function of $b_{red}$ in Fig.~\ref{fig:imbratio_bred_Z1}, again obtained from Fig.~\ref{fig:imbratio_pred_Z1} by an $x$-axis rescaling based on AMD+GEMINI++. After the rescaling, the stronger isospin equilibration effect on the reactions at lower beam energy can be more easily noticed.
 
 In conclusion, all of the plots here presented point towards similar conclusions concerning the process of isospin diffusion between asymmetric projectile and target.
 To our knowledge, such rich and regular complementary results for the isospin transport ratio based on both the $\langle N/Z\rangle$ of the QP remnant and of the isospin content of the evaporated particles are not common to find in the literature.

\section{Summary and conclusions}
We have presented an analysis of the first experimental data collected with the coupled INDRA-FAZIA apparatus: the four reactions $^{58,64}$Ni+$^{58,64}$Ni have been investigated at two incident energies, $32\,$MeV/nucleon and $52\,$MeV/nucleon, to study
the isospin diffusion occurring during the interaction between asymmetric projectile and target in two supposedly different dynamics conditions related to two beam energies.
With the available dataset, we have been able to compare the properties of the products of the two asymmetric reactions with those of both the neutron rich and the neutron poorer symmetric systems; we exploited the isospin transport ratio technique to stress the signature of the isospin equilibration with respect to other effects, such as fast emissions and statistical deexcitation of the primary products, acting similarly in the four systems but possibly spoiling the genuine sought after effect. 

The present analysis focuses on the peripheral and semi-peripheral events resulting in a binary exit channel. 
We have selected the QP evaporation channel by requiring the presence of a single heavy fragment accompanied only by LCPs and IMFs.
Though the main scope of this work is not an extensive comparison of the model predictions with the experimental data, the charge and velocity characteristics of the QP remnant resulting from the AMD+GEMINI++ simulations, filtered with a software replica of the experimental apparatus, have been compared to the QP remnant in the collected events.
The AMD+GEMINI++ simulated data have also been exploited to verify the  reliability of the reduced QP momentum $p_{red}$ as a centrality estimator. Within the model, $p_{red}$ results to be well correlated with the impact parameter, almost independent of the reaction and only slightly dependent on the beam energy.

Thanks to the wealth of information on each event collected by the INDRA-FAZIA apparatus, the isospin equilibration has been followed by means of the average neutron to proton ratio of the QP remnant, obtained from its direct identification, and of the chemical composition of the complex particles (LCPs and IMFs with $A>1$) forward emitted with respect to the QP residue, as messengers of the preceding degree of equilibration achieved just at the end of the interaction. 
Both these observables have been exploited to build the isospin transport ratio, shown as a function of $p_{red}$ and also as a function of $b_{red}$, after rescaling the $x$-axis exploiting the AMD+GEMINI++ predictions for the $b_{red}$ vs $p_{red}$ correlations. A clear evidence of the relaxation of the projectile-target isospin asymmetry towards the equilibrium for increasing reaction centrality has been found equally signaled by both the selected observables. 
Moreover, a stronger trend towards isospin equilibration is observed at the lower bombarding energy. 
This finding, in agreement with previous works at comparable beam energies, can be qualitatively justified in terms of the longer timescales for the lower energy case, allowing for a more effective diffusion process.
The observations here reported refer to the QP evaporation channel, which, according to the model, for the considered $b_{red}$ interval  constitutes the majority of the dataset selected by the preliminary conditions. However, we aim to complete the study by further extending the isospin diffusion analysis to other output channels (e.g., the QP breakup channel for which similar signals have already been evidenced for this dataset \cite{Ciampi2021}).

The analysis presented in this work is essentially model independent, with the exception of the isospin transport ratio plots reported as a function of $b_{red}$. However, we aim to further investigate the simulated dataset in order to extract more precise information on the dynamical processes taking place in the selected events. In this respect, the good quality data presented here are a bench test for further theoretical investigations in terms of sophisticated reaction models through which information on the details of the NEoS and/or the relevant timescales involved in the microscopic processes could be extracted.

The rich information of the INDRA-FAZIA E789 dataset is far from drained, and the analysis is still in progress with the investigation of other topics, such as, for instance, the study of the QP breakup and the particle correlations. 
Finally, the performance achieved in the experiment presented in this paper confirms the excellent capabilities expected from the INDRA-FAZIA experimental apparatus and is promising in view of the forthcoming experiments planned also at higher energies.

\begin{acknowledgments}
This work was partially supported by the National Research Foundation of Korea (NRF; Grant No. 2018R1A5A1025563) and by the Spanish Ministerio de Economía y Empresa (PGC2018-096994-B-C22).
We acknowledge support from Région Normandie under Réseau d’Intérêt Normand FIDNEOS (RIN/FIDNEOS).
Many thanks are due to the accelerator staff of GANIL for delivering a very good quality beam and to the technical staff for the continuous support.
\end{acknowledgments}


\bibliography{E789_QP_iso_equil}

\providecommand{\noopsort}[1]{}
\begin{thebibliography}{51}%
\makeatletter
\providecommand \@ifxundefined [1]{%
 \@ifx{#1\undefined}
}%
\providecommand \@ifnum [1]{%
 \ifnum #1\expandafter \@firstoftwo
 \else \expandafter \@secondoftwo
 \fi
}%
\providecommand \@ifx [1]{%
 \ifx #1\expandafter \@firstoftwo
 \else \expandafter \@secondoftwo
 \fi
}%
\providecommand \natexlab [1]{#1}%
\providecommand \enquote  [1]{``#1''}%
\providecommand \bibnamefont  [1]{#1}%
\providecommand \bibfnamefont [1]{#1}%
\providecommand \citenamefont [1]{#1}%
\providecommand \href@noop [0]{\@secondoftwo}%
\providecommand \href [0]{\begingroup \@sanitize@url \@href}%
\providecommand \@href[1]{\@@startlink{#1}\@@href}%
\providecommand \@@href[1]{\endgroup#1\@@endlink}%
\providecommand \@sanitize@url [0]{\catcode `\\12\catcode `\$12\catcode
  `\&12\catcode `\#12\catcode `\^12\catcode `\_12\catcode `\%12\relax}%
\providecommand \@@startlink[1]{}%
\providecommand \@@endlink[0]{}%
\providecommand \url  [0]{\begingroup\@sanitize@url \@url }%
\providecommand \@url [1]{\endgroup\@href {#1}{\urlprefix }}%
\providecommand \urlprefix  [0]{URL }%
\providecommand \Eprint [0]{\href }%
\providecommand \doibase [0]{https://doi.org/}%
\providecommand \selectlanguage [0]{\@gobble}%
\providecommand \bibinfo  [0]{\@secondoftwo}%
\providecommand \bibfield  [0]{\@secondoftwo}%
\providecommand \translation [1]{[#1]}%
\providecommand \BibitemOpen [0]{}%
\providecommand \bibitemStop [0]{}%
\providecommand \bibitemNoStop [0]{.\EOS\space}%
\providecommand \EOS [0]{\spacefactor3000\relax}%
\providecommand \BibitemShut  [1]{\csname bibitem#1\endcsname}%
\let\auto@bib@innerbib\@empty
\bibitem [{\citenamefont {Li}\ \emph {et~al.}(1998)\citenamefont {Li},
  \citenamefont {Ko},\ and\ \citenamefont {Bauer}}]{Li1998}%
  \BibitemOpen
  \bibfield  {author} {\bibinfo {author} {\bibfnamefont {B.-A.}\ \bibnamefont
  {Li}}, \bibinfo {author} {\bibfnamefont {C.~M.}\ \bibnamefont {Ko}},\ and\
  \bibinfo {author} {\bibfnamefont {W.}~\bibnamefont {Bauer}},\ }\href
  {https://doi.org/10.1142/S0218301398000087} {\bibfield  {journal} {\bibinfo
  {journal} {Int. J. Mod. Phys. E}\ }\textbf {\bibinfo {volume} {07}},\
  \bibinfo {pages} {147} (\bibinfo {year} {1998})}\BibitemShut {NoStop}%
\bibitem [{\citenamefont {{Di Toro}}\ \emph {et~al.}(2010)\citenamefont {{Di
  Toro}}, \citenamefont {Baran}, \citenamefont {Colonna},\ and\ \citenamefont
  {Greco}}]{DiToro2010}%
  \BibitemOpen
  \bibfield  {author} {\bibinfo {author} {\bibfnamefont {M.}~\bibnamefont {{Di
  Toro}}}, \bibinfo {author} {\bibfnamefont {V.}~\bibnamefont {Baran}},
  \bibinfo {author} {\bibfnamefont {M.}~\bibnamefont {Colonna}},\ and\ \bibinfo
  {author} {\bibfnamefont {V.}~\bibnamefont {Greco}},\ }\href
  {https://doi.org/10.1088/0954-3899/37/8/083101} {\bibfield  {journal}
  {\bibinfo  {journal} {J. Phys. G: Nucl. Part. Phys.}\ }\textbf {\bibinfo
  {volume} {37}},\ \bibinfo {pages} {083101} (\bibinfo {year}
  {2010})}\BibitemShut {NoStop}%
\bibitem [{\citenamefont {\L{}ukasik}\ \emph {et~al.}(1997)\citenamefont
  {\L{}ukasik}, \citenamefont {Benlliure}, \citenamefont {M\'etivier},
  \citenamefont {Plagnol}, \citenamefont {Tamain}, \citenamefont {Assenard},
  \citenamefont {Auger}, \citenamefont {Bacri}, \citenamefont {Bisquer},
  \citenamefont {Borderie}, \citenamefont {Bougault}, \citenamefont {Brou},
  \citenamefont {Buchet}, \citenamefont {Charvet}, \citenamefont {Chbihi},
  \citenamefont {Colin}, \citenamefont {Cussol}, \citenamefont {Dayras},
  \citenamefont {Demeyer}, \citenamefont {Dor\'e}, \citenamefont {Durand},
  \citenamefont {Gerlic}, \citenamefont {Germain}, \citenamefont {Gourio},
  \citenamefont {Guinet}, \citenamefont {Lautesse}, \citenamefont {Laville},
  \citenamefont {Lecolley}, \citenamefont {Le~F\`evre}, \citenamefont {Lefort},
  \citenamefont {Legrain}, \citenamefont {Lopez}, \citenamefont {Louvel},
  \citenamefont {Marie}, \citenamefont {Nalpas}, \citenamefont {Parlog},
  \citenamefont {P\'eter}, \citenamefont {Politi}, \citenamefont {Rahmani},
  \citenamefont {Reposeur}, \citenamefont {Rivet}, \citenamefont {Rosato},
  \citenamefont {Saint-Laurent}, \citenamefont {Squalli}, \citenamefont
  {Steckmeyer}, \citenamefont {Stern}, \citenamefont {Tassan-Got},
  \citenamefont {Vient}, \citenamefont {Volant}, \citenamefont {Wieleczko},
  \citenamefont {Colonna}, \citenamefont {Haddad}, \citenamefont {Eudes},
  \citenamefont {Sami},\ and\ \citenamefont {Sebille}}]{Lukasik1997}%
  \BibitemOpen
  \bibfield  {author} {\bibinfo {author} {\bibfnamefont {J.}~\bibnamefont
  {\L{}ukasik}}, \bibinfo {author} {\bibfnamefont {J.}~\bibnamefont
  {Benlliure}}, \bibinfo {author} {\bibfnamefont {V.}~\bibnamefont
  {M\'etivier}}, \bibinfo {author} {\bibfnamefont {E.}~\bibnamefont {Plagnol}},
  \bibinfo {author} {\bibfnamefont {B.}~\bibnamefont {Tamain}}, \bibinfo
  {author} {\bibfnamefont {M.}~\bibnamefont {Assenard}}, \bibinfo {author}
  {\bibfnamefont {G.}~\bibnamefont {Auger}}, \bibinfo {author} {\bibfnamefont
  {C.~O.}\ \bibnamefont {Bacri}}, \bibinfo {author} {\bibfnamefont
  {E.}~\bibnamefont {Bisquer}}, \bibinfo {author} {\bibfnamefont
  {B.}~\bibnamefont {Borderie}}, \bibinfo {author} {\bibfnamefont
  {R.}~\bibnamefont {Bougault}}, \bibinfo {author} {\bibfnamefont
  {R.}~\bibnamefont {Brou}}, \bibinfo {author} {\bibfnamefont {P.}~\bibnamefont
  {Buchet}}, \bibinfo {author} {\bibfnamefont {J.~L.}\ \bibnamefont {Charvet}},
  \bibinfo {author} {\bibfnamefont {A.}~\bibnamefont {Chbihi}}, \bibinfo
  {author} {\bibfnamefont {J.}~\bibnamefont {Colin}}, \bibinfo {author}
  {\bibfnamefont {D.}~\bibnamefont {Cussol}}, \bibinfo {author} {\bibfnamefont
  {R.}~\bibnamefont {Dayras}}, \bibinfo {author} {\bibfnamefont
  {A.}~\bibnamefont {Demeyer}}, \bibinfo {author} {\bibfnamefont
  {D.}~\bibnamefont {Dor\'e}}, \bibinfo {author} {\bibfnamefont
  {D.}~\bibnamefont {Durand}}, \bibinfo {author} {\bibfnamefont
  {E.}~\bibnamefont {Gerlic}}, \bibinfo {author} {\bibfnamefont
  {S.}~\bibnamefont {Germain}}, \bibinfo {author} {\bibfnamefont
  {D.}~\bibnamefont {Gourio}}, \bibinfo {author} {\bibfnamefont
  {D.}~\bibnamefont {Guinet}}, \bibinfo {author} {\bibfnamefont
  {P.}~\bibnamefont {Lautesse}}, \bibinfo {author} {\bibfnamefont {J.~L.}\
  \bibnamefont {Laville}}, \bibinfo {author} {\bibfnamefont {J.~F.}\
  \bibnamefont {Lecolley}}, \bibinfo {author} {\bibfnamefont {A.}~\bibnamefont
  {Le~F\`evre}}, \bibinfo {author} {\bibfnamefont {T.}~\bibnamefont {Lefort}},
  \bibinfo {author} {\bibfnamefont {R.}~\bibnamefont {Legrain}}, \bibinfo
  {author} {\bibfnamefont {O.}~\bibnamefont {Lopez}}, \bibinfo {author}
  {\bibfnamefont {M.}~\bibnamefont {Louvel}}, \bibinfo {author} {\bibfnamefont
  {N.}~\bibnamefont {Marie}}, \bibinfo {author} {\bibfnamefont
  {L.}~\bibnamefont {Nalpas}}, \bibinfo {author} {\bibfnamefont
  {M.}~\bibnamefont {Parlog}}, \bibinfo {author} {\bibfnamefont
  {J.}~\bibnamefont {P\'eter}}, \bibinfo {author} {\bibfnamefont
  {O.}~\bibnamefont {Politi}}, \bibinfo {author} {\bibfnamefont
  {A.}~\bibnamefont {Rahmani}}, \bibinfo {author} {\bibfnamefont
  {T.}~\bibnamefont {Reposeur}}, \bibinfo {author} {\bibfnamefont {M.~F.}\
  \bibnamefont {Rivet}}, \bibinfo {author} {\bibfnamefont {E.}~\bibnamefont
  {Rosato}}, \bibinfo {author} {\bibfnamefont {F.}~\bibnamefont
  {Saint-Laurent}}, \bibinfo {author} {\bibfnamefont {M.}~\bibnamefont
  {Squalli}}, \bibinfo {author} {\bibfnamefont {J.~C.}\ \bibnamefont
  {Steckmeyer}}, \bibinfo {author} {\bibfnamefont {M.}~\bibnamefont {Stern}},
  \bibinfo {author} {\bibfnamefont {L.}~\bibnamefont {Tassan-Got}}, \bibinfo
  {author} {\bibfnamefont {E.}~\bibnamefont {Vient}}, \bibinfo {author}
  {\bibfnamefont {C.}~\bibnamefont {Volant}}, \bibinfo {author} {\bibfnamefont
  {J.~P.}\ \bibnamefont {Wieleczko}}, \bibinfo {author} {\bibfnamefont
  {M.}~\bibnamefont {Colonna}}, \bibinfo {author} {\bibfnamefont
  {F.}~\bibnamefont {Haddad}}, \bibinfo {author} {\bibfnamefont
  {P.}~\bibnamefont {Eudes}}, \bibinfo {author} {\bibfnamefont
  {T.}~\bibnamefont {Sami}},\ and\ \bibinfo {author} {\bibfnamefont
  {F.}~\bibnamefont {Sebille}},\ }\href
  {https://doi.org/10.1103/PhysRevC.55.1906} {\bibfield  {journal} {\bibinfo
  {journal} {Phys. Rev. C}\ }\textbf {\bibinfo {volume} {55}},\ \bibinfo
  {pages} {1906} (\bibinfo {year} {1997})}\BibitemShut {NoStop}%
\bibitem [{\citenamefont {Lionti}\ \emph {et~al.}(2005)\citenamefont {Lionti},
  \citenamefont {Baran}, \citenamefont {Colonna},\ and\ \citenamefont {{Di
  Toro}}}]{Lionti2005}%
  \BibitemOpen
  \bibfield  {author} {\bibinfo {author} {\bibfnamefont {R.}~\bibnamefont
  {Lionti}}, \bibinfo {author} {\bibfnamefont {V.}~\bibnamefont {Baran}},
  \bibinfo {author} {\bibfnamefont {M.}~\bibnamefont {Colonna}},\ and\ \bibinfo
  {author} {\bibfnamefont {M.}~\bibnamefont {{Di Toro}}},\ }\href
  {https://doi.org/https://doi.org/10.1016/j.physletb.2005.08.044} {\bibfield
  {journal} {\bibinfo  {journal} {Phys. Lett. B}\ }\textbf {\bibinfo {volume}
  {625}},\ \bibinfo {pages} {33} (\bibinfo {year} {2005})}\BibitemShut
  {NoStop}%
\bibitem [{\citenamefont {Baran}\ \emph {et~al.}(2005)\citenamefont {Baran},
  \citenamefont {Colonna}, \citenamefont {{Di Toro}}, \citenamefont
  {Zielinska-Pfab\'e},\ and\ \citenamefont {Wolter}}]{Baran2005}%
  \BibitemOpen
  \bibfield  {author} {\bibinfo {author} {\bibfnamefont {V.}~\bibnamefont
  {Baran}}, \bibinfo {author} {\bibfnamefont {M.}~\bibnamefont {Colonna}},
  \bibinfo {author} {\bibfnamefont {M.}~\bibnamefont {{Di Toro}}}, \bibinfo
  {author} {\bibfnamefont {M.}~\bibnamefont {Zielinska-Pfab\'e}},\ and\
  \bibinfo {author} {\bibfnamefont {H.~H.}\ \bibnamefont {Wolter}},\ }\href
  {https://doi.org/10.1103/PhysRevC.72.064620} {\bibfield  {journal} {\bibinfo
  {journal} {Phys. Rev. C}\ }\textbf {\bibinfo {volume} {72}},\ \bibinfo
  {pages} {064620} (\bibinfo {year} {2005})}\BibitemShut {NoStop}%
\bibitem [{\citenamefont {Hong}\ \emph {et~al.}(2002)\citenamefont {Hong},
  \citenamefont {Kim}, \citenamefont {Kang}, \citenamefont {Leifels},
  \citenamefont {Rami}, \citenamefont {de~Schauenburg}, \citenamefont {Sim},
  \citenamefont {Alard}, \citenamefont {Andronic}, \citenamefont {Barret},
  \citenamefont {Basrak}, \citenamefont {Bastid}, \citenamefont {Berek},
  \citenamefont {\ifmmode~\check{C}\else \v{C}\fi{}aplar}, \citenamefont
  {Crochet}, \citenamefont {Devismes}, \citenamefont {Dupieux}, \citenamefont
  {D\ifmmode~\check{z}\else \v{z}\fi{}elalija}, \citenamefont {Finck},
  \citenamefont {Fodor}, \citenamefont {Gobbi}, \citenamefont {Grishkin},
  \citenamefont {Hartmann}, \citenamefont {Herrmann}, \citenamefont
  {Hildenbrand}, \citenamefont {Kecskemeti}, \citenamefont {Kirejczyk},
  \citenamefont {Koczon}, \citenamefont {Korolija}, \citenamefont {Kotte},
  \citenamefont {Kress}, \citenamefont {Kutsche}, \citenamefont {Lebedev},
  \citenamefont {Lopez}, \citenamefont {Neubert}, \citenamefont {Pelte},
  \citenamefont {Petrovici}, \citenamefont {Reisdorf}, \citenamefont
  {Sch\ifmmode~\mbox{\H{u}}\else \H{u}\fi{}ll}, \citenamefont {Seres},
  \citenamefont {Sikora}, \citenamefont {Simion}, \citenamefont
  {Siwek-Wilczy\ifmmode~\acute{n}\else \'{n}\fi{}ska}, \citenamefont
  {Smolyankin}, \citenamefont {Stockmeier}, \citenamefont {Stoicea},
  \citenamefont {Wagner}, \citenamefont {Wi\ifmmode~\acute{s}\else
  \'{s}\fi{}niewski}, \citenamefont {Wohlfarth}, \citenamefont {Yushmanov},\
  and\ \citenamefont {Zhilin}}]{Hong2002}%
  \BibitemOpen
  \bibfield  {author} {\bibinfo {author} {\bibfnamefont {B.}~\bibnamefont
  {Hong}}, \bibinfo {author} {\bibfnamefont {Y.~J.}\ \bibnamefont {Kim}},
  \bibinfo {author} {\bibfnamefont {D.~H.}\ \bibnamefont {Kang}}, \bibinfo
  {author} {\bibfnamefont {Y.}~\bibnamefont {Leifels}}, \bibinfo {author}
  {\bibfnamefont {F.}~\bibnamefont {Rami}}, \bibinfo {author} {\bibfnamefont
  {B.}~\bibnamefont {de~Schauenburg}}, \bibinfo {author} {\bibfnamefont
  {K.~S.}\ \bibnamefont {Sim}}, \bibinfo {author} {\bibfnamefont {J.~P.}\
  \bibnamefont {Alard}}, \bibinfo {author} {\bibfnamefont {A.}~\bibnamefont
  {Andronic}}, \bibinfo {author} {\bibfnamefont {V.}~\bibnamefont {Barret}},
  \bibinfo {author} {\bibfnamefont {Z.}~\bibnamefont {Basrak}}, \bibinfo
  {author} {\bibfnamefont {N.}~\bibnamefont {Bastid}}, \bibinfo {author}
  {\bibfnamefont {G.}~\bibnamefont {Berek}}, \bibinfo {author} {\bibfnamefont
  {R.}~\bibnamefont {\ifmmode~\check{C}\else \v{C}\fi{}aplar}}, \bibinfo
  {author} {\bibfnamefont {P.}~\bibnamefont {Crochet}}, \bibinfo {author}
  {\bibfnamefont {A.}~\bibnamefont {Devismes}}, \bibinfo {author}
  {\bibfnamefont {P.}~\bibnamefont {Dupieux}}, \bibinfo {author} {\bibfnamefont
  {M.}~\bibnamefont {D\ifmmode~\check{z}\else \v{z}\fi{}elalija}}, \bibinfo
  {author} {\bibfnamefont {C.}~\bibnamefont {Finck}}, \bibinfo {author}
  {\bibfnamefont {Z.}~\bibnamefont {Fodor}}, \bibinfo {author} {\bibfnamefont
  {A.}~\bibnamefont {Gobbi}}, \bibinfo {author} {\bibfnamefont
  {Y.}~\bibnamefont {Grishkin}}, \bibinfo {author} {\bibfnamefont {O.~N.}\
  \bibnamefont {Hartmann}}, \bibinfo {author} {\bibfnamefont {N.}~\bibnamefont
  {Herrmann}}, \bibinfo {author} {\bibfnamefont {K.~D.}\ \bibnamefont
  {Hildenbrand}}, \bibinfo {author} {\bibfnamefont {J.}~\bibnamefont
  {Kecskemeti}}, \bibinfo {author} {\bibfnamefont {M.}~\bibnamefont
  {Kirejczyk}}, \bibinfo {author} {\bibfnamefont {P.}~\bibnamefont {Koczon}},
  \bibinfo {author} {\bibfnamefont {M.}~\bibnamefont {Korolija}}, \bibinfo
  {author} {\bibfnamefont {R.}~\bibnamefont {Kotte}}, \bibinfo {author}
  {\bibfnamefont {T.}~\bibnamefont {Kress}}, \bibinfo {author} {\bibfnamefont
  {R.}~\bibnamefont {Kutsche}}, \bibinfo {author} {\bibfnamefont
  {A.}~\bibnamefont {Lebedev}}, \bibinfo {author} {\bibfnamefont
  {X.}~\bibnamefont {Lopez}}, \bibinfo {author} {\bibfnamefont
  {W.}~\bibnamefont {Neubert}}, \bibinfo {author} {\bibfnamefont
  {D.}~\bibnamefont {Pelte}}, \bibinfo {author} {\bibfnamefont
  {M.}~\bibnamefont {Petrovici}}, \bibinfo {author} {\bibfnamefont
  {W.}~\bibnamefont {Reisdorf}}, \bibinfo {author} {\bibfnamefont
  {D.}~\bibnamefont {Sch\ifmmode~\mbox{\H{u}}\else \H{u}\fi{}ll}}, \bibinfo
  {author} {\bibfnamefont {Z.}~\bibnamefont {Seres}}, \bibinfo {author}
  {\bibfnamefont {B.}~\bibnamefont {Sikora}}, \bibinfo {author} {\bibfnamefont
  {V.}~\bibnamefont {Simion}}, \bibinfo {author} {\bibfnamefont
  {K.}~\bibnamefont {Siwek-Wilczy\ifmmode~\acute{n}\else \'{n}\fi{}ska}},
  \bibinfo {author} {\bibfnamefont {V.}~\bibnamefont {Smolyankin}}, \bibinfo
  {author} {\bibfnamefont {M.~R.}\ \bibnamefont {Stockmeier}}, \bibinfo
  {author} {\bibfnamefont {G.}~\bibnamefont {Stoicea}}, \bibinfo {author}
  {\bibfnamefont {P.}~\bibnamefont {Wagner}}, \bibinfo {author} {\bibfnamefont
  {K.}~\bibnamefont {Wi\ifmmode~\acute{s}\else \'{s}\fi{}niewski}}, \bibinfo
  {author} {\bibfnamefont {D.}~\bibnamefont {Wohlfarth}}, \bibinfo {author}
  {\bibfnamefont {I.}~\bibnamefont {Yushmanov}},\ and\ \bibinfo {author}
  {\bibfnamefont {A.}~\bibnamefont {Zhilin}} (\bibinfo {collaboration} {FOPI
  Collaboration}),\ }\href {https://doi.org/10.1103/PhysRevC.66.034901}
  {\bibfield  {journal} {\bibinfo  {journal} {Phys. Rev. C}\ }\textbf {\bibinfo
  {volume} {66}},\ \bibinfo {pages} {034901} (\bibinfo {year}
  {2002})}\BibitemShut {NoStop}%
\bibitem [{\citenamefont {Tsang}\ \emph {et~al.}(2004)\citenamefont {Tsang},
  \citenamefont {Liu}, \citenamefont {Shi}, \citenamefont {Danielewicz},
  \citenamefont {Gelbke}, \citenamefont {Liu}, \citenamefont {Lynch},
  \citenamefont {Tan}, \citenamefont {Verde}, \citenamefont {Wagner},
  \citenamefont {Xu}, \citenamefont {Friedman}, \citenamefont {Beaulieu},
  \citenamefont {Davin}, \citenamefont {de~Souza}, \citenamefont {Larochelle},
  \citenamefont {Lefort}, \citenamefont {Yanez}, \citenamefont {Viola},
  \citenamefont {Charity},\ and\ \citenamefont {Sobotka}}]{Tsang2004}%
  \BibitemOpen
  \bibfield  {author} {\bibinfo {author} {\bibfnamefont {M.~B.}\ \bibnamefont
  {Tsang}}, \bibinfo {author} {\bibfnamefont {T.~X.}\ \bibnamefont {Liu}},
  \bibinfo {author} {\bibfnamefont {L.}~\bibnamefont {Shi}}, \bibinfo {author}
  {\bibfnamefont {P.}~\bibnamefont {Danielewicz}}, \bibinfo {author}
  {\bibfnamefont {C.~K.}\ \bibnamefont {Gelbke}}, \bibinfo {author}
  {\bibfnamefont {X.~D.}\ \bibnamefont {Liu}}, \bibinfo {author} {\bibfnamefont
  {W.~G.}\ \bibnamefont {Lynch}}, \bibinfo {author} {\bibfnamefont {W.~P.}\
  \bibnamefont {Tan}}, \bibinfo {author} {\bibfnamefont {G.}~\bibnamefont
  {Verde}}, \bibinfo {author} {\bibfnamefont {A.}~\bibnamefont {Wagner}},
  \bibinfo {author} {\bibfnamefont {H.~S.}\ \bibnamefont {Xu}}, \bibinfo
  {author} {\bibfnamefont {W.~A.}\ \bibnamefont {Friedman}}, \bibinfo {author}
  {\bibfnamefont {L.}~\bibnamefont {Beaulieu}}, \bibinfo {author}
  {\bibfnamefont {B.}~\bibnamefont {Davin}}, \bibinfo {author} {\bibfnamefont
  {R.~T.}\ \bibnamefont {de~Souza}}, \bibinfo {author} {\bibfnamefont
  {Y.}~\bibnamefont {Larochelle}}, \bibinfo {author} {\bibfnamefont
  {T.}~\bibnamefont {Lefort}}, \bibinfo {author} {\bibfnamefont
  {R.}~\bibnamefont {Yanez}}, \bibinfo {author} {\bibfnamefont {V.~E.}\
  \bibnamefont {Viola}}, \bibinfo {author} {\bibfnamefont {R.~J.}\ \bibnamefont
  {Charity}},\ and\ \bibinfo {author} {\bibfnamefont {L.~G.}\ \bibnamefont
  {Sobotka}},\ }\href {https://doi.org/10.1103/PhysRevLett.92.062701}
  {\bibfield  {journal} {\bibinfo  {journal} {Phys. Rev. Lett.}\ }\textbf
  {\bibinfo {volume} {92}},\ \bibinfo {pages} {062701} (\bibinfo {year}
  {2004})}\BibitemShut {NoStop}%
\bibitem [{\citenamefont {Liu}\ \emph {et~al.}(2007)\citenamefont {Liu},
  \citenamefont {Lynch}, \citenamefont {Tsang}, \citenamefont {Liu},
  \citenamefont {Shomin}, \citenamefont {Tan}, \citenamefont {Verde},
  \citenamefont {Wagner}, \citenamefont {Xi}, \citenamefont {Xu}, \citenamefont
  {Davin}, \citenamefont {Larochelle}, \citenamefont {Souza}, \citenamefont
  {Charity},\ and\ \citenamefont {Sobotka}}]{Liu2007}%
  \BibitemOpen
  \bibfield  {author} {\bibinfo {author} {\bibfnamefont {T.~X.}\ \bibnamefont
  {Liu}}, \bibinfo {author} {\bibfnamefont {W.~G.}\ \bibnamefont {Lynch}},
  \bibinfo {author} {\bibfnamefont {M.~B.}\ \bibnamefont {Tsang}}, \bibinfo
  {author} {\bibfnamefont {X.~D.}\ \bibnamefont {Liu}}, \bibinfo {author}
  {\bibfnamefont {R.}~\bibnamefont {Shomin}}, \bibinfo {author} {\bibfnamefont
  {W.~P.}\ \bibnamefont {Tan}}, \bibinfo {author} {\bibfnamefont
  {G.}~\bibnamefont {Verde}}, \bibinfo {author} {\bibfnamefont
  {A.}~\bibnamefont {Wagner}}, \bibinfo {author} {\bibfnamefont {H.~F.}\
  \bibnamefont {Xi}}, \bibinfo {author} {\bibfnamefont {H.~S.}\ \bibnamefont
  {Xu}}, \bibinfo {author} {\bibfnamefont {B.}~\bibnamefont {Davin}}, \bibinfo
  {author} {\bibfnamefont {Y.}~\bibnamefont {Larochelle}}, \bibinfo {author}
  {\bibfnamefont {R.~T.~d.}\ \bibnamefont {Souza}}, \bibinfo {author}
  {\bibfnamefont {R.~J.}\ \bibnamefont {Charity}},\ and\ \bibinfo {author}
  {\bibfnamefont {L.~G.}\ \bibnamefont {Sobotka}},\ }\href
  {https://doi.org/10.1103/PhysRevC.76.034603} {\bibfield  {journal} {\bibinfo
  {journal} {Phys. Rev. C}\ }\textbf {\bibinfo {volume} {76}},\ \bibinfo
  {pages} {034603} (\bibinfo {year} {2007})}\BibitemShut {NoStop}%
\bibitem [{\citenamefont {Galichet}\ \emph
  {et~al.}(2009{\natexlab{a}})\citenamefont {Galichet}, \citenamefont {Rivet},
  \citenamefont {Borderie}, \citenamefont {Colonna}, \citenamefont {Bougault},
  \citenamefont {Chbihi}, \citenamefont {Dayras}, \citenamefont {Durand},
  \citenamefont {Frankland}, \citenamefont {Guinet}, \citenamefont {Lautesse},
  \citenamefont {Neindre}, \citenamefont {Lopez}, \citenamefont {Manduci},
  \citenamefont {P\^arlog}, \citenamefont {Rosato}, \citenamefont {Tamain},
  \citenamefont {Vient}, \citenamefont {Volant},\ and\ \citenamefont
  {Wieleczko}}]{Galichet2009}%
  \BibitemOpen
  \bibfield  {author} {\bibinfo {author} {\bibfnamefont {E.}~\bibnamefont
  {Galichet}}, \bibinfo {author} {\bibfnamefont {M.~F.}\ \bibnamefont {Rivet}},
  \bibinfo {author} {\bibfnamefont {B.}~\bibnamefont {Borderie}}, \bibinfo
  {author} {\bibfnamefont {M.}~\bibnamefont {Colonna}}, \bibinfo {author}
  {\bibfnamefont {R.}~\bibnamefont {Bougault}}, \bibinfo {author}
  {\bibfnamefont {A.}~\bibnamefont {Chbihi}}, \bibinfo {author} {\bibfnamefont
  {R.}~\bibnamefont {Dayras}}, \bibinfo {author} {\bibfnamefont
  {D.}~\bibnamefont {Durand}}, \bibinfo {author} {\bibfnamefont {J.~D.}\
  \bibnamefont {Frankland}}, \bibinfo {author} {\bibfnamefont {D.~C.~R.}\
  \bibnamefont {Guinet}}, \bibinfo {author} {\bibfnamefont {P.}~\bibnamefont
  {Lautesse}}, \bibinfo {author} {\bibfnamefont {N.~L.}\ \bibnamefont
  {Neindre}}, \bibinfo {author} {\bibfnamefont {O.}~\bibnamefont {Lopez}},
  \bibinfo {author} {\bibfnamefont {L.}~\bibnamefont {Manduci}}, \bibinfo
  {author} {\bibfnamefont {M.}~\bibnamefont {P\^arlog}}, \bibinfo {author}
  {\bibfnamefont {E.}~\bibnamefont {Rosato}}, \bibinfo {author} {\bibfnamefont
  {B.}~\bibnamefont {Tamain}}, \bibinfo {author} {\bibfnamefont
  {E.}~\bibnamefont {Vient}}, \bibinfo {author} {\bibfnamefont
  {C.}~\bibnamefont {Volant}},\ and\ \bibinfo {author} {\bibfnamefont {J.~P.}\
  \bibnamefont {Wieleczko}} (\bibinfo {collaboration} {INDRA Collaboration}),\
  }\href {https://doi.org/10.1103/PhysRevC.79.064614} {\bibfield  {journal}
  {\bibinfo  {journal} {Phys. Rev. C}\ }\textbf {\bibinfo {volume} {79}},\
  \bibinfo {pages} {064614} (\bibinfo {year} {2009}{\natexlab{a}})}\BibitemShut
  {NoStop}%
\bibitem [{\citenamefont {Lombardo}\ \emph {et~al.}(2010)\citenamefont
  {Lombardo}, \citenamefont {Agodi}, \citenamefont {Alba}, \citenamefont
  {Amorini}, \citenamefont {Anzalone}, \citenamefont {Berceanu}, \citenamefont
  {Cardella}, \citenamefont {Cavallaro}, \citenamefont {Chatterjee},
  \citenamefont {De~Filippo}, \citenamefont {Di~Pietro}, \citenamefont
  {Figuera}, \citenamefont {Geraci}, \citenamefont {Giuliani}, \citenamefont
  {Grassi}, \citenamefont {Grzeszczuk}, \citenamefont {Han}, \citenamefont
  {La~Guidara}, \citenamefont {Lanzalone}, \citenamefont {Le~Neindre},
  \citenamefont {Maiolino}, \citenamefont {Pagano}, \citenamefont {Papa},
  \citenamefont {Pirrone}, \citenamefont {Politi}, \citenamefont {Pop},
  \citenamefont {Porto}, \citenamefont {Rizzo}, \citenamefont {Russotto},
  \citenamefont {Santonocito},\ and\ \citenamefont {Verde}}]{Lombardo2010}%
  \BibitemOpen
  \bibfield  {author} {\bibinfo {author} {\bibfnamefont {I.}~\bibnamefont
  {Lombardo}}, \bibinfo {author} {\bibfnamefont {C.}~\bibnamefont {Agodi}},
  \bibinfo {author} {\bibfnamefont {R.}~\bibnamefont {Alba}}, \bibinfo {author}
  {\bibfnamefont {F.}~\bibnamefont {Amorini}}, \bibinfo {author} {\bibfnamefont
  {A.}~\bibnamefont {Anzalone}}, \bibinfo {author} {\bibfnamefont
  {I.}~\bibnamefont {Berceanu}}, \bibinfo {author} {\bibfnamefont
  {G.}~\bibnamefont {Cardella}}, \bibinfo {author} {\bibfnamefont
  {S.}~\bibnamefont {Cavallaro}}, \bibinfo {author} {\bibfnamefont {M.~B.}\
  \bibnamefont {Chatterjee}}, \bibinfo {author} {\bibfnamefont
  {E.}~\bibnamefont {De~Filippo}}, \bibinfo {author} {\bibfnamefont
  {A.}~\bibnamefont {Di~Pietro}}, \bibinfo {author} {\bibfnamefont
  {P.}~\bibnamefont {Figuera}}, \bibinfo {author} {\bibfnamefont
  {E.}~\bibnamefont {Geraci}}, \bibinfo {author} {\bibfnamefont
  {G.}~\bibnamefont {Giuliani}}, \bibinfo {author} {\bibfnamefont
  {L.}~\bibnamefont {Grassi}}, \bibinfo {author} {\bibfnamefont
  {A.}~\bibnamefont {Grzeszczuk}}, \bibinfo {author} {\bibfnamefont
  {J.}~\bibnamefont {Han}}, \bibinfo {author} {\bibfnamefont {E.}~\bibnamefont
  {La~Guidara}}, \bibinfo {author} {\bibfnamefont {G.}~\bibnamefont
  {Lanzalone}}, \bibinfo {author} {\bibfnamefont {N.}~\bibnamefont
  {Le~Neindre}}, \bibinfo {author} {\bibfnamefont {C.}~\bibnamefont
  {Maiolino}}, \bibinfo {author} {\bibfnamefont {A.}~\bibnamefont {Pagano}},
  \bibinfo {author} {\bibfnamefont {M.}~\bibnamefont {Papa}}, \bibinfo {author}
  {\bibfnamefont {S.}~\bibnamefont {Pirrone}}, \bibinfo {author} {\bibfnamefont
  {G.}~\bibnamefont {Politi}}, \bibinfo {author} {\bibfnamefont
  {A.}~\bibnamefont {Pop}}, \bibinfo {author} {\bibfnamefont {F.}~\bibnamefont
  {Porto}}, \bibinfo {author} {\bibfnamefont {F.}~\bibnamefont {Rizzo}},
  \bibinfo {author} {\bibfnamefont {P.}~\bibnamefont {Russotto}}, \bibinfo
  {author} {\bibfnamefont {D.}~\bibnamefont {Santonocito}},\ and\ \bibinfo
  {author} {\bibfnamefont {G.}~\bibnamefont {Verde}},\ }\href
  {https://doi.org/10.1103/PhysRevC.82.014608} {\bibfield  {journal} {\bibinfo
  {journal} {Phys. Rev. C}\ }\textbf {\bibinfo {volume} {82}},\ \bibinfo
  {pages} {014608} (\bibinfo {year} {2010})}\BibitemShut {NoStop}%
\bibitem [{\citenamefont {Barlini}\ \emph {et~al.}(2013)\citenamefont
  {Barlini}, \citenamefont {Piantelli}, \citenamefont {Casini}, \citenamefont
  {Maurenzig}, \citenamefont {Olmi}, \citenamefont {Bini}, \citenamefont
  {Carboni}, \citenamefont {Pasquali}, \citenamefont {Poggi}, \citenamefont
  {Stefanini}, \citenamefont {Bougault}, \citenamefont {Bonnet}, \citenamefont
  {Borderie}, \citenamefont {Chbihi}, \citenamefont {Frankland}, \citenamefont
  {Gruyer}, \citenamefont {Lopez}, \citenamefont {Le~Neindre}, \citenamefont
  {P\^arlog}, \citenamefont {Rivet}, \citenamefont {Vient}, \citenamefont
  {Rosato}, \citenamefont {Spadaccini}, \citenamefont {Vigilante},
  \citenamefont {Bruno}, \citenamefont {Marchi}, \citenamefont {Morelli},
  \citenamefont {Cinausero}, \citenamefont {Degerlier}, \citenamefont
  {Gramegna}, \citenamefont {Kozik}, \citenamefont {Twar\'og}, \citenamefont
  {Alba}, \citenamefont {Maiolino},\ and\ \citenamefont
  {Santonocito}}]{Barlini2013}%
  \BibitemOpen
  \bibfield  {author} {\bibinfo {author} {\bibfnamefont {S.}~\bibnamefont
  {Barlini}}, \bibinfo {author} {\bibfnamefont {S.}~\bibnamefont {Piantelli}},
  \bibinfo {author} {\bibfnamefont {G.}~\bibnamefont {Casini}}, \bibinfo
  {author} {\bibfnamefont {P.~R.}\ \bibnamefont {Maurenzig}}, \bibinfo {author}
  {\bibfnamefont {A.}~\bibnamefont {Olmi}}, \bibinfo {author} {\bibfnamefont
  {M.}~\bibnamefont {Bini}}, \bibinfo {author} {\bibfnamefont {S.}~\bibnamefont
  {Carboni}}, \bibinfo {author} {\bibfnamefont {G.}~\bibnamefont {Pasquali}},
  \bibinfo {author} {\bibfnamefont {G.}~\bibnamefont {Poggi}}, \bibinfo
  {author} {\bibfnamefont {A.~A.}\ \bibnamefont {Stefanini}}, \bibinfo {author}
  {\bibfnamefont {R.}~\bibnamefont {Bougault}}, \bibinfo {author}
  {\bibfnamefont {E.}~\bibnamefont {Bonnet}}, \bibinfo {author} {\bibfnamefont
  {B.}~\bibnamefont {Borderie}}, \bibinfo {author} {\bibfnamefont
  {A.}~\bibnamefont {Chbihi}}, \bibinfo {author} {\bibfnamefont {J.~D.}\
  \bibnamefont {Frankland}}, \bibinfo {author} {\bibfnamefont {D.}~\bibnamefont
  {Gruyer}}, \bibinfo {author} {\bibfnamefont {O.}~\bibnamefont {Lopez}},
  \bibinfo {author} {\bibfnamefont {N.}~\bibnamefont {Le~Neindre}}, \bibinfo
  {author} {\bibfnamefont {M.}~\bibnamefont {P\^arlog}}, \bibinfo {author}
  {\bibfnamefont {M.~F.}\ \bibnamefont {Rivet}}, \bibinfo {author}
  {\bibfnamefont {E.}~\bibnamefont {Vient}}, \bibinfo {author} {\bibfnamefont
  {E.}~\bibnamefont {Rosato}}, \bibinfo {author} {\bibfnamefont
  {G.}~\bibnamefont {Spadaccini}}, \bibinfo {author} {\bibfnamefont
  {M.}~\bibnamefont {Vigilante}}, \bibinfo {author} {\bibfnamefont
  {M.}~\bibnamefont {Bruno}}, \bibinfo {author} {\bibfnamefont
  {T.}~\bibnamefont {Marchi}}, \bibinfo {author} {\bibfnamefont
  {L.}~\bibnamefont {Morelli}}, \bibinfo {author} {\bibfnamefont
  {M.}~\bibnamefont {Cinausero}}, \bibinfo {author} {\bibfnamefont
  {M.}~\bibnamefont {Degerlier}}, \bibinfo {author} {\bibfnamefont
  {F.}~\bibnamefont {Gramegna}}, \bibinfo {author} {\bibfnamefont
  {T.}~\bibnamefont {Kozik}}, \bibinfo {author} {\bibfnamefont
  {T.}~\bibnamefont {Twar\'og}}, \bibinfo {author} {\bibfnamefont
  {R.}~\bibnamefont {Alba}}, \bibinfo {author} {\bibfnamefont {C.}~\bibnamefont
  {Maiolino}},\ and\ \bibinfo {author} {\bibfnamefont {D.}~\bibnamefont
  {Santonocito}} (\bibinfo {collaboration} {FAZIA Collaboration}),\ }\href
  {https://doi.org/10.1103/PhysRevC.87.054607} {\bibfield  {journal} {\bibinfo
  {journal} {Phys. Rev. C}\ }\textbf {\bibinfo {volume} {87}},\ \bibinfo
  {pages} {054607} (\bibinfo {year} {2013})}\BibitemShut {NoStop}%
\bibitem [{\citenamefont {Keksis}\ \emph {et~al.}(2010)\citenamefont {Keksis},
  \citenamefont {May}, \citenamefont {Souliotis}, \citenamefont {Veselsky},
  \citenamefont {Galanopoulos}, \citenamefont {Kohley}, \citenamefont {Shetty},
  \citenamefont {Soisson}, \citenamefont {Stein}, \citenamefont {Tripathi},
  \citenamefont {Wuenschel}, \citenamefont {Yennello},\ and\ \citenamefont
  {Li}}]{Keksis2010}%
  \BibitemOpen
  \bibfield  {author} {\bibinfo {author} {\bibfnamefont {A.~L.}\ \bibnamefont
  {Keksis}}, \bibinfo {author} {\bibfnamefont {L.~W.}\ \bibnamefont {May}},
  \bibinfo {author} {\bibfnamefont {G.~A.}\ \bibnamefont {Souliotis}}, \bibinfo
  {author} {\bibfnamefont {M.}~\bibnamefont {Veselsky}}, \bibinfo {author}
  {\bibfnamefont {S.}~\bibnamefont {Galanopoulos}}, \bibinfo {author}
  {\bibfnamefont {Z.}~\bibnamefont {Kohley}}, \bibinfo {author} {\bibfnamefont
  {D.~V.}\ \bibnamefont {Shetty}}, \bibinfo {author} {\bibfnamefont {S.~N.}\
  \bibnamefont {Soisson}}, \bibinfo {author} {\bibfnamefont {B.~C.}\
  \bibnamefont {Stein}}, \bibinfo {author} {\bibfnamefont {R.}~\bibnamefont
  {Tripathi}}, \bibinfo {author} {\bibfnamefont {S.}~\bibnamefont {Wuenschel}},
  \bibinfo {author} {\bibfnamefont {S.~J.}\ \bibnamefont {Yennello}},\ and\
  \bibinfo {author} {\bibfnamefont {B.~A.}\ \bibnamefont {Li}},\ }\href
  {https://doi.org/10.1103/PhysRevC.81.054602} {\bibfield  {journal} {\bibinfo
  {journal} {Phys. Rev. C}\ }\textbf {\bibinfo {volume} {81}},\ \bibinfo
  {pages} {054602} (\bibinfo {year} {2010})}\BibitemShut {NoStop}%
\bibitem [{\citenamefont {May}\ \emph {et~al.}(2018)\citenamefont {May},
  \citenamefont {Wakhle}, \citenamefont {McIntosh}, \citenamefont {Kohley},
  \citenamefont {Behling}, \citenamefont {Bonasera}, \citenamefont {Bonasera},
  \citenamefont {Cammarata}, \citenamefont {Hagel}, \citenamefont {Heilborn},
  \citenamefont {Jedele}, \citenamefont {Raphelt}, \citenamefont {Manso},
  \citenamefont {Souliotis}, \citenamefont {Tripathi}, \citenamefont {Youngs},
  \citenamefont {Zarrella},\ and\ \citenamefont {Yennello}}]{May2018}%
  \BibitemOpen
  \bibfield  {author} {\bibinfo {author} {\bibfnamefont {L.~W.}\ \bibnamefont
  {May}}, \bibinfo {author} {\bibfnamefont {A.}~\bibnamefont {Wakhle}},
  \bibinfo {author} {\bibfnamefont {A.~B.}\ \bibnamefont {McIntosh}}, \bibinfo
  {author} {\bibfnamefont {Z.}~\bibnamefont {Kohley}}, \bibinfo {author}
  {\bibfnamefont {S.}~\bibnamefont {Behling}}, \bibinfo {author} {\bibfnamefont
  {A.}~\bibnamefont {Bonasera}}, \bibinfo {author} {\bibfnamefont
  {G.}~\bibnamefont {Bonasera}}, \bibinfo {author} {\bibfnamefont
  {P.}~\bibnamefont {Cammarata}}, \bibinfo {author} {\bibfnamefont
  {K.}~\bibnamefont {Hagel}}, \bibinfo {author} {\bibfnamefont
  {L.}~\bibnamefont {Heilborn}}, \bibinfo {author} {\bibfnamefont
  {A.}~\bibnamefont {Jedele}}, \bibinfo {author} {\bibfnamefont
  {A.}~\bibnamefont {Raphelt}}, \bibinfo {author} {\bibfnamefont {A.~R.}\
  \bibnamefont {Manso}}, \bibinfo {author} {\bibfnamefont {G.}~\bibnamefont
  {Souliotis}}, \bibinfo {author} {\bibfnamefont {R.}~\bibnamefont {Tripathi}},
  \bibinfo {author} {\bibfnamefont {M.~D.}\ \bibnamefont {Youngs}}, \bibinfo
  {author} {\bibfnamefont {A.}~\bibnamefont {Zarrella}},\ and\ \bibinfo
  {author} {\bibfnamefont {S.~J.}\ \bibnamefont {Yennello}},\ }\href
  {https://doi.org/10.1103/PhysRevC.98.044602} {\bibfield  {journal} {\bibinfo
  {journal} {Phys. Rev. C}\ }\textbf {\bibinfo {volume} {98}},\ \bibinfo
  {pages} {044602} (\bibinfo {year} {2018})}\BibitemShut {NoStop}%
\bibitem [{\citenamefont {Fable}(2018)}]{Fable2018}%
  \BibitemOpen
  \bibfield  {author} {\bibinfo {author} {\bibfnamefont {Q.}~\bibnamefont
  {Fable}},\ }\href@noop {} {\bibinfo {type} {{Ph.D.} thesis}},\ \bibinfo
  {school} {{Université de Caen Normandie}} (\bibinfo {year}
  {2018})\BibitemShut {NoStop}%
\bibitem [{\citenamefont {Chbihi}\ \emph {et~al.}(2020)\citenamefont {Chbihi},
  \citenamefont {Fable},\ and\ \citenamefont {Verde}}]{Chbihi2020}%
  \BibitemOpen
  \bibfield  {author} {\bibinfo {author} {\bibfnamefont {A.}~\bibnamefont
  {Chbihi}}, \bibinfo {author} {\bibfnamefont {Q.}~\bibnamefont {Fable}},\ and\
  \bibinfo {author} {\bibfnamefont {G.}~\bibnamefont {Verde}},\ }\href
  {https://doi.org/10.7566/JPSCP.32.010075} {\bibfield  {journal} {\bibinfo
  {journal} {JPS Conf. Proc.}\ }\textbf {\bibinfo {volume} {32}},\ \bibinfo
  {pages} {010075} (\bibinfo {year} {2020})}\BibitemShut {NoStop}%
\bibitem [{\citenamefont {Piantelli}\ \emph {et~al.}(2021)\citenamefont
  {Piantelli}, \citenamefont {Casini}, \citenamefont {Ono}, \citenamefont
  {Poggi}, \citenamefont {Pastore}, \citenamefont {Barlini}, \citenamefont
  {Bini}, \citenamefont {Boiano}, \citenamefont {Bonnet}, \citenamefont
  {Borderie}, \citenamefont {Bougault}, \citenamefont {Bruno}, \citenamefont
  {Buccola}, \citenamefont {Camaiani}, \citenamefont {Chbihi}, \citenamefont
  {Ciampi}, \citenamefont {Cicerchia}, \citenamefont {Cinausero}, \citenamefont
  {Degerlier}, \citenamefont {Due\~nas}, \citenamefont {Fable}, \citenamefont
  {Fabris}, \citenamefont {Frankland}, \citenamefont {Frosin}, \citenamefont
  {Gramegna}, \citenamefont {Gruyer}, \citenamefont {Kordyasz}, \citenamefont
  {Kozik}, \citenamefont {Lemari\'e}, \citenamefont {Le~Neindre}, \citenamefont
  {Lombardo}, \citenamefont {Lopez}, \citenamefont {Mantovani}, \citenamefont
  {Marchi}, \citenamefont {Henri}, \citenamefont {Morelli}, \citenamefont
  {Olmi}, \citenamefont {Ottanelli}, \citenamefont {P\^arlog}, \citenamefont
  {Pasquali}, \citenamefont {Quicray}, \citenamefont {Stefanini}, \citenamefont
  {Tortone}, \citenamefont {Upadhyaya}, \citenamefont {Valdr\'e}, \citenamefont
  {Verde}, \citenamefont {Vient},\ and\ \citenamefont
  {Vigilante}}]{Piantelli2021}%
  \BibitemOpen
  \bibfield  {author} {\bibinfo {author} {\bibfnamefont {S.}~\bibnamefont
  {Piantelli}}, \bibinfo {author} {\bibfnamefont {G.}~\bibnamefont {Casini}},
  \bibinfo {author} {\bibfnamefont {A.}~\bibnamefont {Ono}}, \bibinfo {author}
  {\bibfnamefont {G.}~\bibnamefont {Poggi}}, \bibinfo {author} {\bibfnamefont
  {G.}~\bibnamefont {Pastore}}, \bibinfo {author} {\bibfnamefont
  {S.}~\bibnamefont {Barlini}}, \bibinfo {author} {\bibfnamefont
  {M.}~\bibnamefont {Bini}}, \bibinfo {author} {\bibfnamefont {A.}~\bibnamefont
  {Boiano}}, \bibinfo {author} {\bibfnamefont {E.}~\bibnamefont {Bonnet}},
  \bibinfo {author} {\bibfnamefont {B.}~\bibnamefont {Borderie}}, \bibinfo
  {author} {\bibfnamefont {R.}~\bibnamefont {Bougault}}, \bibinfo {author}
  {\bibfnamefont {M.}~\bibnamefont {Bruno}}, \bibinfo {author} {\bibfnamefont
  {A.}~\bibnamefont {Buccola}}, \bibinfo {author} {\bibfnamefont
  {A.}~\bibnamefont {Camaiani}}, \bibinfo {author} {\bibfnamefont
  {A.}~\bibnamefont {Chbihi}}, \bibinfo {author} {\bibfnamefont
  {C.}~\bibnamefont {Ciampi}}, \bibinfo {author} {\bibfnamefont
  {M.}~\bibnamefont {Cicerchia}}, \bibinfo {author} {\bibfnamefont
  {M.}~\bibnamefont {Cinausero}}, \bibinfo {author} {\bibfnamefont
  {M.}~\bibnamefont {Degerlier}}, \bibinfo {author} {\bibfnamefont {J.~A.}\
  \bibnamefont {Due\~nas}}, \bibinfo {author} {\bibfnamefont {Q.}~\bibnamefont
  {Fable}}, \bibinfo {author} {\bibfnamefont {D.}~\bibnamefont {Fabris}},
  \bibinfo {author} {\bibfnamefont {J.~D.}\ \bibnamefont {Frankland}}, \bibinfo
  {author} {\bibfnamefont {C.}~\bibnamefont {Frosin}}, \bibinfo {author}
  {\bibfnamefont {F.}~\bibnamefont {Gramegna}}, \bibinfo {author}
  {\bibfnamefont {D.}~\bibnamefont {Gruyer}}, \bibinfo {author} {\bibfnamefont
  {A.}~\bibnamefont {Kordyasz}}, \bibinfo {author} {\bibfnamefont
  {T.}~\bibnamefont {Kozik}}, \bibinfo {author} {\bibfnamefont
  {J.}~\bibnamefont {Lemari\'e}}, \bibinfo {author} {\bibfnamefont
  {N.}~\bibnamefont {Le~Neindre}}, \bibinfo {author} {\bibfnamefont
  {I.}~\bibnamefont {Lombardo}}, \bibinfo {author} {\bibfnamefont
  {O.}~\bibnamefont {Lopez}}, \bibinfo {author} {\bibfnamefont
  {G.}~\bibnamefont {Mantovani}}, \bibinfo {author} {\bibfnamefont
  {T.}~\bibnamefont {Marchi}}, \bibinfo {author} {\bibfnamefont
  {M.}~\bibnamefont {Henri}}, \bibinfo {author} {\bibfnamefont
  {L.}~\bibnamefont {Morelli}}, \bibinfo {author} {\bibfnamefont
  {A.}~\bibnamefont {Olmi}}, \bibinfo {author} {\bibfnamefont {P.}~\bibnamefont
  {Ottanelli}}, \bibinfo {author} {\bibfnamefont {M.}~\bibnamefont {P\^arlog}},
  \bibinfo {author} {\bibfnamefont {G.}~\bibnamefont {Pasquali}}, \bibinfo
  {author} {\bibfnamefont {J.}~\bibnamefont {Quicray}}, \bibinfo {author}
  {\bibfnamefont {A.~A.}\ \bibnamefont {Stefanini}}, \bibinfo {author}
  {\bibfnamefont {G.}~\bibnamefont {Tortone}}, \bibinfo {author} {\bibfnamefont
  {S.}~\bibnamefont {Upadhyaya}}, \bibinfo {author} {\bibfnamefont
  {S.}~\bibnamefont {Valdr\'e}}, \bibinfo {author} {\bibfnamefont
  {G.}~\bibnamefont {Verde}}, \bibinfo {author} {\bibfnamefont
  {E.}~\bibnamefont {Vient}},\ and\ \bibinfo {author} {\bibfnamefont
  {M.}~\bibnamefont {Vigilante}} (\bibinfo {collaboration} {FAZIA
  Collaboration}),\ }\href {https://doi.org/10.1103/PhysRevC.103.014603}
  {\bibfield  {journal} {\bibinfo  {journal} {Phys. Rev. C}\ }\textbf {\bibinfo
  {volume} {103}},\ \bibinfo {pages} {014603} (\bibinfo {year}
  {2021})}\BibitemShut {NoStop}%
\bibitem [{\citenamefont {Camaiani}\ \emph {et~al.}(2021)\citenamefont
  {Camaiani}, \citenamefont {Casini}, \citenamefont {Piantelli}, \citenamefont
  {Ono}, \citenamefont {Bonnet}, \citenamefont {Alba}, \citenamefont {Barlini},
  \citenamefont {Borderie}, \citenamefont {Bougault}, \citenamefont {Ciampi},
  \citenamefont {Chbihi}, \citenamefont {Cicerchia}, \citenamefont {Cinausero},
  \citenamefont {Dueñas}, \citenamefont {Dell'Aquila}, \citenamefont {Fable},
  \citenamefont {Fabris}, \citenamefont {Frosin}, \citenamefont {Frankland},
  \citenamefont {Gramegna}, \citenamefont {Gruyer}, \citenamefont {Hahn},
  \citenamefont {Henri}, \citenamefont {Hong}, \citenamefont {Kim},
  \citenamefont {Kordyasz}, \citenamefont {Kweon}, \citenamefont {Lee},
  \citenamefont {Lemarié}, \citenamefont {{Le Neindre}}, \citenamefont
  {Lombardo}, \citenamefont {Lopez}, \citenamefont {Marchi}, \citenamefont
  {Nam}, \citenamefont {Ottanelli}, \citenamefont {Parlog}, \citenamefont
  {Pasquali}, \citenamefont {Poggi}, \citenamefont {Quicray}, \citenamefont
  {Stefanini}, \citenamefont {Upadhyaya}, \citenamefont {Valdré}, ,\ and\
  \citenamefont {Vient}}]{Camaiani2021}%
  \BibitemOpen
  \bibfield  {author} {\bibinfo {author} {\bibfnamefont {A.}~\bibnamefont
  {Camaiani}}, \bibinfo {author} {\bibfnamefont {G.}~\bibnamefont {Casini}},
  \bibinfo {author} {\bibfnamefont {S.}~\bibnamefont {Piantelli}}, \bibinfo
  {author} {\bibfnamefont {A.}~\bibnamefont {Ono}}, \bibinfo {author}
  {\bibfnamefont {E.}~\bibnamefont {Bonnet}}, \bibinfo {author} {\bibfnamefont
  {R.}~\bibnamefont {Alba}}, \bibinfo {author} {\bibfnamefont {S.}~\bibnamefont
  {Barlini}}, \bibinfo {author} {\bibfnamefont {B.}~\bibnamefont {Borderie}},
  \bibinfo {author} {\bibfnamefont {R.}~\bibnamefont {Bougault}}, \bibinfo
  {author} {\bibfnamefont {C.}~\bibnamefont {Ciampi}}, \bibinfo {author}
  {\bibfnamefont {A.}~\bibnamefont {Chbihi}}, \bibinfo {author} {\bibfnamefont
  {M.}~\bibnamefont {Cicerchia}}, \bibinfo {author} {\bibfnamefont
  {M.}~\bibnamefont {Cinausero}}, \bibinfo {author} {\bibfnamefont {J.~A.}\
  \bibnamefont {Dueñas}}, \bibinfo {author} {\bibfnamefont {D.}~\bibnamefont
  {Dell'Aquila}}, \bibinfo {author} {\bibfnamefont {Q.}~\bibnamefont {Fable}},
  \bibinfo {author} {\bibfnamefont {D.}~\bibnamefont {Fabris}}, \bibinfo
  {author} {\bibfnamefont {C.}~\bibnamefont {Frosin}}, \bibinfo {author}
  {\bibfnamefont {J.~D.}\ \bibnamefont {Frankland}}, \bibinfo {author}
  {\bibfnamefont {F.}~\bibnamefont {Gramegna}}, \bibinfo {author}
  {\bibfnamefont {D.}~\bibnamefont {Gruyer}}, \bibinfo {author} {\bibfnamefont
  {K.~I.}\ \bibnamefont {Hahn}}, \bibinfo {author} {\bibfnamefont
  {M.}~\bibnamefont {Henri}}, \bibinfo {author} {\bibfnamefont
  {B.}~\bibnamefont {Hong}}, \bibinfo {author} {\bibfnamefont {S.}~\bibnamefont
  {Kim}}, \bibinfo {author} {\bibfnamefont {A.}~\bibnamefont {Kordyasz}},
  \bibinfo {author} {\bibfnamefont {M.~J.}\ \bibnamefont {Kweon}}, \bibinfo
  {author} {\bibfnamefont {H.~J.}\ \bibnamefont {Lee}}, \bibinfo {author}
  {\bibfnamefont {J.}~\bibnamefont {Lemarié}}, \bibinfo {author}
  {\bibfnamefont {N.}~\bibnamefont {{Le Neindre}}}, \bibinfo {author}
  {\bibfnamefont {I.}~\bibnamefont {Lombardo}}, \bibinfo {author}
  {\bibfnamefont {O.}~\bibnamefont {Lopez}}, \bibinfo {author} {\bibfnamefont
  {T.}~\bibnamefont {Marchi}}, \bibinfo {author} {\bibfnamefont {S.~H.}\
  \bibnamefont {Nam}}, \bibinfo {author} {\bibfnamefont {P.}~\bibnamefont
  {Ottanelli}}, \bibinfo {author} {\bibfnamefont {M.}~\bibnamefont {Parlog}},
  \bibinfo {author} {\bibfnamefont {G.}~\bibnamefont {Pasquali}}, \bibinfo
  {author} {\bibfnamefont {G.}~\bibnamefont {Poggi}}, \bibinfo {author}
  {\bibfnamefont {J.}~\bibnamefont {Quicray}}, \bibinfo {author} {\bibfnamefont
  {A.~A.}\ \bibnamefont {Stefanini}}, \bibinfo {author} {\bibfnamefont
  {S.}~\bibnamefont {Upadhyaya}}, \bibinfo {author} {\bibfnamefont
  {S.}~\bibnamefont {Valdré}}, ,\ and\ \bibinfo {author} {\bibfnamefont
  {E.}~\bibnamefont {Vient}},\ }\href
  {https://doi.org/10.1103/PhysRevC.103.014605} {\bibfield  {journal} {\bibinfo
   {journal} {Phys. Rev. C}\ }\textbf {\bibinfo {volume} {103}},\ \bibinfo
  {pages} {014605} (\bibinfo {year} {2021})}\BibitemShut {NoStop}%
\bibitem [{\citenamefont {Larochelle}\ \emph {et~al.}(2000)\citenamefont
  {Larochelle}, \citenamefont {Gingras}, \citenamefont {Ball}, \citenamefont
  {Beaulieu}, \citenamefont {Gagn\'e}, \citenamefont {Hagberg}, \citenamefont
  {He}, \citenamefont {Horn}, \citenamefont {Laforest}, \citenamefont {Roy},\
  and\ \citenamefont {St-Pierre}}]{Larochelle2000}%
  \BibitemOpen
  \bibfield  {author} {\bibinfo {author} {\bibfnamefont {Y.}~\bibnamefont
  {Larochelle}}, \bibinfo {author} {\bibfnamefont {L.}~\bibnamefont {Gingras}},
  \bibinfo {author} {\bibfnamefont {G.~C.}\ \bibnamefont {Ball}}, \bibinfo
  {author} {\bibfnamefont {L.}~\bibnamefont {Beaulieu}}, \bibinfo {author}
  {\bibfnamefont {P.}~\bibnamefont {Gagn\'e}}, \bibinfo {author} {\bibfnamefont
  {E.}~\bibnamefont {Hagberg}}, \bibinfo {author} {\bibfnamefont {Z.~Y.}\
  \bibnamefont {He}}, \bibinfo {author} {\bibfnamefont {D.}~\bibnamefont
  {Horn}}, \bibinfo {author} {\bibfnamefont {R.}~\bibnamefont {Laforest}},
  \bibinfo {author} {\bibfnamefont {R.}~\bibnamefont {Roy}},\ and\ \bibinfo
  {author} {\bibfnamefont {C.}~\bibnamefont {St-Pierre}},\ }\href
  {https://doi.org/10.1103/PhysRevC.62.051602} {\bibfield  {journal} {\bibinfo
  {journal} {Phys. Rev. C}\ }\textbf {\bibinfo {volume} {62}},\ \bibinfo
  {pages} {051602} (\bibinfo {year} {2000})}\BibitemShut {NoStop}%
\bibitem [{\citenamefont {Milazzo}\ \emph {et~al.}(2002)\citenamefont
  {Milazzo}, \citenamefont {Vannini}, \citenamefont {Agodi}, \citenamefont
  {Alba}, \citenamefont {Bellia}, \citenamefont {Bruno}, \citenamefont
  {Colonna}, \citenamefont {Colonna}, \citenamefont {Coniglione}, \citenamefont
  {D'Agostino}, \citenamefont {{Del Zoppo}}, \citenamefont {Fabbietti},
  \citenamefont {Finocchiaro}, \citenamefont {Gramegna}, \citenamefont {Iori},
  \citenamefont {Loukachine}, \citenamefont {Maiolino}, \citenamefont
  {Margagliotti}, \citenamefont {Mastinu}, \citenamefont {Migneco},
  \citenamefont {Moroni}, \citenamefont {Piattelli}, \citenamefont {Rui},
  \citenamefont {Santonocito}, \citenamefont {Sapienza},\ and\ \citenamefont
  {Sisto}}]{Milazzo2002}%
  \BibitemOpen
  \bibfield  {author} {\bibinfo {author} {\bibfnamefont {P.}~\bibnamefont
  {Milazzo}}, \bibinfo {author} {\bibfnamefont {G.}~\bibnamefont {Vannini}},
  \bibinfo {author} {\bibfnamefont {C.}~\bibnamefont {Agodi}}, \bibinfo
  {author} {\bibfnamefont {R.}~\bibnamefont {Alba}}, \bibinfo {author}
  {\bibfnamefont {G.}~\bibnamefont {Bellia}}, \bibinfo {author} {\bibfnamefont
  {M.}~\bibnamefont {Bruno}}, \bibinfo {author} {\bibfnamefont
  {M.}~\bibnamefont {Colonna}}, \bibinfo {author} {\bibfnamefont
  {N.}~\bibnamefont {Colonna}}, \bibinfo {author} {\bibfnamefont
  {R.}~\bibnamefont {Coniglione}}, \bibinfo {author} {\bibfnamefont
  {M.}~\bibnamefont {D'Agostino}}, \bibinfo {author} {\bibfnamefont
  {A.}~\bibnamefont {{Del Zoppo}}}, \bibinfo {author} {\bibfnamefont
  {L.}~\bibnamefont {Fabbietti}}, \bibinfo {author} {\bibfnamefont
  {P.}~\bibnamefont {Finocchiaro}}, \bibinfo {author} {\bibfnamefont
  {F.}~\bibnamefont {Gramegna}}, \bibinfo {author} {\bibfnamefont
  {I.}~\bibnamefont {Iori}}, \bibinfo {author} {\bibfnamefont {K.}~\bibnamefont
  {Loukachine}}, \bibinfo {author} {\bibfnamefont {C.}~\bibnamefont
  {Maiolino}}, \bibinfo {author} {\bibfnamefont {G.}~\bibnamefont
  {Margagliotti}}, \bibinfo {author} {\bibfnamefont {P.}~\bibnamefont
  {Mastinu}}, \bibinfo {author} {\bibfnamefont {E.}~\bibnamefont {Migneco}},
  \bibinfo {author} {\bibfnamefont {A.}~\bibnamefont {Moroni}}, \bibinfo
  {author} {\bibfnamefont {P.}~\bibnamefont {Piattelli}}, \bibinfo {author}
  {\bibfnamefont {R.}~\bibnamefont {Rui}}, \bibinfo {author} {\bibfnamefont
  {D.}~\bibnamefont {Santonocito}}, \bibinfo {author} {\bibfnamefont
  {P.}~\bibnamefont {Sapienza}},\ and\ \bibinfo {author} {\bibfnamefont
  {M.}~\bibnamefont {Sisto}},\ }\href
  {https://doi.org/https://doi.org/10.1016/S0375-9474(01)01337-9} {\bibfield
  {journal} {\bibinfo  {journal} {Nucl. Phys. A}\ }\textbf {\bibinfo {volume}
  {703}},\ \bibinfo {pages} {466} (\bibinfo {year} {2002})}\BibitemShut
  {NoStop}%
\bibitem [{\citenamefont {Th\'eriault}\ \emph {et~al.}(2006)\citenamefont
  {Th\'eriault}, \citenamefont {Gauthier}, \citenamefont {Grenier},
  \citenamefont {Moisan}, \citenamefont {St-Pierre}, \citenamefont {Roy},
  \citenamefont {Davin}, \citenamefont {Hudan}, \citenamefont {Paduszynski},
  \citenamefont {Souza}, \citenamefont {Bell}, \citenamefont {Garey},
  \citenamefont {Iglio}, \citenamefont {Keksis}, \citenamefont {Parketon},
  \citenamefont {Richers}, \citenamefont {Shetty}, \citenamefont {Soisson},
  \citenamefont {Souliotis}, \citenamefont {Stein},\ and\ \citenamefont
  {Yennello}}]{Theriault2006}%
  \BibitemOpen
  \bibfield  {author} {\bibinfo {author} {\bibfnamefont {D.}~\bibnamefont
  {Th\'eriault}}, \bibinfo {author} {\bibfnamefont {J.}~\bibnamefont
  {Gauthier}}, \bibinfo {author} {\bibfnamefont {F.}~\bibnamefont {Grenier}},
  \bibinfo {author} {\bibfnamefont {F.}~\bibnamefont {Moisan}}, \bibinfo
  {author} {\bibfnamefont {C.}~\bibnamefont {St-Pierre}}, \bibinfo {author}
  {\bibfnamefont {R.}~\bibnamefont {Roy}}, \bibinfo {author} {\bibfnamefont
  {B.}~\bibnamefont {Davin}}, \bibinfo {author} {\bibfnamefont
  {S.}~\bibnamefont {Hudan}}, \bibinfo {author} {\bibfnamefont
  {T.}~\bibnamefont {Paduszynski}}, \bibinfo {author} {\bibfnamefont
  {R.~T.~d.}\ \bibnamefont {Souza}}, \bibinfo {author} {\bibfnamefont
  {E.}~\bibnamefont {Bell}}, \bibinfo {author} {\bibfnamefont {J.}~\bibnamefont
  {Garey}}, \bibinfo {author} {\bibfnamefont {J.}~\bibnamefont {Iglio}},
  \bibinfo {author} {\bibfnamefont {A.~L.}\ \bibnamefont {Keksis}}, \bibinfo
  {author} {\bibfnamefont {S.}~\bibnamefont {Parketon}}, \bibinfo {author}
  {\bibfnamefont {C.}~\bibnamefont {Richers}}, \bibinfo {author} {\bibfnamefont
  {D.~V.}\ \bibnamefont {Shetty}}, \bibinfo {author} {\bibfnamefont {S.~N.}\
  \bibnamefont {Soisson}}, \bibinfo {author} {\bibfnamefont {G.~A.}\
  \bibnamefont {Souliotis}}, \bibinfo {author} {\bibfnamefont {B.~C.}\
  \bibnamefont {Stein}},\ and\ \bibinfo {author} {\bibfnamefont {S.~J.}\
  \bibnamefont {Yennello}},\ }\href
  {https://doi.org/10.1103/PhysRevC.74.051602} {\bibfield  {journal} {\bibinfo
  {journal} {Phys. Rev. C}\ }\textbf {\bibinfo {volume} {74}},\ \bibinfo
  {pages} {051602} (\bibinfo {year} {2006})}\BibitemShut {NoStop}%
\bibitem [{\citenamefont {Colonna}\ and\ \citenamefont
  {Tsang}(2006)}]{Colonna2006}%
  \BibitemOpen
  \bibfield  {author} {\bibinfo {author} {\bibfnamefont {M.}~\bibnamefont
  {Colonna}}\ and\ \bibinfo {author} {\bibfnamefont {M.~B.}\ \bibnamefont
  {Tsang}},\ }\href
  {https://doi.org/https://doi.org/10.1140/epja/i2006-10114-9} {\bibfield
  {journal} {\bibinfo  {journal} {Eur. Phys. J. A}\ }\textbf {\bibinfo {volume}
  {30}},\ \bibinfo {pages} {165–182} (\bibinfo {year} {2006})}\BibitemShut
  {NoStop}%
\bibitem [{\citenamefont {M\"uller}\ and\ \citenamefont
  {Serot}(1995)}]{Muller1995}%
  \BibitemOpen
  \bibfield  {author} {\bibinfo {author} {\bibfnamefont {H.}~\bibnamefont
  {M\"uller}}\ and\ \bibinfo {author} {\bibfnamefont {B.~D.}\ \bibnamefont
  {Serot}},\ }\href {https://doi.org/10.1103/PhysRevC.52.2072} {\bibfield
  {journal} {\bibinfo  {journal} {Phys. Rev. C}\ }\textbf {\bibinfo {volume}
  {52}},\ \bibinfo {pages} {2072} (\bibinfo {year} {1995})}\BibitemShut
  {NoStop}%
\bibitem [{\citenamefont {Ono}\ \emph {et~al.}(2003)\citenamefont {Ono},
  \citenamefont {Danielewicz}, \citenamefont {Friedman}, \citenamefont
  {Lynch},\ and\ \citenamefont {Tsang}}]{Ono2003}%
  \BibitemOpen
  \bibfield  {author} {\bibinfo {author} {\bibfnamefont {A.}~\bibnamefont
  {Ono}}, \bibinfo {author} {\bibfnamefont {P.}~\bibnamefont {Danielewicz}},
  \bibinfo {author} {\bibfnamefont {W.~A.}\ \bibnamefont {Friedman}}, \bibinfo
  {author} {\bibfnamefont {W.~G.}\ \bibnamefont {Lynch}},\ and\ \bibinfo
  {author} {\bibfnamefont {M.~B.}\ \bibnamefont {Tsang}},\ }\href
  {https://doi.org/10.1103/PhysRevC.68.051601} {\bibfield  {journal} {\bibinfo
  {journal} {Phys. Rev. C}\ }\textbf {\bibinfo {volume} {68}},\ \bibinfo
  {pages} {051601} (\bibinfo {year} {2003})}\BibitemShut {NoStop}%
\bibitem [{\citenamefont {Ono}\ \emph {et~al.}(2004)\citenamefont {Ono},
  \citenamefont {Danielewicz}, \citenamefont {Friedman}, \citenamefont
  {Lynch},\ and\ \citenamefont {Tsang}}]{Ono2004}%
  \BibitemOpen
  \bibfield  {author} {\bibinfo {author} {\bibfnamefont {A.}~\bibnamefont
  {Ono}}, \bibinfo {author} {\bibfnamefont {P.}~\bibnamefont {Danielewicz}},
  \bibinfo {author} {\bibfnamefont {W.~A.}\ \bibnamefont {Friedman}}, \bibinfo
  {author} {\bibfnamefont {W.~G.}\ \bibnamefont {Lynch}},\ and\ \bibinfo
  {author} {\bibfnamefont {M.~B.}\ \bibnamefont {Tsang}},\ }\href
  {https://doi.org/10.1103/PhysRevC.70.041604} {\bibfield  {journal} {\bibinfo
  {journal} {Phys. Rev. C}\ }\textbf {\bibinfo {volume} {70}},\ \bibinfo
  {pages} {041604} (\bibinfo {year} {2004})}\BibitemShut {NoStop}%
\bibitem [{\citenamefont {Colonna}(2020)}]{Colonna2020}%
  \BibitemOpen
  \bibfield  {author} {\bibinfo {author} {\bibfnamefont {M.}~\bibnamefont
  {Colonna}},\ }\href
  {https://doi.org/https://doi.org/10.1016/j.ppnp.2020.103775} {\bibfield
  {journal} {\bibinfo  {journal} {Prog. Part. Nucl. Phys.}\ }\textbf {\bibinfo
  {volume} {113}},\ \bibinfo {pages} {103775} (\bibinfo {year}
  {2020})}\BibitemShut {NoStop}%
\bibitem [{\citenamefont {Souliotis}\ \emph {et~al.}(2011)\citenamefont
  {Souliotis}, \citenamefont {Veselsky}, \citenamefont {Galanopoulos},
  \citenamefont {Jandel}, \citenamefont {Kohley}, \citenamefont {May},
  \citenamefont {Shetty}, \citenamefont {Stein},\ and\ \citenamefont
  {Yennello}}]{Souliotis2011}%
  \BibitemOpen
  \bibfield  {author} {\bibinfo {author} {\bibfnamefont {G.~A.}\ \bibnamefont
  {Souliotis}}, \bibinfo {author} {\bibfnamefont {M.}~\bibnamefont {Veselsky}},
  \bibinfo {author} {\bibfnamefont {S.}~\bibnamefont {Galanopoulos}}, \bibinfo
  {author} {\bibfnamefont {M.}~\bibnamefont {Jandel}}, \bibinfo {author}
  {\bibfnamefont {Z.}~\bibnamefont {Kohley}}, \bibinfo {author} {\bibfnamefont
  {L.~W.}\ \bibnamefont {May}}, \bibinfo {author} {\bibfnamefont {D.~V.}\
  \bibnamefont {Shetty}}, \bibinfo {author} {\bibfnamefont {B.~C.}\
  \bibnamefont {Stein}},\ and\ \bibinfo {author} {\bibfnamefont {S.~J.}\
  \bibnamefont {Yennello}},\ }\href
  {https://doi.org/10.1103/PhysRevC.84.064607} {\bibfield  {journal} {\bibinfo
  {journal} {Phys. Rev. C}\ }\textbf {\bibinfo {volume} {84}},\ \bibinfo
  {pages} {064607} (\bibinfo {year} {2011})}\BibitemShut {NoStop}%
\bibitem [{\citenamefont {Lopez}(2018)}]{Lopez2018}%
  \BibitemOpen
  \bibfield  {author} {\bibinfo {author} {\bibfnamefont {O.}~\bibnamefont
  {Lopez}},\ }\href {https://doi.org/10.1393/ncc/i2018-18165-9} {\bibfield
  {journal} {\bibinfo  {journal} {Nuovo Cim. C}\ }\textbf {\bibinfo {volume}
  {41}},\ \bibinfo {pages} {165} (\bibinfo {year} {2018})}\BibitemShut
  {NoStop}%
\bibitem [{\citenamefont {Piantelli}\ \emph {et~al.}(2017)\citenamefont
  {Piantelli}, \citenamefont {Valdr\'e}, \citenamefont {Barlini}, \citenamefont
  {Casini}, \citenamefont {Colonna}, \citenamefont {Baiocco}, \citenamefont
  {Bini}, \citenamefont {Bruno}, \citenamefont {Camaiani}, \citenamefont
  {Carboni}, \citenamefont {Cicerchia}, \citenamefont {Cinausero},
  \citenamefont {D'Agostino}, \citenamefont {Degerlier}, \citenamefont
  {Fabris}, \citenamefont {Gelli}, \citenamefont {Gramegna}, \citenamefont
  {Gruyer}, \citenamefont {Kravchuk}, \citenamefont {Mabiala}, \citenamefont
  {Marchi}, \citenamefont {Morelli}, \citenamefont {Olmi}, \citenamefont
  {Ottanelli}, \citenamefont {Pasquali},\ and\ \citenamefont
  {Pastore}}]{Piantelli2017}%
  \BibitemOpen
  \bibfield  {author} {\bibinfo {author} {\bibfnamefont {S.}~\bibnamefont
  {Piantelli}}, \bibinfo {author} {\bibfnamefont {S.}~\bibnamefont {Valdr\'e}},
  \bibinfo {author} {\bibfnamefont {S.}~\bibnamefont {Barlini}}, \bibinfo
  {author} {\bibfnamefont {G.}~\bibnamefont {Casini}}, \bibinfo {author}
  {\bibfnamefont {M.}~\bibnamefont {Colonna}}, \bibinfo {author} {\bibfnamefont
  {G.}~\bibnamefont {Baiocco}}, \bibinfo {author} {\bibfnamefont
  {M.}~\bibnamefont {Bini}}, \bibinfo {author} {\bibfnamefont {M.}~\bibnamefont
  {Bruno}}, \bibinfo {author} {\bibfnamefont {A.}~\bibnamefont {Camaiani}},
  \bibinfo {author} {\bibfnamefont {S.}~\bibnamefont {Carboni}}, \bibinfo
  {author} {\bibfnamefont {M.}~\bibnamefont {Cicerchia}}, \bibinfo {author}
  {\bibfnamefont {M.}~\bibnamefont {Cinausero}}, \bibinfo {author}
  {\bibfnamefont {M.}~\bibnamefont {D'Agostino}}, \bibinfo {author}
  {\bibfnamefont {M.}~\bibnamefont {Degerlier}}, \bibinfo {author}
  {\bibfnamefont {D.}~\bibnamefont {Fabris}}, \bibinfo {author} {\bibfnamefont
  {N.}~\bibnamefont {Gelli}}, \bibinfo {author} {\bibfnamefont
  {F.}~\bibnamefont {Gramegna}}, \bibinfo {author} {\bibfnamefont
  {D.}~\bibnamefont {Gruyer}}, \bibinfo {author} {\bibfnamefont {V.~L.}\
  \bibnamefont {Kravchuk}}, \bibinfo {author} {\bibfnamefont {J.}~\bibnamefont
  {Mabiala}}, \bibinfo {author} {\bibfnamefont {T.}~\bibnamefont {Marchi}},
  \bibinfo {author} {\bibfnamefont {L.}~\bibnamefont {Morelli}}, \bibinfo
  {author} {\bibfnamefont {A.}~\bibnamefont {Olmi}}, \bibinfo {author}
  {\bibfnamefont {P.}~\bibnamefont {Ottanelli}}, \bibinfo {author}
  {\bibfnamefont {G.}~\bibnamefont {Pasquali}},\ and\ \bibinfo {author}
  {\bibfnamefont {G.}~\bibnamefont {Pastore}},\ }\href
  {https://doi.org/10.1103/PhysRevC.96.034622} {\bibfield  {journal} {\bibinfo
  {journal} {Phys. Rev. C}\ }\textbf {\bibinfo {volume} {96}},\ \bibinfo
  {pages} {034622} (\bibinfo {year} {2017})}\BibitemShut {NoStop}%
\bibitem [{\citenamefont {Wuenschel}\ \emph {et~al.}(2009)\citenamefont
  {Wuenschel}, \citenamefont {Hagel}, \citenamefont {Wada}, \citenamefont
  {Natowitz}, \citenamefont {Yennello}, \citenamefont {Kohley}, \citenamefont
  {Bottosso}, \citenamefont {May}, \citenamefont {Smith}, \citenamefont
  {Shetty}, \citenamefont {Stein}, \citenamefont {Soisson},\ and\ \citenamefont
  {Prete}}]{Wuenschel2009}%
  \BibitemOpen
  \bibfield  {author} {\bibinfo {author} {\bibfnamefont {S.}~\bibnamefont
  {Wuenschel}}, \bibinfo {author} {\bibfnamefont {K.}~\bibnamefont {Hagel}},
  \bibinfo {author} {\bibfnamefont {R.}~\bibnamefont {Wada}}, \bibinfo {author}
  {\bibfnamefont {J.}~\bibnamefont {Natowitz}}, \bibinfo {author}
  {\bibfnamefont {S.}~\bibnamefont {Yennello}}, \bibinfo {author}
  {\bibfnamefont {Z.}~\bibnamefont {Kohley}}, \bibinfo {author} {\bibfnamefont
  {C.}~\bibnamefont {Bottosso}}, \bibinfo {author} {\bibfnamefont
  {L.}~\bibnamefont {May}}, \bibinfo {author} {\bibfnamefont {W.}~\bibnamefont
  {Smith}}, \bibinfo {author} {\bibfnamefont {D.}~\bibnamefont {Shetty}},
  \bibinfo {author} {\bibfnamefont {B.}~\bibnamefont {Stein}}, \bibinfo
  {author} {\bibfnamefont {S.}~\bibnamefont {Soisson}},\ and\ \bibinfo {author}
  {\bibfnamefont {G.}~\bibnamefont {Prete}},\ }\href
  {https://doi.org/https://doi.org/10.1016/j.nima.2009.03.187} {\bibfield
  {journal} {\bibinfo  {journal} {Nucl. Instr. and Meth. in Phys. Res. A}\
  }\textbf {\bibinfo {volume} {604}},\ \bibinfo {pages} {578} (\bibinfo {year}
  {2009})}\BibitemShut {NoStop}%
\bibitem [{\citenamefont {Rami}\ \emph {et~al.}(2000)\citenamefont {Rami},
  \citenamefont {Leifels}, \citenamefont {de~Schauenburg}, \citenamefont
  {Gobbi}, \citenamefont {Hong}, \citenamefont {Alard}, \citenamefont
  {Andronic}, \citenamefont {Averbeck}, \citenamefont {Barret}, \citenamefont
  {Basrak}, \citenamefont {Bastid}, \citenamefont {Belyaev}, \citenamefont
  {Bendarag}, \citenamefont {Berek}, \citenamefont {\v{C}aplar}, \citenamefont
  {Cindro}, \citenamefont {Crochet}, \citenamefont {Devismes}, \citenamefont
  {Dupieux}, \citenamefont {D\v{z}elalija}, \citenamefont {Eskef},
  \citenamefont {Finck}, \citenamefont {Fodor}, \citenamefont {Folger},
  \citenamefont {Fraysse}, \citenamefont {Genoux-Lubain}, \citenamefont
  {Grigorian}, \citenamefont {Grishkin}, \citenamefont {Herrmann},
  \citenamefont {Hildenbrand}, \citenamefont {Kecskemeti}, \citenamefont {Kim},
  \citenamefont {Koczon}, \citenamefont {Kirejczyk}, \citenamefont {Korolija},
  \citenamefont {Kotte}, \citenamefont {Kowalczyk}, \citenamefont {Kress},
  \citenamefont {Kutsche}, \citenamefont {Lebedev}, \citenamefont {Lee},
  \citenamefont {Manko}, \citenamefont {Merlitz}, \citenamefont {Mohren},
  \citenamefont {Moisa}, \citenamefont {Mösner}, \citenamefont {Neubert},
  \citenamefont {Nianine}, \citenamefont {Pelte}, \citenamefont {Petrovici},
  \citenamefont {Pinkenburg}, \citenamefont {Plettner}, \citenamefont
  {Reisdorf}, \citenamefont {Ritman}, \citenamefont {Schüll}, \citenamefont
  {Seres}, \citenamefont {Sikora}, \citenamefont {Sim}, \citenamefont {Simion},
  \citenamefont {Siwek-Wilczy\'{n}ska}, \citenamefont {Somov}, \citenamefont
  {Stockmeier}, \citenamefont {Stoicea}, \citenamefont {Vasiliev},
  \citenamefont {Wagner}, \citenamefont {Wi\'{s}niewski}, \citenamefont
  {Wohlfarth}, \citenamefont {Yang}, \citenamefont {Yushmanov},\ and\
  \citenamefont {Zhilin}}]{Rami2000}%
  \BibitemOpen
  \bibfield  {author} {\bibinfo {author} {\bibfnamefont {F.}~\bibnamefont
  {Rami}}, \bibinfo {author} {\bibfnamefont {Y.}~\bibnamefont {Leifels}},
  \bibinfo {author} {\bibfnamefont {B.}~\bibnamefont {de~Schauenburg}},
  \bibinfo {author} {\bibfnamefont {A.}~\bibnamefont {Gobbi}}, \bibinfo
  {author} {\bibfnamefont {B.}~\bibnamefont {Hong}}, \bibinfo {author}
  {\bibfnamefont {J.~P.}\ \bibnamefont {Alard}}, \bibinfo {author}
  {\bibfnamefont {A.}~\bibnamefont {Andronic}}, \bibinfo {author}
  {\bibfnamefont {R.}~\bibnamefont {Averbeck}}, \bibinfo {author}
  {\bibfnamefont {V.}~\bibnamefont {Barret}}, \bibinfo {author} {\bibfnamefont
  {Z.}~\bibnamefont {Basrak}}, \bibinfo {author} {\bibfnamefont
  {N.}~\bibnamefont {Bastid}}, \bibinfo {author} {\bibfnamefont
  {I.}~\bibnamefont {Belyaev}}, \bibinfo {author} {\bibfnamefont
  {A.}~\bibnamefont {Bendarag}}, \bibinfo {author} {\bibfnamefont
  {G.}~\bibnamefont {Berek}}, \bibinfo {author} {\bibfnamefont
  {R.}~\bibnamefont {\v{C}aplar}}, \bibinfo {author} {\bibfnamefont
  {N.}~\bibnamefont {Cindro}}, \bibinfo {author} {\bibfnamefont
  {P.}~\bibnamefont {Crochet}}, \bibinfo {author} {\bibfnamefont
  {A.}~\bibnamefont {Devismes}}, \bibinfo {author} {\bibfnamefont
  {P.}~\bibnamefont {Dupieux}}, \bibinfo {author} {\bibfnamefont
  {M.}~\bibnamefont {D\v{z}elalija}}, \bibinfo {author} {\bibfnamefont
  {M.}~\bibnamefont {Eskef}}, \bibinfo {author} {\bibfnamefont
  {C.}~\bibnamefont {Finck}}, \bibinfo {author} {\bibfnamefont
  {Z.}~\bibnamefont {Fodor}}, \bibinfo {author} {\bibfnamefont
  {H.}~\bibnamefont {Folger}}, \bibinfo {author} {\bibfnamefont
  {L.}~\bibnamefont {Fraysse}}, \bibinfo {author} {\bibfnamefont
  {A.}~\bibnamefont {Genoux-Lubain}}, \bibinfo {author} {\bibfnamefont
  {Y.}~\bibnamefont {Grigorian}}, \bibinfo {author} {\bibfnamefont
  {Y.}~\bibnamefont {Grishkin}}, \bibinfo {author} {\bibfnamefont
  {N.}~\bibnamefont {Herrmann}}, \bibinfo {author} {\bibfnamefont {K.~D.}\
  \bibnamefont {Hildenbrand}}, \bibinfo {author} {\bibfnamefont
  {J.}~\bibnamefont {Kecskemeti}}, \bibinfo {author} {\bibfnamefont {Y.~J.}\
  \bibnamefont {Kim}}, \bibinfo {author} {\bibfnamefont {P.}~\bibnamefont
  {Koczon}}, \bibinfo {author} {\bibfnamefont {M.}~\bibnamefont {Kirejczyk}},
  \bibinfo {author} {\bibfnamefont {M.}~\bibnamefont {Korolija}}, \bibinfo
  {author} {\bibfnamefont {R.}~\bibnamefont {Kotte}}, \bibinfo {author}
  {\bibfnamefont {M.}~\bibnamefont {Kowalczyk}}, \bibinfo {author}
  {\bibfnamefont {T.}~\bibnamefont {Kress}}, \bibinfo {author} {\bibfnamefont
  {R.}~\bibnamefont {Kutsche}}, \bibinfo {author} {\bibfnamefont
  {A.}~\bibnamefont {Lebedev}}, \bibinfo {author} {\bibfnamefont {K.~S.}\
  \bibnamefont {Lee}}, \bibinfo {author} {\bibfnamefont {V.}~\bibnamefont
  {Manko}}, \bibinfo {author} {\bibfnamefont {H.}~\bibnamefont {Merlitz}},
  \bibinfo {author} {\bibfnamefont {S.}~\bibnamefont {Mohren}}, \bibinfo
  {author} {\bibfnamefont {D.}~\bibnamefont {Moisa}}, \bibinfo {author}
  {\bibfnamefont {J.}~\bibnamefont {Mösner}}, \bibinfo {author} {\bibfnamefont
  {W.}~\bibnamefont {Neubert}}, \bibinfo {author} {\bibfnamefont
  {A.}~\bibnamefont {Nianine}}, \bibinfo {author} {\bibfnamefont
  {D.}~\bibnamefont {Pelte}}, \bibinfo {author} {\bibfnamefont
  {M.}~\bibnamefont {Petrovici}}, \bibinfo {author} {\bibfnamefont
  {C.}~\bibnamefont {Pinkenburg}}, \bibinfo {author} {\bibfnamefont
  {C.}~\bibnamefont {Plettner}}, \bibinfo {author} {\bibfnamefont
  {W.}~\bibnamefont {Reisdorf}}, \bibinfo {author} {\bibfnamefont
  {J.}~\bibnamefont {Ritman}}, \bibinfo {author} {\bibfnamefont
  {D.}~\bibnamefont {Schüll}}, \bibinfo {author} {\bibfnamefont
  {Z.}~\bibnamefont {Seres}}, \bibinfo {author} {\bibfnamefont
  {B.}~\bibnamefont {Sikora}}, \bibinfo {author} {\bibfnamefont {K.~S.}\
  \bibnamefont {Sim}}, \bibinfo {author} {\bibfnamefont {V.}~\bibnamefont
  {Simion}}, \bibinfo {author} {\bibfnamefont {K.}~\bibnamefont
  {Siwek-Wilczy\'{n}ska}}, \bibinfo {author} {\bibfnamefont {A.}~\bibnamefont
  {Somov}}, \bibinfo {author} {\bibfnamefont {M.~R.}\ \bibnamefont
  {Stockmeier}}, \bibinfo {author} {\bibfnamefont {G.}~\bibnamefont {Stoicea}},
  \bibinfo {author} {\bibfnamefont {M.}~\bibnamefont {Vasiliev}}, \bibinfo
  {author} {\bibfnamefont {P.}~\bibnamefont {Wagner}}, \bibinfo {author}
  {\bibfnamefont {K.}~\bibnamefont {Wi\'{s}niewski}}, \bibinfo {author}
  {\bibfnamefont {D.}~\bibnamefont {Wohlfarth}}, \bibinfo {author}
  {\bibfnamefont {J.~T.}\ \bibnamefont {Yang}}, \bibinfo {author}
  {\bibfnamefont {I.}~\bibnamefont {Yushmanov}},\ and\ \bibinfo {author}
  {\bibfnamefont {A.}~\bibnamefont {Zhilin}} (\bibinfo {collaboration} {FOPI
  Collaboration}),\ }\href
  {https://doi.org/https://doi.org/10.1103/PhysRevLett.84.1120} {\bibfield
  {journal} {\bibinfo  {journal} {Phys. Rev. Lett.}\ }\textbf {\bibinfo
  {volume} {84}},\ \bibinfo {pages} {1120} (\bibinfo {year}
  {2000})}\BibitemShut {NoStop}%
\bibitem [{\citenamefont {Camaiani}\ \emph {et~al.}(2020)\citenamefont
  {Camaiani}, \citenamefont {Piantelli}, \citenamefont {Ono}, \citenamefont
  {Casini}, \citenamefont {Borderie}, \citenamefont {Bougault}, \citenamefont
  {Ciampi}, \citenamefont {Dueñas}, \citenamefont {Frosin}, \citenamefont
  {Frankland}, \citenamefont {Gruyer}, \citenamefont {LeNeindre}, \citenamefont
  {Lombardo}, \citenamefont {Mantovani}, \citenamefont {Ottanelli},
  \citenamefont {Parlog}, \citenamefont {Pasquali}, \citenamefont {Upadhyaya},
  \citenamefont {Valdré}, \citenamefont {Verde}, ,\ and\ \citenamefont
  {Vient}}]{Camaiani2020}%
  \BibitemOpen
  \bibfield  {author} {\bibinfo {author} {\bibfnamefont {A.}~\bibnamefont
  {Camaiani}}, \bibinfo {author} {\bibfnamefont {S.}~\bibnamefont {Piantelli}},
  \bibinfo {author} {\bibfnamefont {A.}~\bibnamefont {Ono}}, \bibinfo {author}
  {\bibfnamefont {G.}~\bibnamefont {Casini}}, \bibinfo {author} {\bibfnamefont
  {B.}~\bibnamefont {Borderie}}, \bibinfo {author} {\bibfnamefont
  {R.}~\bibnamefont {Bougault}}, \bibinfo {author} {\bibfnamefont
  {C.}~\bibnamefont {Ciampi}}, \bibinfo {author} {\bibfnamefont {J.~A.}\
  \bibnamefont {Dueñas}}, \bibinfo {author} {\bibfnamefont {C.}~\bibnamefont
  {Frosin}}, \bibinfo {author} {\bibfnamefont {J.~D.}\ \bibnamefont
  {Frankland}}, \bibinfo {author} {\bibfnamefont {D.}~\bibnamefont {Gruyer}},
  \bibinfo {author} {\bibfnamefont {N.}~\bibnamefont {LeNeindre}}, \bibinfo
  {author} {\bibfnamefont {I.}~\bibnamefont {Lombardo}}, \bibinfo {author}
  {\bibfnamefont {G.}~\bibnamefont {Mantovani}}, \bibinfo {author}
  {\bibfnamefont {P.}~\bibnamefont {Ottanelli}}, \bibinfo {author}
  {\bibfnamefont {M.}~\bibnamefont {Parlog}}, \bibinfo {author} {\bibfnamefont
  {G.}~\bibnamefont {Pasquali}}, \bibinfo {author} {\bibfnamefont
  {S.}~\bibnamefont {Upadhyaya}}, \bibinfo {author} {\bibfnamefont
  {S.}~\bibnamefont {Valdré}}, \bibinfo {author} {\bibfnamefont
  {G.}~\bibnamefont {Verde}}, ,\ and\ \bibinfo {author} {\bibfnamefont
  {E.}~\bibnamefont {Vient}},\ }\href
  {https://doi.org/10.1103/PhysRevC.102.044607} {\bibfield  {journal} {\bibinfo
   {journal} {Phys. Rev. C}\ }\textbf {\bibinfo {volume} {102}},\ \bibinfo
  {pages} {044607} (\bibinfo {year} {2020})}\BibitemShut {NoStop}%
\bibitem [{\citenamefont {Tsang}\ \emph {et~al.}(2009)\citenamefont {Tsang},
  \citenamefont {Zhang}, \citenamefont {Danielewicz}, \citenamefont {Famiano},
  \citenamefont {Li}, \citenamefont {Lynch}, ,\ and\ \citenamefont
  {Steiner}}]{Tsang2009}%
  \BibitemOpen
  \bibfield  {author} {\bibinfo {author} {\bibfnamefont {M.~B.}\ \bibnamefont
  {Tsang}}, \bibinfo {author} {\bibfnamefont {Y.}~\bibnamefont {Zhang}},
  \bibinfo {author} {\bibfnamefont {P.}~\bibnamefont {Danielewicz}}, \bibinfo
  {author} {\bibfnamefont {M.}~\bibnamefont {Famiano}}, \bibinfo {author}
  {\bibfnamefont {Z.}~\bibnamefont {Li}}, \bibinfo {author} {\bibfnamefont
  {W.~G.}\ \bibnamefont {Lynch}}, ,\ and\ \bibinfo {author} {\bibfnamefont
  {A.~W.}\ \bibnamefont {Steiner}},\ }\href
  {https://doi.org/10.1103/PhysRevLett.102.122701} {\bibfield  {journal}
  {\bibinfo  {journal} {Phys. Rev. Lett.}\ }\textbf {\bibinfo {volume} {102}},\
  \bibinfo {pages} {122701} (\bibinfo {year} {2009})}\BibitemShut {NoStop}%
\bibitem [{\citenamefont {Mallik}\ \emph {et~al.}(2021)\citenamefont {Mallik},
  \citenamefont {Gulminelli},\ and\ \citenamefont {Gruyer}}]{Mallik2022}%
  \BibitemOpen
  \bibfield  {author} {\bibinfo {author} {\bibfnamefont {S.}~\bibnamefont
  {Mallik}}, \bibinfo {author} {\bibfnamefont {F.}~\bibnamefont {Gulminelli}},\
  and\ \bibinfo {author} {\bibfnamefont {D.}~\bibnamefont {Gruyer}},\ }\href
  {https://doi.org/doi.org/10.1088/1361-6471/ac3473} {\bibfield  {journal}
  {\bibinfo  {journal} {J. Phys. G: Nucl. Part. Phys.}\ }\textbf {\bibinfo
  {volume} {49}},\ \bibinfo {pages} {015102} (\bibinfo {year}
  {2021})}\BibitemShut {NoStop}%
\bibitem [{\citenamefont {Sun}\ \emph {et~al.}(2010)\citenamefont {Sun},
  \citenamefont {Tsang}, \citenamefont {Lynch}, \citenamefont {Verde},
  \citenamefont {Amorini}, \citenamefont {Andronenko}, \citenamefont
  {Andronenko}, \citenamefont {Cardella}, \citenamefont {Chatterje},
  \citenamefont {Danielewicz}, \citenamefont {De~Filippo}, \citenamefont
  {Dinh}, \citenamefont {Galichet}, \citenamefont {Geraci}, \citenamefont
  {Hua}, \citenamefont {La~Guidara}, \citenamefont {Lanzalone}, \citenamefont
  {Liu}, \citenamefont {Lu}, \citenamefont {Lukyanov}, \citenamefont
  {Maiolino}, \citenamefont {Pagano}, \citenamefont {Piantelli}, \citenamefont
  {Papa}, \citenamefont {Pirrone}, \citenamefont {Politi}, \citenamefont
  {Porto}, \citenamefont {Rizzo}, \citenamefont {Russotto}, \citenamefont
  {Santonocito},\ and\ \citenamefont {Zhang}}]{Sun2010}%
  \BibitemOpen
  \bibfield  {author} {\bibinfo {author} {\bibfnamefont {Z.~Y.}\ \bibnamefont
  {Sun}}, \bibinfo {author} {\bibfnamefont {M.~B.}\ \bibnamefont {Tsang}},
  \bibinfo {author} {\bibfnamefont {W.~G.}\ \bibnamefont {Lynch}}, \bibinfo
  {author} {\bibfnamefont {G.}~\bibnamefont {Verde}}, \bibinfo {author}
  {\bibfnamefont {F.}~\bibnamefont {Amorini}}, \bibinfo {author} {\bibfnamefont
  {L.}~\bibnamefont {Andronenko}}, \bibinfo {author} {\bibfnamefont
  {M.}~\bibnamefont {Andronenko}}, \bibinfo {author} {\bibfnamefont
  {G.}~\bibnamefont {Cardella}}, \bibinfo {author} {\bibfnamefont
  {M.}~\bibnamefont {Chatterje}}, \bibinfo {author} {\bibfnamefont
  {P.}~\bibnamefont {Danielewicz}}, \bibinfo {author} {\bibfnamefont
  {E.}~\bibnamefont {De~Filippo}}, \bibinfo {author} {\bibfnamefont
  {P.}~\bibnamefont {Dinh}}, \bibinfo {author} {\bibfnamefont {E.}~\bibnamefont
  {Galichet}}, \bibinfo {author} {\bibfnamefont {E.}~\bibnamefont {Geraci}},
  \bibinfo {author} {\bibfnamefont {H.}~\bibnamefont {Hua}}, \bibinfo {author}
  {\bibfnamefont {E.}~\bibnamefont {La~Guidara}}, \bibinfo {author}
  {\bibfnamefont {G.}~\bibnamefont {Lanzalone}}, \bibinfo {author}
  {\bibfnamefont {H.}~\bibnamefont {Liu}}, \bibinfo {author} {\bibfnamefont
  {F.}~\bibnamefont {Lu}}, \bibinfo {author} {\bibfnamefont {S.}~\bibnamefont
  {Lukyanov}}, \bibinfo {author} {\bibfnamefont {C.}~\bibnamefont {Maiolino}},
  \bibinfo {author} {\bibfnamefont {A.}~\bibnamefont {Pagano}}, \bibinfo
  {author} {\bibfnamefont {S.}~\bibnamefont {Piantelli}}, \bibinfo {author}
  {\bibfnamefont {M.}~\bibnamefont {Papa}}, \bibinfo {author} {\bibfnamefont
  {S.}~\bibnamefont {Pirrone}}, \bibinfo {author} {\bibfnamefont
  {G.}~\bibnamefont {Politi}}, \bibinfo {author} {\bibfnamefont
  {F.}~\bibnamefont {Porto}}, \bibinfo {author} {\bibfnamefont
  {F.}~\bibnamefont {Rizzo}}, \bibinfo {author} {\bibfnamefont
  {P.}~\bibnamefont {Russotto}}, \bibinfo {author} {\bibfnamefont
  {D.}~\bibnamefont {Santonocito}},\ and\ \bibinfo {author} {\bibfnamefont
  {Y.~X.}\ \bibnamefont {Zhang}},\ }\href
  {https://doi.org/10.1103/PhysRevC.82.051603} {\bibfield  {journal} {\bibinfo
  {journal} {Phys. Rev. C}\ }\textbf {\bibinfo {volume} {82}},\ \bibinfo
  {pages} {051603} (\bibinfo {year} {2010})}\BibitemShut {NoStop}%
\bibitem [{\citenamefont {Napolitani}\ \emph {et~al.}(2010)\citenamefont
  {Napolitani}, \citenamefont {Colonna}, \citenamefont {Gulminelli},
  \citenamefont {Galichet}, \citenamefont {Piantelli}, \citenamefont {Verde},\
  and\ \citenamefont {Vient}}]{Napolitani2010}%
  \BibitemOpen
  \bibfield  {author} {\bibinfo {author} {\bibfnamefont {P.}~\bibnamefont
  {Napolitani}}, \bibinfo {author} {\bibfnamefont {M.}~\bibnamefont {Colonna}},
  \bibinfo {author} {\bibfnamefont {F.}~\bibnamefont {Gulminelli}}, \bibinfo
  {author} {\bibfnamefont {E.}~\bibnamefont {Galichet}}, \bibinfo {author}
  {\bibfnamefont {S.}~\bibnamefont {Piantelli}}, \bibinfo {author}
  {\bibfnamefont {G.}~\bibnamefont {Verde}},\ and\ \bibinfo {author}
  {\bibfnamefont {E.}~\bibnamefont {Vient}},\ }\href
  {https://doi.org/10.1103/PhysRevC.81.044619} {\bibfield  {journal} {\bibinfo
  {journal} {Phys. Rev. C}\ }\textbf {\bibinfo {volume} {81}},\ \bibinfo
  {pages} {044619} (\bibinfo {year} {2010})}\BibitemShut {NoStop}%
\bibitem [{\citenamefont {Ono}\ \emph {et~al.}(1992)\citenamefont {Ono},
  \citenamefont {Horiuchi}, \citenamefont {Maruyama},\ and\ \citenamefont
  {Ohnishi}}]{Ono1992}%
  \BibitemOpen
  \bibfield  {author} {\bibinfo {author} {\bibfnamefont {A.}~\bibnamefont
  {Ono}}, \bibinfo {author} {\bibfnamefont {H.}~\bibnamefont {Horiuchi}},
  \bibinfo {author} {\bibfnamefont {T.}~\bibnamefont {Maruyama}},\ and\
  \bibinfo {author} {\bibfnamefont {A.}~\bibnamefont {Ohnishi}},\ }\href
  {https://doi.org/10.1143/ptp/87.5.1185} {\bibfield  {journal} {\bibinfo
  {journal} {Prog. Theor. Phys.}\ }\textbf {\bibinfo {volume} {87}},\ \bibinfo
  {pages} {1185} (\bibinfo {year} {1992})}\BibitemShut {NoStop}%
\bibitem [{\citenamefont {Charity}(2010)}]{Charity2010}%
  \BibitemOpen
  \bibfield  {author} {\bibinfo {author} {\bibfnamefont {R.~J.}\ \bibnamefont
  {Charity}},\ }\href {https://doi.org/10.1103/PhysRevC.82.014610} {\bibfield
  {journal} {\bibinfo  {journal} {Phys. Rev. C}\ }\textbf {\bibinfo {volume}
  {82}},\ \bibinfo {pages} {014610} (\bibinfo {year} {2010})}\BibitemShut
  {NoStop}%
\bibitem [{\citenamefont {Bougault}\ \emph {et~al.}(2014)\citenamefont
  {Bougault}, \citenamefont {Poggi}, \citenamefont {Barlini}, \citenamefont
  {Borderie}, \citenamefont {Casini}, \citenamefont {Chbihi}, \citenamefont
  {Neindre}, \citenamefont {P\^{a}rlog}, \citenamefont {Pasquali},
  \citenamefont {Piantelli}, \citenamefont {Sosin}, \citenamefont {Ademard},
  \citenamefont {Alba}, \citenamefont {Anastasio}, \citenamefont {Barbey},
  \citenamefont {Bardelli}, \citenamefont {Bini}, \citenamefont {Boiano},
  \citenamefont {Boisjoli}, \citenamefont {Bonnet}, \citenamefont {Borcea},
  \citenamefont {Bougard}, \citenamefont {Brulin}, \citenamefont {Bruno},
  \citenamefont {Carboni}, \citenamefont {Cassese}, \citenamefont {Cassese},
  \citenamefont {Cinausero}, \citenamefont {Ciolacu}, \citenamefont {Cruceru},
  \citenamefont {Cruceru}, \citenamefont {D'Aquino}, \citenamefont {Fazio},
  \citenamefont {Degerlier}, \citenamefont {Desrues}, \citenamefont {Meo},
  \citenamefont {Dueñas}, \citenamefont {Edelbruck}, \citenamefont {Energico},
  \citenamefont {Falorsi}, \citenamefont {Frankland}, \citenamefont {Galichet},
  \citenamefont {Gasior}, \citenamefont {Gramegna}, \citenamefont {Giordano},
  \citenamefont {Gruyer}, \citenamefont {Grzeszczuk}, \citenamefont {Guerzoni},
  \citenamefont {Hamrita}, \citenamefont {Huss}, \citenamefont {Kajetanowicz},
  \citenamefont {Korcyl}, \citenamefont {Kordyasz}, \citenamefont {Kozik},
  \citenamefont {Kulig}, \citenamefont {Lavergne}, \citenamefont {Legou\'{e}e},
  \citenamefont {Lopez}, \citenamefont {\L{}ukasik}, \citenamefont {Maiolino},
  \citenamefont {Marchi}, \citenamefont {Marini}, \citenamefont {Martel},
  \citenamefont {Masone}, \citenamefont {Meoli}, \citenamefont {Merrer},
  \citenamefont {Morelli}, \citenamefont {Negoita}, \citenamefont {Olmi},
  \citenamefont {Ordine}, \citenamefont {Paduano}, \citenamefont {Pain},
  \citenamefont {Pa\l{}ka}, \citenamefont {Passeggio}, \citenamefont {Pastore},
  \citenamefont {Paw\l{}owski}, \citenamefont {Petcu}, \citenamefont
  {Petrascu}, \citenamefont {Piasecki}, \citenamefont {Pontoriere},
  \citenamefont {Rauly}, \citenamefont {Rivet}, \citenamefont {Rocco},
  \citenamefont {Rosato}, \citenamefont {Roscilli}, \citenamefont {Scarlini},
  \citenamefont {Salomon}, \citenamefont {Santonocito}, \citenamefont
  {Seredov}, \citenamefont {Serra}, \citenamefont {Sierpowski}, \citenamefont
  {Spadaccini}, \citenamefont {Spitaels}, \citenamefont {Stefanini},
  \citenamefont {Tobia}, \citenamefont {Tortone}, \citenamefont {Twar\'{o}g},
  \citenamefont {Valdr\'{e}}, \citenamefont {Vanzanella}, \citenamefont
  {Vanzanella}, \citenamefont {Vient}, \citenamefont {Vigilante}, \citenamefont
  {Vitiello}, \citenamefont {Wanlin}, \citenamefont {Wieloch},\ and\
  \citenamefont {Zipper}}]{Bougault2014}%
  \BibitemOpen
  \bibfield  {author} {\bibinfo {author} {\bibfnamefont {R.}~\bibnamefont
  {Bougault}}, \bibinfo {author} {\bibfnamefont {G.}~\bibnamefont {Poggi}},
  \bibinfo {author} {\bibfnamefont {S.}~\bibnamefont {Barlini}}, \bibinfo
  {author} {\bibfnamefont {B.}~\bibnamefont {Borderie}}, \bibinfo {author}
  {\bibfnamefont {G.}~\bibnamefont {Casini}}, \bibinfo {author} {\bibfnamefont
  {A.}~\bibnamefont {Chbihi}}, \bibinfo {author} {\bibfnamefont {N.~L.}\
  \bibnamefont {Neindre}}, \bibinfo {author} {\bibfnamefont {M.}~\bibnamefont
  {P\^{a}rlog}}, \bibinfo {author} {\bibfnamefont {G.}~\bibnamefont
  {Pasquali}}, \bibinfo {author} {\bibfnamefont {S.}~\bibnamefont {Piantelli}},
  \bibinfo {author} {\bibfnamefont {Z.}~\bibnamefont {Sosin}}, \bibinfo
  {author} {\bibfnamefont {G.}~\bibnamefont {Ademard}}, \bibinfo {author}
  {\bibfnamefont {R.}~\bibnamefont {Alba}}, \bibinfo {author} {\bibfnamefont
  {A.}~\bibnamefont {Anastasio}}, \bibinfo {author} {\bibfnamefont
  {S.}~\bibnamefont {Barbey}}, \bibinfo {author} {\bibfnamefont
  {L.}~\bibnamefont {Bardelli}}, \bibinfo {author} {\bibfnamefont
  {M.}~\bibnamefont {Bini}}, \bibinfo {author} {\bibfnamefont {A.}~\bibnamefont
  {Boiano}}, \bibinfo {author} {\bibfnamefont {M.}~\bibnamefont {Boisjoli}},
  \bibinfo {author} {\bibfnamefont {E.}~\bibnamefont {Bonnet}}, \bibinfo
  {author} {\bibfnamefont {R.}~\bibnamefont {Borcea}}, \bibinfo {author}
  {\bibfnamefont {B.}~\bibnamefont {Bougard}}, \bibinfo {author} {\bibfnamefont
  {G.}~\bibnamefont {Brulin}}, \bibinfo {author} {\bibfnamefont
  {M.}~\bibnamefont {Bruno}}, \bibinfo {author} {\bibfnamefont
  {S.}~\bibnamefont {Carboni}}, \bibinfo {author} {\bibfnamefont
  {C.}~\bibnamefont {Cassese}}, \bibinfo {author} {\bibfnamefont
  {F.}~\bibnamefont {Cassese}}, \bibinfo {author} {\bibfnamefont
  {M.}~\bibnamefont {Cinausero}}, \bibinfo {author} {\bibfnamefont
  {L.}~\bibnamefont {Ciolacu}}, \bibinfo {author} {\bibfnamefont
  {I.}~\bibnamefont {Cruceru}}, \bibinfo {author} {\bibfnamefont
  {M.}~\bibnamefont {Cruceru}}, \bibinfo {author} {\bibfnamefont
  {B.}~\bibnamefont {D'Aquino}}, \bibinfo {author} {\bibfnamefont {B.~D.}\
  \bibnamefont {Fazio}}, \bibinfo {author} {\bibfnamefont {M.}~\bibnamefont
  {Degerlier}}, \bibinfo {author} {\bibfnamefont {P.}~\bibnamefont {Desrues}},
  \bibinfo {author} {\bibfnamefont {P.~D.}\ \bibnamefont {Meo}}, \bibinfo
  {author} {\bibfnamefont {J.~A.}\ \bibnamefont {Dueñas}}, \bibinfo {author}
  {\bibfnamefont {P.}~\bibnamefont {Edelbruck}}, \bibinfo {author}
  {\bibfnamefont {S.}~\bibnamefont {Energico}}, \bibinfo {author}
  {\bibfnamefont {M.}~\bibnamefont {Falorsi}}, \bibinfo {author} {\bibfnamefont
  {J.~D.}\ \bibnamefont {Frankland}}, \bibinfo {author} {\bibfnamefont
  {E.}~\bibnamefont {Galichet}}, \bibinfo {author} {\bibfnamefont
  {K.}~\bibnamefont {Gasior}}, \bibinfo {author} {\bibfnamefont
  {F.}~\bibnamefont {Gramegna}}, \bibinfo {author} {\bibfnamefont
  {R.}~\bibnamefont {Giordano}}, \bibinfo {author} {\bibfnamefont
  {D.}~\bibnamefont {Gruyer}}, \bibinfo {author} {\bibfnamefont
  {A.}~\bibnamefont {Grzeszczuk}}, \bibinfo {author} {\bibfnamefont
  {M.}~\bibnamefont {Guerzoni}}, \bibinfo {author} {\bibfnamefont
  {H.}~\bibnamefont {Hamrita}}, \bibinfo {author} {\bibfnamefont
  {C.}~\bibnamefont {Huss}}, \bibinfo {author} {\bibfnamefont {M.}~\bibnamefont
  {Kajetanowicz}}, \bibinfo {author} {\bibfnamefont {K.}~\bibnamefont
  {Korcyl}}, \bibinfo {author} {\bibfnamefont {A.}~\bibnamefont {Kordyasz}},
  \bibinfo {author} {\bibfnamefont {T.}~\bibnamefont {Kozik}}, \bibinfo
  {author} {\bibfnamefont {P.}~\bibnamefont {Kulig}}, \bibinfo {author}
  {\bibfnamefont {L.}~\bibnamefont {Lavergne}}, \bibinfo {author}
  {\bibfnamefont {E.}~\bibnamefont {Legou\'{e}e}}, \bibinfo {author}
  {\bibfnamefont {O.}~\bibnamefont {Lopez}}, \bibinfo {author} {\bibfnamefont
  {J.}~\bibnamefont {\L{}ukasik}}, \bibinfo {author} {\bibfnamefont
  {C.}~\bibnamefont {Maiolino}}, \bibinfo {author} {\bibfnamefont
  {T.}~\bibnamefont {Marchi}}, \bibinfo {author} {\bibfnamefont
  {P.}~\bibnamefont {Marini}}, \bibinfo {author} {\bibfnamefont
  {I.}~\bibnamefont {Martel}}, \bibinfo {author} {\bibfnamefont
  {V.}~\bibnamefont {Masone}}, \bibinfo {author} {\bibfnamefont
  {A.}~\bibnamefont {Meoli}}, \bibinfo {author} {\bibfnamefont
  {Y.}~\bibnamefont {Merrer}}, \bibinfo {author} {\bibfnamefont
  {L.}~\bibnamefont {Morelli}}, \bibinfo {author} {\bibfnamefont
  {F.}~\bibnamefont {Negoita}}, \bibinfo {author} {\bibfnamefont
  {A.}~\bibnamefont {Olmi}}, \bibinfo {author} {\bibfnamefont {A.}~\bibnamefont
  {Ordine}}, \bibinfo {author} {\bibfnamefont {G.}~\bibnamefont {Paduano}},
  \bibinfo {author} {\bibfnamefont {C.}~\bibnamefont {Pain}}, \bibinfo {author}
  {\bibfnamefont {M.}~\bibnamefont {Pa\l{}ka}}, \bibinfo {author}
  {\bibfnamefont {G.}~\bibnamefont {Passeggio}}, \bibinfo {author}
  {\bibfnamefont {G.}~\bibnamefont {Pastore}}, \bibinfo {author} {\bibfnamefont
  {P.}~\bibnamefont {Paw\l{}owski}}, \bibinfo {author} {\bibfnamefont
  {M.}~\bibnamefont {Petcu}}, \bibinfo {author} {\bibfnamefont
  {H.}~\bibnamefont {Petrascu}}, \bibinfo {author} {\bibfnamefont
  {E.}~\bibnamefont {Piasecki}}, \bibinfo {author} {\bibfnamefont
  {G.}~\bibnamefont {Pontoriere}}, \bibinfo {author} {\bibfnamefont
  {E.}~\bibnamefont {Rauly}}, \bibinfo {author} {\bibfnamefont {M.~F.}\
  \bibnamefont {Rivet}}, \bibinfo {author} {\bibfnamefont {R.}~\bibnamefont
  {Rocco}}, \bibinfo {author} {\bibfnamefont {E.}~\bibnamefont {Rosato}},
  \bibinfo {author} {\bibfnamefont {L.}~\bibnamefont {Roscilli}}, \bibinfo
  {author} {\bibfnamefont {E.}~\bibnamefont {Scarlini}}, \bibinfo {author}
  {\bibfnamefont {F.}~\bibnamefont {Salomon}}, \bibinfo {author} {\bibfnamefont
  {D.}~\bibnamefont {Santonocito}}, \bibinfo {author} {\bibfnamefont
  {V.}~\bibnamefont {Seredov}}, \bibinfo {author} {\bibfnamefont
  {S.}~\bibnamefont {Serra}}, \bibinfo {author} {\bibfnamefont
  {D.}~\bibnamefont {Sierpowski}}, \bibinfo {author} {\bibfnamefont
  {G.}~\bibnamefont {Spadaccini}}, \bibinfo {author} {\bibfnamefont
  {C.}~\bibnamefont {Spitaels}}, \bibinfo {author} {\bibfnamefont {A.~A.}\
  \bibnamefont {Stefanini}}, \bibinfo {author} {\bibfnamefont {G.}~\bibnamefont
  {Tobia}}, \bibinfo {author} {\bibfnamefont {G.}~\bibnamefont {Tortone}},
  \bibinfo {author} {\bibfnamefont {T.}~\bibnamefont {Twar\'{o}g}}, \bibinfo
  {author} {\bibfnamefont {S.}~\bibnamefont {Valdr\'{e}}}, \bibinfo {author}
  {\bibfnamefont {A.}~\bibnamefont {Vanzanella}}, \bibinfo {author}
  {\bibfnamefont {E.}~\bibnamefont {Vanzanella}}, \bibinfo {author}
  {\bibfnamefont {E.}~\bibnamefont {Vient}}, \bibinfo {author} {\bibfnamefont
  {M.}~\bibnamefont {Vigilante}}, \bibinfo {author} {\bibfnamefont
  {G.}~\bibnamefont {Vitiello}}, \bibinfo {author} {\bibfnamefont
  {E.}~\bibnamefont {Wanlin}}, \bibinfo {author} {\bibfnamefont
  {A.}~\bibnamefont {Wieloch}},\ and\ \bibinfo {author} {\bibfnamefont
  {W.}~\bibnamefont {Zipper}},\ }\href
  {https://doi.org/https://doi.org/10.1140/epja/i2014-14047-4} {\bibfield
  {journal} {\bibinfo  {journal} {Eur. Phys. J. A}\ }\textbf {\bibinfo {volume}
  {50}},\ \bibinfo {pages} {47} (\bibinfo {year} {2014})}\BibitemShut {NoStop}%
\bibitem [{\citenamefont {Valdr\'{e}}\ \emph {et~al.}(2019)\citenamefont
  {Valdr\'{e}}, \citenamefont {Casini}, \citenamefont {{Le Neindre}},
  \citenamefont {Bini}, \citenamefont {Boiano}, \citenamefont {Borderie},
  \citenamefont {Edelbruck}, \citenamefont {Poggi}, \citenamefont {Salomon},
  \citenamefont {Tortone}, \citenamefont {Alba}, \citenamefont {Barlini},
  \citenamefont {Bonnet}, \citenamefont {Bougard}, \citenamefont {Bougault},
  \citenamefont {Brulin}, \citenamefont {Bruno}, \citenamefont {Buccola},
  \citenamefont {Camaiani}, \citenamefont {Chbihi}, \citenamefont {Ciampi},
  \citenamefont {Cicerchia}, \citenamefont {Cinausero}, \citenamefont
  {Dell’Aquila}, \citenamefont {Desrues}, \citenamefont {Dueñas},
  \citenamefont {Fabris}, \citenamefont {Falorsi}, \citenamefont {Frankland},
  \citenamefont {Frosin}, \citenamefont {Galichet}, \citenamefont {Giordano},
  \citenamefont {Gramegna}, \citenamefont {Grassi}, \citenamefont {Gruyer},
  \citenamefont {Guerzoni}, \citenamefont {Henri}, \citenamefont
  {Kajetanowicz}, \citenamefont {Korcyl}, \citenamefont {Kordyasz},
  \citenamefont {Kozik}, \citenamefont {Lecomte}, \citenamefont {Lombardo},
  \citenamefont {Lopez}, \citenamefont {Maiolino}, \citenamefont {Mantovani},
  \citenamefont {Marchi}, \citenamefont {Margotti}, \citenamefont {Merrer},
  \citenamefont {Morelli}, \citenamefont {Olmi}, \citenamefont {Ordine},
  \citenamefont {Ottanelli}, \citenamefont {Pain}, \citenamefont {Pałka},
  \citenamefont {Pârlog}, \citenamefont {Pasquali}, \citenamefont {Pastore},
  \citenamefont {Piantelli}, \citenamefont {{de Préaumont}}, \citenamefont
  {Revenko}, \citenamefont {Richard}, \citenamefont {Rivet}, \citenamefont
  {Ropert}, \citenamefont {Rosato}, \citenamefont {Saillant}, \citenamefont
  {Santonocito}, \citenamefont {Scarlini}, \citenamefont {Serra}, \citenamefont
  {Soulet}, \citenamefont {Spadaccini}, \citenamefont {Stefanini},
  \citenamefont {Tobia}, \citenamefont {Upadhyaya}, \citenamefont {Vanzanella},
  \citenamefont {Verde}, \citenamefont {Vient}, \citenamefont {Vigilante},
  \citenamefont {Wanlin}, \citenamefont {Wittwer},\ and\ \citenamefont
  {Zucchini}}]{Valdre2019}%
  \BibitemOpen
  \bibfield  {author} {\bibinfo {author} {\bibfnamefont {S.}~\bibnamefont
  {Valdr\'{e}}}, \bibinfo {author} {\bibfnamefont {G.}~\bibnamefont {Casini}},
  \bibinfo {author} {\bibfnamefont {N.}~\bibnamefont {{Le Neindre}}}, \bibinfo
  {author} {\bibfnamefont {M.}~\bibnamefont {Bini}}, \bibinfo {author}
  {\bibfnamefont {A.}~\bibnamefont {Boiano}}, \bibinfo {author} {\bibfnamefont
  {B.}~\bibnamefont {Borderie}}, \bibinfo {author} {\bibfnamefont
  {P.}~\bibnamefont {Edelbruck}}, \bibinfo {author} {\bibfnamefont
  {G.}~\bibnamefont {Poggi}}, \bibinfo {author} {\bibfnamefont
  {F.}~\bibnamefont {Salomon}}, \bibinfo {author} {\bibfnamefont
  {G.}~\bibnamefont {Tortone}}, \bibinfo {author} {\bibfnamefont
  {R.}~\bibnamefont {Alba}}, \bibinfo {author} {\bibfnamefont {S.}~\bibnamefont
  {Barlini}}, \bibinfo {author} {\bibfnamefont {E.}~\bibnamefont {Bonnet}},
  \bibinfo {author} {\bibfnamefont {B.}~\bibnamefont {Bougard}}, \bibinfo
  {author} {\bibfnamefont {R.}~\bibnamefont {Bougault}}, \bibinfo {author}
  {\bibfnamefont {G.}~\bibnamefont {Brulin}}, \bibinfo {author} {\bibfnamefont
  {M.}~\bibnamefont {Bruno}}, \bibinfo {author} {\bibfnamefont
  {A.}~\bibnamefont {Buccola}}, \bibinfo {author} {\bibfnamefont
  {A.}~\bibnamefont {Camaiani}}, \bibinfo {author} {\bibfnamefont
  {A.}~\bibnamefont {Chbihi}}, \bibinfo {author} {\bibfnamefont
  {C.}~\bibnamefont {Ciampi}}, \bibinfo {author} {\bibfnamefont
  {M.}~\bibnamefont {Cicerchia}}, \bibinfo {author} {\bibfnamefont
  {M.}~\bibnamefont {Cinausero}}, \bibinfo {author} {\bibfnamefont
  {D.}~\bibnamefont {Dell’Aquila}}, \bibinfo {author} {\bibfnamefont
  {P.}~\bibnamefont {Desrues}}, \bibinfo {author} {\bibfnamefont
  {J.}~\bibnamefont {Dueñas}}, \bibinfo {author} {\bibfnamefont
  {D.}~\bibnamefont {Fabris}}, \bibinfo {author} {\bibfnamefont
  {M.}~\bibnamefont {Falorsi}}, \bibinfo {author} {\bibfnamefont
  {J.}~\bibnamefont {Frankland}}, \bibinfo {author} {\bibfnamefont
  {C.}~\bibnamefont {Frosin}}, \bibinfo {author} {\bibfnamefont
  {E.}~\bibnamefont {Galichet}}, \bibinfo {author} {\bibfnamefont
  {R.}~\bibnamefont {Giordano}}, \bibinfo {author} {\bibfnamefont
  {F.}~\bibnamefont {Gramegna}}, \bibinfo {author} {\bibfnamefont
  {L.}~\bibnamefont {Grassi}}, \bibinfo {author} {\bibfnamefont
  {D.}~\bibnamefont {Gruyer}}, \bibinfo {author} {\bibfnamefont
  {M.}~\bibnamefont {Guerzoni}}, \bibinfo {author} {\bibfnamefont
  {M.}~\bibnamefont {Henri}}, \bibinfo {author} {\bibfnamefont
  {M.}~\bibnamefont {Kajetanowicz}}, \bibinfo {author} {\bibfnamefont
  {K.}~\bibnamefont {Korcyl}}, \bibinfo {author} {\bibfnamefont
  {A.}~\bibnamefont {Kordyasz}}, \bibinfo {author} {\bibfnamefont
  {T.}~\bibnamefont {Kozik}}, \bibinfo {author} {\bibfnamefont
  {P.}~\bibnamefont {Lecomte}}, \bibinfo {author} {\bibfnamefont
  {I.}~\bibnamefont {Lombardo}}, \bibinfo {author} {\bibfnamefont
  {O.}~\bibnamefont {Lopez}}, \bibinfo {author} {\bibfnamefont
  {C.}~\bibnamefont {Maiolino}}, \bibinfo {author} {\bibfnamefont
  {G.}~\bibnamefont {Mantovani}}, \bibinfo {author} {\bibfnamefont
  {T.}~\bibnamefont {Marchi}}, \bibinfo {author} {\bibfnamefont
  {A.}~\bibnamefont {Margotti}}, \bibinfo {author} {\bibfnamefont
  {Y.}~\bibnamefont {Merrer}}, \bibinfo {author} {\bibfnamefont
  {L.}~\bibnamefont {Morelli}}, \bibinfo {author} {\bibfnamefont
  {A.}~\bibnamefont {Olmi}}, \bibinfo {author} {\bibfnamefont {A.}~\bibnamefont
  {Ordine}}, \bibinfo {author} {\bibfnamefont {P.}~\bibnamefont {Ottanelli}},
  \bibinfo {author} {\bibfnamefont {C.}~\bibnamefont {Pain}}, \bibinfo {author}
  {\bibfnamefont {M.}~\bibnamefont {Pałka}}, \bibinfo {author} {\bibfnamefont
  {M.}~\bibnamefont {Pârlog}}, \bibinfo {author} {\bibfnamefont
  {G.}~\bibnamefont {Pasquali}}, \bibinfo {author} {\bibfnamefont
  {G.}~\bibnamefont {Pastore}}, \bibinfo {author} {\bibfnamefont
  {S.}~\bibnamefont {Piantelli}}, \bibinfo {author} {\bibfnamefont
  {H.}~\bibnamefont {{de Préaumont}}}, \bibinfo {author} {\bibfnamefont
  {R.}~\bibnamefont {Revenko}}, \bibinfo {author} {\bibfnamefont
  {A.}~\bibnamefont {Richard}}, \bibinfo {author} {\bibfnamefont
  {M.}~\bibnamefont {Rivet}}, \bibinfo {author} {\bibfnamefont
  {J.}~\bibnamefont {Ropert}}, \bibinfo {author} {\bibfnamefont
  {E.}~\bibnamefont {Rosato}}, \bibinfo {author} {\bibfnamefont
  {F.}~\bibnamefont {Saillant}}, \bibinfo {author} {\bibfnamefont
  {D.}~\bibnamefont {Santonocito}}, \bibinfo {author} {\bibfnamefont
  {E.}~\bibnamefont {Scarlini}}, \bibinfo {author} {\bibfnamefont
  {S.}~\bibnamefont {Serra}}, \bibinfo {author} {\bibfnamefont
  {C.}~\bibnamefont {Soulet}}, \bibinfo {author} {\bibfnamefont
  {G.}~\bibnamefont {Spadaccini}}, \bibinfo {author} {\bibfnamefont
  {A.}~\bibnamefont {Stefanini}}, \bibinfo {author} {\bibfnamefont
  {G.}~\bibnamefont {Tobia}}, \bibinfo {author} {\bibfnamefont
  {S.}~\bibnamefont {Upadhyaya}}, \bibinfo {author} {\bibfnamefont
  {A.}~\bibnamefont {Vanzanella}}, \bibinfo {author} {\bibfnamefont
  {G.}~\bibnamefont {Verde}}, \bibinfo {author} {\bibfnamefont
  {E.}~\bibnamefont {Vient}}, \bibinfo {author} {\bibfnamefont
  {M.}~\bibnamefont {Vigilante}}, \bibinfo {author} {\bibfnamefont
  {E.}~\bibnamefont {Wanlin}}, \bibinfo {author} {\bibfnamefont
  {G.}~\bibnamefont {Wittwer}},\ and\ \bibinfo {author} {\bibfnamefont
  {A.}~\bibnamefont {Zucchini}},\ }\href
  {https://doi.org/https://doi.org/10.1016/j.nima.2019.03.082} {\bibfield
  {journal} {\bibinfo  {journal} {Nucl. Instr. and Meth. in Phys. Res. A}\
  }\textbf {\bibinfo {volume} {930}},\ \bibinfo {pages} {27} (\bibinfo {year}
  {2019})}\BibitemShut {NoStop}%
\bibitem [{\citenamefont {Carboni}\ \emph {et~al.}(2012)\citenamefont
  {Carboni}, \citenamefont {Barlini}, \citenamefont {Bardelli}, \citenamefont
  {{Le Neindre}}, \citenamefont {Bini}, \citenamefont {Borderie}, \citenamefont
  {Bougault}, \citenamefont {Casini}, \citenamefont {Edelbruck}, \citenamefont
  {Olmi}, \citenamefont {Pasquali}, \citenamefont {Poggi}, \citenamefont
  {Rivet}, \citenamefont {Stefanini}, \citenamefont {Baiocco}, \citenamefont
  {Berjillos}, \citenamefont {Bonnet}, \citenamefont {Bruno}, \citenamefont
  {Chbihi}, \citenamefont {Cruceru}, \citenamefont {Degerlier}, \citenamefont
  {Dueñas}, \citenamefont {Galichet}, \citenamefont {Gramegna}, \citenamefont
  {Kordyasz}, \citenamefont {Kozik}, \citenamefont {Kravchuk}, \citenamefont
  {Lopez}, \citenamefont {Marchi}, \citenamefont {Martel}, \citenamefont
  {Morelli}, \citenamefont {Parlog}, \citenamefont {Petrascu}, \citenamefont
  {Rosato}, \citenamefont {Seredov}, \citenamefont {Vient}, \citenamefont
  {Vigilante}, \citenamefont {Alba}, \citenamefont {Santonocito},\ and\
  \citenamefont {Maiolino}}]{Carboni2012}%
  \BibitemOpen
  \bibfield  {author} {\bibinfo {author} {\bibfnamefont {S.}~\bibnamefont
  {Carboni}}, \bibinfo {author} {\bibfnamefont {S.}~\bibnamefont {Barlini}},
  \bibinfo {author} {\bibfnamefont {L.}~\bibnamefont {Bardelli}}, \bibinfo
  {author} {\bibfnamefont {N.}~\bibnamefont {{Le Neindre}}}, \bibinfo {author}
  {\bibfnamefont {M.}~\bibnamefont {Bini}}, \bibinfo {author} {\bibfnamefont
  {B.}~\bibnamefont {Borderie}}, \bibinfo {author} {\bibfnamefont
  {R.}~\bibnamefont {Bougault}}, \bibinfo {author} {\bibfnamefont
  {G.}~\bibnamefont {Casini}}, \bibinfo {author} {\bibfnamefont
  {P.}~\bibnamefont {Edelbruck}}, \bibinfo {author} {\bibfnamefont
  {A.}~\bibnamefont {Olmi}}, \bibinfo {author} {\bibfnamefont {G.}~\bibnamefont
  {Pasquali}}, \bibinfo {author} {\bibfnamefont {G.}~\bibnamefont {Poggi}},
  \bibinfo {author} {\bibfnamefont {M.}~\bibnamefont {Rivet}}, \bibinfo
  {author} {\bibfnamefont {A.}~\bibnamefont {Stefanini}}, \bibinfo {author}
  {\bibfnamefont {G.}~\bibnamefont {Baiocco}}, \bibinfo {author} {\bibfnamefont
  {R.}~\bibnamefont {Berjillos}}, \bibinfo {author} {\bibfnamefont
  {E.}~\bibnamefont {Bonnet}}, \bibinfo {author} {\bibfnamefont
  {M.}~\bibnamefont {Bruno}}, \bibinfo {author} {\bibfnamefont
  {A.}~\bibnamefont {Chbihi}}, \bibinfo {author} {\bibfnamefont
  {I.}~\bibnamefont {Cruceru}}, \bibinfo {author} {\bibfnamefont
  {M.}~\bibnamefont {Degerlier}}, \bibinfo {author} {\bibfnamefont
  {J.}~\bibnamefont {Dueñas}}, \bibinfo {author} {\bibfnamefont
  {E.}~\bibnamefont {Galichet}}, \bibinfo {author} {\bibfnamefont
  {F.}~\bibnamefont {Gramegna}}, \bibinfo {author} {\bibfnamefont
  {A.}~\bibnamefont {Kordyasz}}, \bibinfo {author} {\bibfnamefont
  {T.}~\bibnamefont {Kozik}}, \bibinfo {author} {\bibfnamefont
  {V.}~\bibnamefont {Kravchuk}}, \bibinfo {author} {\bibfnamefont
  {O.}~\bibnamefont {Lopez}}, \bibinfo {author} {\bibfnamefont
  {T.}~\bibnamefont {Marchi}}, \bibinfo {author} {\bibfnamefont
  {I.}~\bibnamefont {Martel}}, \bibinfo {author} {\bibfnamefont
  {L.}~\bibnamefont {Morelli}}, \bibinfo {author} {\bibfnamefont
  {M.}~\bibnamefont {Parlog}}, \bibinfo {author} {\bibfnamefont
  {H.}~\bibnamefont {Petrascu}}, \bibinfo {author} {\bibfnamefont
  {E.}~\bibnamefont {Rosato}}, \bibinfo {author} {\bibfnamefont
  {V.}~\bibnamefont {Seredov}}, \bibinfo {author} {\bibfnamefont
  {E.}~\bibnamefont {Vient}}, \bibinfo {author} {\bibfnamefont
  {M.}~\bibnamefont {Vigilante}}, \bibinfo {author} {\bibfnamefont
  {R.}~\bibnamefont {Alba}}, \bibinfo {author} {\bibfnamefont {D.}~\bibnamefont
  {Santonocito}},\ and\ \bibinfo {author} {\bibfnamefont {C.}~\bibnamefont
  {Maiolino}},\ }\href
  {https://doi.org/https://doi.org/10.1016/j.nima.2011.10.061} {\bibfield
  {journal} {\bibinfo  {journal} {Nucl. Instr. and Meth. in Phys. Res. A}\
  }\textbf {\bibinfo {volume} {664}},\ \bibinfo {pages} {251} (\bibinfo {year}
  {2012})}\BibitemShut {NoStop}%
\bibitem [{\citenamefont {Pastore}\ \emph {et~al.}(2017)\citenamefont
  {Pastore}, \citenamefont {Gruyer}, \citenamefont {Ottanelli}, \citenamefont
  {{Le Neindre}}, \citenamefont {Pasquali}, \citenamefont {Alba}, \citenamefont
  {Barlini}, \citenamefont {Bini}, \citenamefont {Bonnet}, \citenamefont
  {Borderie}, \citenamefont {Bougault}, \citenamefont {Bruno}, \citenamefont
  {Casini}, \citenamefont {Chbihi}, \citenamefont {Dell'Aquila}, \citenamefont
  {Dueñas}, \citenamefont {Fabris}, \citenamefont {Francalanza}, \citenamefont
  {Frankland}, \citenamefont {Gramegna}, \citenamefont {Henri}, \citenamefont
  {Kordyasz}, \citenamefont {Kozik}, \citenamefont {Lombardo}, \citenamefont
  {Lopez}, \citenamefont {Morelli}, \citenamefont {Olmi}, \citenamefont
  {P\^{a}rlog}, \citenamefont {Piantelli}, \citenamefont {Poggi}, \citenamefont
  {Santonocito}, \citenamefont {Stefanini}, \citenamefont {Valdr\'{e}},
  \citenamefont {Verde}, \citenamefont {Vient},\ and\ \citenamefont
  {Vigilante}}]{Pastore2017}%
  \BibitemOpen
  \bibfield  {author} {\bibinfo {author} {\bibfnamefont {G.}~\bibnamefont
  {Pastore}}, \bibinfo {author} {\bibfnamefont {D.}~\bibnamefont {Gruyer}},
  \bibinfo {author} {\bibfnamefont {P.}~\bibnamefont {Ottanelli}}, \bibinfo
  {author} {\bibfnamefont {N.}~\bibnamefont {{Le Neindre}}}, \bibinfo {author}
  {\bibfnamefont {G.}~\bibnamefont {Pasquali}}, \bibinfo {author}
  {\bibfnamefont {R.}~\bibnamefont {Alba}}, \bibinfo {author} {\bibfnamefont
  {S.}~\bibnamefont {Barlini}}, \bibinfo {author} {\bibfnamefont
  {M.}~\bibnamefont {Bini}}, \bibinfo {author} {\bibfnamefont {E.}~\bibnamefont
  {Bonnet}}, \bibinfo {author} {\bibfnamefont {B.}~\bibnamefont {Borderie}},
  \bibinfo {author} {\bibfnamefont {R.}~\bibnamefont {Bougault}}, \bibinfo
  {author} {\bibfnamefont {M.}~\bibnamefont {Bruno}}, \bibinfo {author}
  {\bibfnamefont {G.}~\bibnamefont {Casini}}, \bibinfo {author} {\bibfnamefont
  {A.}~\bibnamefont {Chbihi}}, \bibinfo {author} {\bibfnamefont
  {D.}~\bibnamefont {Dell'Aquila}}, \bibinfo {author} {\bibfnamefont
  {J.}~\bibnamefont {Dueñas}}, \bibinfo {author} {\bibfnamefont
  {D.}~\bibnamefont {Fabris}}, \bibinfo {author} {\bibfnamefont
  {L.}~\bibnamefont {Francalanza}}, \bibinfo {author} {\bibfnamefont
  {J.}~\bibnamefont {Frankland}}, \bibinfo {author} {\bibfnamefont
  {F.}~\bibnamefont {Gramegna}}, \bibinfo {author} {\bibfnamefont
  {M.}~\bibnamefont {Henri}}, \bibinfo {author} {\bibfnamefont
  {A.}~\bibnamefont {Kordyasz}}, \bibinfo {author} {\bibfnamefont
  {T.}~\bibnamefont {Kozik}}, \bibinfo {author} {\bibfnamefont
  {I.}~\bibnamefont {Lombardo}}, \bibinfo {author} {\bibfnamefont
  {O.}~\bibnamefont {Lopez}}, \bibinfo {author} {\bibfnamefont
  {L.}~\bibnamefont {Morelli}}, \bibinfo {author} {\bibfnamefont
  {A.}~\bibnamefont {Olmi}}, \bibinfo {author} {\bibfnamefont {M.}~\bibnamefont
  {P\^{a}rlog}}, \bibinfo {author} {\bibfnamefont {S.}~\bibnamefont
  {Piantelli}}, \bibinfo {author} {\bibfnamefont {G.}~\bibnamefont {Poggi}},
  \bibinfo {author} {\bibfnamefont {D.}~\bibnamefont {Santonocito}}, \bibinfo
  {author} {\bibfnamefont {A.}~\bibnamefont {Stefanini}}, \bibinfo {author}
  {\bibfnamefont {S.}~\bibnamefont {Valdr\'{e}}}, \bibinfo {author}
  {\bibfnamefont {G.}~\bibnamefont {Verde}}, \bibinfo {author} {\bibfnamefont
  {E.}~\bibnamefont {Vient}},\ and\ \bibinfo {author} {\bibfnamefont
  {M.}~\bibnamefont {Vigilante}},\ }\href
  {https://doi.org/https://doi.org/10.1016/j.nima.2017.01.048} {\bibfield
  {journal} {\bibinfo  {journal} {Nucl. Instr. and Meth. in Phys. Res. A}\
  }\textbf {\bibinfo {volume} {860}},\ \bibinfo {pages} {42} (\bibinfo {year}
  {2017})}\BibitemShut {NoStop}%
\bibitem [{\citenamefont {Piantelli}\ \emph {et~al.}(2020)\citenamefont
  {Piantelli}, \citenamefont {Casini}, \citenamefont {Ono}, \citenamefont
  {Poggi}, \citenamefont {Pastore}, \citenamefont {Barlini}, \citenamefont
  {Boiano}, \citenamefont {Bonnet}, \citenamefont {Borderie}, \citenamefont
  {Bougault}, \citenamefont {Bruno}, \citenamefont {Buccola}, \citenamefont
  {Camaiani}, \citenamefont {Chbihi}, \citenamefont {Cicerchia}, \citenamefont
  {Cinausero}, \citenamefont {D'Agostino}, \citenamefont {Degerlier},
  \citenamefont {Due\~nas}, \citenamefont {Fable}, \citenamefont {Fabris},
  \citenamefont {Frankland}, \citenamefont {Frosin}, \citenamefont {Gramegna},
  \citenamefont {Gruyer}, \citenamefont {Henri}, \citenamefont {Kordyasz},
  \citenamefont {Kozik}, \citenamefont {Le~Neindre}, \citenamefont {Lombardo},
  \citenamefont {Lopez}, \citenamefont {Mantovani}, \citenamefont {Marchi},
  \citenamefont {Morelli}, \citenamefont {Olmi}, \citenamefont {Ottanelli},
  \citenamefont {P\^arlog}, \citenamefont {Pasquali}, \citenamefont
  {Stefanini}, \citenamefont {Tortone}, \citenamefont {Upadhyaya},
  \citenamefont {Valdr\'e}, \citenamefont {Verde}, \citenamefont {Vient},
  \citenamefont {Vigilante}, \citenamefont {Alba},\ and\ \citenamefont
  {Maiolino}}]{Piantelli2020}%
  \BibitemOpen
  \bibfield  {author} {\bibinfo {author} {\bibfnamefont {S.}~\bibnamefont
  {Piantelli}}, \bibinfo {author} {\bibfnamefont {G.}~\bibnamefont {Casini}},
  \bibinfo {author} {\bibfnamefont {A.}~\bibnamefont {Ono}}, \bibinfo {author}
  {\bibfnamefont {G.}~\bibnamefont {Poggi}}, \bibinfo {author} {\bibfnamefont
  {G.}~\bibnamefont {Pastore}}, \bibinfo {author} {\bibfnamefont
  {S.}~\bibnamefont {Barlini}}, \bibinfo {author} {\bibfnamefont
  {A.}~\bibnamefont {Boiano}}, \bibinfo {author} {\bibfnamefont
  {E.}~\bibnamefont {Bonnet}}, \bibinfo {author} {\bibfnamefont
  {B.}~\bibnamefont {Borderie}}, \bibinfo {author} {\bibfnamefont
  {R.}~\bibnamefont {Bougault}}, \bibinfo {author} {\bibfnamefont
  {M.}~\bibnamefont {Bruno}}, \bibinfo {author} {\bibfnamefont
  {A.}~\bibnamefont {Buccola}}, \bibinfo {author} {\bibfnamefont
  {A.}~\bibnamefont {Camaiani}}, \bibinfo {author} {\bibfnamefont
  {A.}~\bibnamefont {Chbihi}}, \bibinfo {author} {\bibfnamefont
  {M.}~\bibnamefont {Cicerchia}}, \bibinfo {author} {\bibfnamefont
  {M.}~\bibnamefont {Cinausero}}, \bibinfo {author} {\bibfnamefont
  {M.}~\bibnamefont {D'Agostino}}, \bibinfo {author} {\bibfnamefont
  {M.}~\bibnamefont {Degerlier}}, \bibinfo {author} {\bibfnamefont {J.~A.}\
  \bibnamefont {Due\~nas}}, \bibinfo {author} {\bibfnamefont {Q.}~\bibnamefont
  {Fable}}, \bibinfo {author} {\bibfnamefont {D.}~\bibnamefont {Fabris}},
  \bibinfo {author} {\bibfnamefont {J.~D.}\ \bibnamefont {Frankland}}, \bibinfo
  {author} {\bibfnamefont {C.}~\bibnamefont {Frosin}}, \bibinfo {author}
  {\bibfnamefont {F.}~\bibnamefont {Gramegna}}, \bibinfo {author}
  {\bibfnamefont {D.}~\bibnamefont {Gruyer}}, \bibinfo {author} {\bibfnamefont
  {M.}~\bibnamefont {Henri}}, \bibinfo {author} {\bibfnamefont
  {A.}~\bibnamefont {Kordyasz}}, \bibinfo {author} {\bibfnamefont
  {T.}~\bibnamefont {Kozik}}, \bibinfo {author} {\bibfnamefont
  {N.}~\bibnamefont {Le~Neindre}}, \bibinfo {author} {\bibfnamefont
  {I.}~\bibnamefont {Lombardo}}, \bibinfo {author} {\bibfnamefont
  {O.}~\bibnamefont {Lopez}}, \bibinfo {author} {\bibfnamefont
  {G.}~\bibnamefont {Mantovani}}, \bibinfo {author} {\bibfnamefont
  {T.}~\bibnamefont {Marchi}}, \bibinfo {author} {\bibfnamefont
  {L.}~\bibnamefont {Morelli}}, \bibinfo {author} {\bibfnamefont
  {A.}~\bibnamefont {Olmi}}, \bibinfo {author} {\bibfnamefont {P.}~\bibnamefont
  {Ottanelli}}, \bibinfo {author} {\bibfnamefont {M.}~\bibnamefont {P\^arlog}},
  \bibinfo {author} {\bibfnamefont {G.}~\bibnamefont {Pasquali}}, \bibinfo
  {author} {\bibfnamefont {A.~A.}\ \bibnamefont {Stefanini}}, \bibinfo {author}
  {\bibfnamefont {G.}~\bibnamefont {Tortone}}, \bibinfo {author} {\bibfnamefont
  {S.}~\bibnamefont {Upadhyaya}}, \bibinfo {author} {\bibfnamefont
  {S.}~\bibnamefont {Valdr\'e}}, \bibinfo {author} {\bibfnamefont
  {G.}~\bibnamefont {Verde}}, \bibinfo {author} {\bibfnamefont
  {E.}~\bibnamefont {Vient}}, \bibinfo {author} {\bibfnamefont
  {M.}~\bibnamefont {Vigilante}}, \bibinfo {author} {\bibfnamefont
  {R.}~\bibnamefont {Alba}},\ and\ \bibinfo {author} {\bibfnamefont
  {C.}~\bibnamefont {Maiolino}},\ }\href
  {https://doi.org/10.1103/PhysRevC.101.034613} {\bibfield  {journal} {\bibinfo
   {journal} {Phys. Rev. C}\ }\textbf {\bibinfo {volume} {101}},\ \bibinfo
  {pages} {034613} (\bibinfo {year} {2020})}\BibitemShut {NoStop}%
\bibitem [{\citenamefont {Pouthas}\ \emph {et~al.}(1995)\citenamefont
  {Pouthas}, \citenamefont {Borderie}, \citenamefont {Dayras}, \citenamefont
  {Plagnol}, \citenamefont {Rivet}, \citenamefont {Saint-Laurent},
  \citenamefont {Steckmeyer}, \citenamefont {Auger}, \citenamefont {Bacri},
  \citenamefont {Barbey}, \citenamefont {Barbier}, \citenamefont {Benkirane},
  \citenamefont {Benlliure}, \citenamefont {Berthier}, \citenamefont
  {Bougamont}, \citenamefont {Bourgault}, \citenamefont {Box}, \citenamefont
  {Bzyl}, \citenamefont {Cahan}, \citenamefont {Cassagnou}, \citenamefont
  {Charlet}, \citenamefont {Charvet}, \citenamefont {Chbihi}, \citenamefont
  {Clerc}, \citenamefont {Copinet}, \citenamefont {Cussol}, \citenamefont
  {Engrand}, \citenamefont {Gautier}, \citenamefont {Huguet}, \citenamefont
  {Jouniaux}, \citenamefont {Laville}, \citenamefont {{Le Botlan}},
  \citenamefont {Leconte}, \citenamefont {Legrain}, \citenamefont {Lelong},
  \citenamefont {{Le Guay}}, \citenamefont {Martina}, \citenamefont {Mazur},
  \citenamefont {Mosrin}, \citenamefont {Olivier}, \citenamefont {Passerieux},
  \citenamefont {Pierre}, \citenamefont {Piquet}, \citenamefont {Plaige},
  \citenamefont {Pollacco}, \citenamefont {Raine}, \citenamefont {Richard},
  \citenamefont {Ropert}, \citenamefont {Spitaels}, \citenamefont {Stab},
  \citenamefont {Sznajderman}, \citenamefont {Tassan-got}, \citenamefont
  {Tillier}, \citenamefont {Tripon}, \citenamefont {Vallerand}, \citenamefont
  {Volant}, \citenamefont {Volkov}, \citenamefont {Wieleczko},\ and\
  \citenamefont {Wittwer}}]{Pouthas1995}%
  \BibitemOpen
  \bibfield  {author} {\bibinfo {author} {\bibfnamefont {J.}~\bibnamefont
  {Pouthas}}, \bibinfo {author} {\bibfnamefont {B.}~\bibnamefont {Borderie}},
  \bibinfo {author} {\bibfnamefont {R.}~\bibnamefont {Dayras}}, \bibinfo
  {author} {\bibfnamefont {E.}~\bibnamefont {Plagnol}}, \bibinfo {author}
  {\bibfnamefont {M.}~\bibnamefont {Rivet}}, \bibinfo {author} {\bibfnamefont
  {F.}~\bibnamefont {Saint-Laurent}}, \bibinfo {author} {\bibfnamefont
  {J.}~\bibnamefont {Steckmeyer}}, \bibinfo {author} {\bibfnamefont
  {G.}~\bibnamefont {Auger}}, \bibinfo {author} {\bibfnamefont
  {C.}~\bibnamefont {Bacri}}, \bibinfo {author} {\bibfnamefont
  {S.}~\bibnamefont {Barbey}}, \bibinfo {author} {\bibfnamefont
  {A.}~\bibnamefont {Barbier}}, \bibinfo {author} {\bibfnamefont
  {A.}~\bibnamefont {Benkirane}}, \bibinfo {author} {\bibfnamefont
  {J.}~\bibnamefont {Benlliure}}, \bibinfo {author} {\bibfnamefont
  {B.}~\bibnamefont {Berthier}}, \bibinfo {author} {\bibfnamefont
  {E.}~\bibnamefont {Bougamont}}, \bibinfo {author} {\bibfnamefont
  {P.}~\bibnamefont {Bourgault}}, \bibinfo {author} {\bibfnamefont
  {P.}~\bibnamefont {Box}}, \bibinfo {author} {\bibfnamefont {R.}~\bibnamefont
  {Bzyl}}, \bibinfo {author} {\bibfnamefont {B.}~\bibnamefont {Cahan}},
  \bibinfo {author} {\bibfnamefont {Y.}~\bibnamefont {Cassagnou}}, \bibinfo
  {author} {\bibfnamefont {D.}~\bibnamefont {Charlet}}, \bibinfo {author}
  {\bibfnamefont {J.}~\bibnamefont {Charvet}}, \bibinfo {author} {\bibfnamefont
  {A.}~\bibnamefont {Chbihi}}, \bibinfo {author} {\bibfnamefont
  {T.}~\bibnamefont {Clerc}}, \bibinfo {author} {\bibfnamefont
  {N.}~\bibnamefont {Copinet}}, \bibinfo {author} {\bibfnamefont
  {D.}~\bibnamefont {Cussol}}, \bibinfo {author} {\bibfnamefont
  {M.}~\bibnamefont {Engrand}}, \bibinfo {author} {\bibfnamefont
  {J.}~\bibnamefont {Gautier}}, \bibinfo {author} {\bibfnamefont
  {Y.}~\bibnamefont {Huguet}}, \bibinfo {author} {\bibfnamefont
  {O.}~\bibnamefont {Jouniaux}}, \bibinfo {author} {\bibfnamefont
  {J.}~\bibnamefont {Laville}}, \bibinfo {author} {\bibfnamefont
  {P.}~\bibnamefont {{Le Botlan}}}, \bibinfo {author} {\bibfnamefont
  {A.}~\bibnamefont {Leconte}}, \bibinfo {author} {\bibfnamefont
  {R.}~\bibnamefont {Legrain}}, \bibinfo {author} {\bibfnamefont
  {P.}~\bibnamefont {Lelong}}, \bibinfo {author} {\bibfnamefont
  {M.}~\bibnamefont {{Le Guay}}}, \bibinfo {author} {\bibfnamefont
  {L.}~\bibnamefont {Martina}}, \bibinfo {author} {\bibfnamefont
  {C.}~\bibnamefont {Mazur}}, \bibinfo {author} {\bibfnamefont
  {P.}~\bibnamefont {Mosrin}}, \bibinfo {author} {\bibfnamefont
  {L.}~\bibnamefont {Olivier}}, \bibinfo {author} {\bibfnamefont
  {J.}~\bibnamefont {Passerieux}}, \bibinfo {author} {\bibfnamefont
  {S.}~\bibnamefont {Pierre}}, \bibinfo {author} {\bibfnamefont
  {B.}~\bibnamefont {Piquet}}, \bibinfo {author} {\bibfnamefont
  {E.}~\bibnamefont {Plaige}}, \bibinfo {author} {\bibfnamefont
  {E.}~\bibnamefont {Pollacco}}, \bibinfo {author} {\bibfnamefont
  {B.}~\bibnamefont {Raine}}, \bibinfo {author} {\bibfnamefont
  {A.}~\bibnamefont {Richard}}, \bibinfo {author} {\bibfnamefont
  {J.}~\bibnamefont {Ropert}}, \bibinfo {author} {\bibfnamefont
  {C.}~\bibnamefont {Spitaels}}, \bibinfo {author} {\bibfnamefont
  {L.}~\bibnamefont {Stab}}, \bibinfo {author} {\bibfnamefont {D.}~\bibnamefont
  {Sznajderman}}, \bibinfo {author} {\bibfnamefont {L.}~\bibnamefont
  {Tassan-got}}, \bibinfo {author} {\bibfnamefont {J.}~\bibnamefont {Tillier}},
  \bibinfo {author} {\bibfnamefont {M.}~\bibnamefont {Tripon}}, \bibinfo
  {author} {\bibfnamefont {P.}~\bibnamefont {Vallerand}}, \bibinfo {author}
  {\bibfnamefont {C.}~\bibnamefont {Volant}}, \bibinfo {author} {\bibfnamefont
  {P.}~\bibnamefont {Volkov}}, \bibinfo {author} {\bibfnamefont
  {J.}~\bibnamefont {Wieleczko}},\ and\ \bibinfo {author} {\bibfnamefont
  {G.}~\bibnamefont {Wittwer}},\ }\href
  {https://doi.org/https://doi.org/10.1016/0168-9002(94)01543-0} {\bibfield
  {journal} {\bibinfo  {journal} {Nucl. Instr. and Meth. in Phys. Res. A}\
  }\textbf {\bibinfo {volume} {357}},\ \bibinfo {pages} {418} (\bibinfo {year}
  {1995})}\BibitemShut {NoStop}%
\bibitem [{\citenamefont {Pouthas}\ \emph {et~al.}(1996)\citenamefont
  {Pouthas}, \citenamefont {Bertaut}, \citenamefont {Borderie}, \citenamefont
  {Bourgault}, \citenamefont {Cahan}, \citenamefont {Carles}, \citenamefont
  {Charlet}, \citenamefont {Cussol}, \citenamefont {Dayras}, \citenamefont
  {Engrand}, \citenamefont {Jouniaux}, \citenamefont {{Le Botlan}},
  \citenamefont {Leconte}, \citenamefont {Lelong}, \citenamefont {Martina},
  \citenamefont {Mosrin}, \citenamefont {Olivier}, \citenamefont {Passerieux},
  \citenamefont {Piquet}, \citenamefont {Plagnol}, \citenamefont {Plaige},
  \citenamefont {Raine}, \citenamefont {Richard}, \citenamefont
  {Saint-Laurent}, \citenamefont {Spitaels}, \citenamefont {Tillier},
  \citenamefont {Tripon}, \citenamefont {Vallerand}, \citenamefont {Volkov},\
  and\ \citenamefont {Wittwer}}]{Pouthas1996}%
  \BibitemOpen
  \bibfield  {author} {\bibinfo {author} {\bibfnamefont {J.}~\bibnamefont
  {Pouthas}}, \bibinfo {author} {\bibfnamefont {A.}~\bibnamefont {Bertaut}},
  \bibinfo {author} {\bibfnamefont {B.}~\bibnamefont {Borderie}}, \bibinfo
  {author} {\bibfnamefont {P.}~\bibnamefont {Bourgault}}, \bibinfo {author}
  {\bibfnamefont {B.}~\bibnamefont {Cahan}}, \bibinfo {author} {\bibfnamefont
  {G.}~\bibnamefont {Carles}}, \bibinfo {author} {\bibfnamefont
  {D.}~\bibnamefont {Charlet}}, \bibinfo {author} {\bibfnamefont
  {D.}~\bibnamefont {Cussol}}, \bibinfo {author} {\bibfnamefont
  {R.}~\bibnamefont {Dayras}}, \bibinfo {author} {\bibfnamefont
  {M.}~\bibnamefont {Engrand}}, \bibinfo {author} {\bibfnamefont
  {O.}~\bibnamefont {Jouniaux}}, \bibinfo {author} {\bibfnamefont
  {P.}~\bibnamefont {{Le Botlan}}}, \bibinfo {author} {\bibfnamefont
  {A.}~\bibnamefont {Leconte}}, \bibinfo {author} {\bibfnamefont
  {P.}~\bibnamefont {Lelong}}, \bibinfo {author} {\bibfnamefont
  {L.}~\bibnamefont {Martina}}, \bibinfo {author} {\bibfnamefont
  {P.}~\bibnamefont {Mosrin}}, \bibinfo {author} {\bibfnamefont
  {L.}~\bibnamefont {Olivier}}, \bibinfo {author} {\bibfnamefont
  {J.}~\bibnamefont {Passerieux}}, \bibinfo {author} {\bibfnamefont
  {B.}~\bibnamefont {Piquet}}, \bibinfo {author} {\bibfnamefont
  {E.}~\bibnamefont {Plagnol}}, \bibinfo {author} {\bibfnamefont
  {E.}~\bibnamefont {Plaige}}, \bibinfo {author} {\bibfnamefont
  {B.}~\bibnamefont {Raine}}, \bibinfo {author} {\bibfnamefont
  {A.}~\bibnamefont {Richard}}, \bibinfo {author} {\bibfnamefont
  {F.}~\bibnamefont {Saint-Laurent}}, \bibinfo {author} {\bibfnamefont
  {C.}~\bibnamefont {Spitaels}}, \bibinfo {author} {\bibfnamefont
  {J.}~\bibnamefont {Tillier}}, \bibinfo {author} {\bibfnamefont
  {M.}~\bibnamefont {Tripon}}, \bibinfo {author} {\bibfnamefont
  {P.}~\bibnamefont {Vallerand}}, \bibinfo {author} {\bibfnamefont
  {P.}~\bibnamefont {Volkov}},\ and\ \bibinfo {author} {\bibfnamefont
  {G.}~\bibnamefont {Wittwer}},\ }\href
  {https://doi.org/https://doi.org/10.1016/0168-9002(95)00770-9} {\bibfield
  {journal} {\bibinfo  {journal} {Nucl. Instr. and Meth. in Phys. Res. A}\
  }\textbf {\bibinfo {volume} {369}},\ \bibinfo {pages} {222} (\bibinfo {year}
  {1996})}\BibitemShut {NoStop}%
\bibitem [{\citenamefont {Wittwer}(2004)}]{CENTRUM_ref}%
  \BibitemOpen
  \bibfield  {author} {\bibinfo {author} {\bibfnamefont {G.}~\bibnamefont
  {Wittwer}},\ }\href@noop {} {\bibinfo {title} {Clock event number transmitter
  receiver universal module, user’s manual}},\ \bibinfo {howpublished}
  {GANIL} (\bibinfo {year} {2004})\BibitemShut {NoStop}%
\bibitem [{\citenamefont {Charity}(1998)}]{Charity1998}%
  \BibitemOpen
  \bibfield  {author} {\bibinfo {author} {\bibfnamefont {R.~J.}\ \bibnamefont
  {Charity}},\ }\href {https://doi.org/10.1103/PhysRevC.58.1073} {\bibfield
  {journal} {\bibinfo  {journal} {Phys. Rev. C}\ }\textbf {\bibinfo {volume}
  {58}},\ \bibinfo {pages} {1073} (\bibinfo {year} {1998})}\BibitemShut
  {NoStop}%
\bibitem [{\citenamefont {Lombardo}\ \emph {et~al.}(2011)\citenamefont
  {Lombardo}, \citenamefont {Agodi}, \citenamefont {Amorini}, \citenamefont
  {Anzalone}, \citenamefont {Auditore}, \citenamefont {Berceanu}, \citenamefont
  {Cardella}, \citenamefont {Cavallaro}, \citenamefont {Chatterjee},
  \citenamefont {De~Filippo}, \citenamefont {Geraci}, \citenamefont {Giuliani},
  \citenamefont {Grassi}, \citenamefont {Han}, \citenamefont {La~Guidara},
  \citenamefont {Loria}, \citenamefont {Lanzalone}, \citenamefont {Maiolino},
  \citenamefont {Pagano}, \citenamefont {Papa}, \citenamefont {Pirrone},
  \citenamefont {Politi}, \citenamefont {Porto}, \citenamefont {Rizzo},
  \citenamefont {Russotto}, \citenamefont {Trifir\`o}, \citenamefont
  {Trimarchi}, \citenamefont {Verde},\ and\ \citenamefont
  {Vigilante}}]{Lombardo2011}%
  \BibitemOpen
  \bibfield  {author} {\bibinfo {author} {\bibfnamefont {I.}~\bibnamefont
  {Lombardo}}, \bibinfo {author} {\bibfnamefont {C.}~\bibnamefont {Agodi}},
  \bibinfo {author} {\bibfnamefont {F.}~\bibnamefont {Amorini}}, \bibinfo
  {author} {\bibfnamefont {A.}~\bibnamefont {Anzalone}}, \bibinfo {author}
  {\bibfnamefont {L.}~\bibnamefont {Auditore}}, \bibinfo {author}
  {\bibfnamefont {I.}~\bibnamefont {Berceanu}}, \bibinfo {author}
  {\bibfnamefont {G.}~\bibnamefont {Cardella}}, \bibinfo {author}
  {\bibfnamefont {S.}~\bibnamefont {Cavallaro}}, \bibinfo {author}
  {\bibfnamefont {M.~B.}\ \bibnamefont {Chatterjee}}, \bibinfo {author}
  {\bibfnamefont {E.}~\bibnamefont {De~Filippo}}, \bibinfo {author}
  {\bibfnamefont {E.}~\bibnamefont {Geraci}}, \bibinfo {author} {\bibfnamefont
  {G.}~\bibnamefont {Giuliani}}, \bibinfo {author} {\bibfnamefont
  {L.}~\bibnamefont {Grassi}}, \bibinfo {author} {\bibfnamefont
  {J.}~\bibnamefont {Han}}, \bibinfo {author} {\bibfnamefont {E.}~\bibnamefont
  {La~Guidara}}, \bibinfo {author} {\bibfnamefont {D.}~\bibnamefont {Loria}},
  \bibinfo {author} {\bibfnamefont {G.}~\bibnamefont {Lanzalone}}, \bibinfo
  {author} {\bibfnamefont {C.}~\bibnamefont {Maiolino}}, \bibinfo {author}
  {\bibfnamefont {A.}~\bibnamefont {Pagano}}, \bibinfo {author} {\bibfnamefont
  {M.}~\bibnamefont {Papa}}, \bibinfo {author} {\bibfnamefont {S.}~\bibnamefont
  {Pirrone}}, \bibinfo {author} {\bibfnamefont {G.}~\bibnamefont {Politi}},
  \bibinfo {author} {\bibfnamefont {F.}~\bibnamefont {Porto}}, \bibinfo
  {author} {\bibfnamefont {F.}~\bibnamefont {Rizzo}}, \bibinfo {author}
  {\bibfnamefont {P.}~\bibnamefont {Russotto}}, \bibinfo {author}
  {\bibfnamefont {A.}~\bibnamefont {Trifir\`o}}, \bibinfo {author}
  {\bibfnamefont {M.}~\bibnamefont {Trimarchi}}, \bibinfo {author}
  {\bibfnamefont {G.}~\bibnamefont {Verde}},\ and\ \bibinfo {author}
  {\bibfnamefont {M.}~\bibnamefont {Vigilante}},\ }\href
  {https://doi.org/10.1103/PhysRevC.84.024613} {\bibfield  {journal} {\bibinfo
  {journal} {Phys. Rev. C}\ }\textbf {\bibinfo {volume} {84}},\ \bibinfo
  {pages} {024613} (\bibinfo {year} {2011})}\BibitemShut {NoStop}%
\bibitem [{\citenamefont {Piantelli}\ \emph {et~al.}(2013)\citenamefont
  {Piantelli}, \citenamefont {Casini}, \citenamefont {Maurenzig}, \citenamefont
  {Olmi}, \citenamefont {Barlini}, \citenamefont {Bini}, \citenamefont
  {Carboni}, \citenamefont {Pasquali}, \citenamefont {Poggi}, \citenamefont
  {Stefanini}, \citenamefont {Valdr\`e}, \citenamefont {Bougault},
  \citenamefont {Bonnet}, \citenamefont {Borderie}, \citenamefont {Chbihi},
  \citenamefont {Frankland}, \citenamefont {Gruyer}, \citenamefont {Lopez},
  \citenamefont {Le~Neindre}, \citenamefont {P\^arlog}, \citenamefont {Rivet},
  \citenamefont {Vient}, \citenamefont {Rosato}, \citenamefont {Spadaccini},
  \citenamefont {Vigilante}, \citenamefont {Bruno}, \citenamefont {Marchi},
  \citenamefont {Morelli}, \citenamefont {Cinausero}, \citenamefont
  {Degerlier}, \citenamefont {Gramegna}, \citenamefont {Kozik}, \citenamefont
  {Twar\'og}, \citenamefont {Alba}, \citenamefont {Maiolino},\ and\
  \citenamefont {Santonocito}}]{Piantelli2013}%
  \BibitemOpen
  \bibfield  {author} {\bibinfo {author} {\bibfnamefont {S.}~\bibnamefont
  {Piantelli}}, \bibinfo {author} {\bibfnamefont {G.}~\bibnamefont {Casini}},
  \bibinfo {author} {\bibfnamefont {P.~R.}\ \bibnamefont {Maurenzig}}, \bibinfo
  {author} {\bibfnamefont {A.}~\bibnamefont {Olmi}}, \bibinfo {author}
  {\bibfnamefont {S.}~\bibnamefont {Barlini}}, \bibinfo {author} {\bibfnamefont
  {M.}~\bibnamefont {Bini}}, \bibinfo {author} {\bibfnamefont {S.}~\bibnamefont
  {Carboni}}, \bibinfo {author} {\bibfnamefont {G.}~\bibnamefont {Pasquali}},
  \bibinfo {author} {\bibfnamefont {G.}~\bibnamefont {Poggi}}, \bibinfo
  {author} {\bibfnamefont {A.~A.}\ \bibnamefont {Stefanini}}, \bibinfo {author}
  {\bibfnamefont {S.}~\bibnamefont {Valdr\`e}}, \bibinfo {author}
  {\bibfnamefont {R.}~\bibnamefont {Bougault}}, \bibinfo {author}
  {\bibfnamefont {E.}~\bibnamefont {Bonnet}}, \bibinfo {author} {\bibfnamefont
  {B.}~\bibnamefont {Borderie}}, \bibinfo {author} {\bibfnamefont
  {A.}~\bibnamefont {Chbihi}}, \bibinfo {author} {\bibfnamefont {J.~D.}\
  \bibnamefont {Frankland}}, \bibinfo {author} {\bibfnamefont {D.}~\bibnamefont
  {Gruyer}}, \bibinfo {author} {\bibfnamefont {O.}~\bibnamefont {Lopez}},
  \bibinfo {author} {\bibfnamefont {N.}~\bibnamefont {Le~Neindre}}, \bibinfo
  {author} {\bibfnamefont {M.}~\bibnamefont {P\^arlog}}, \bibinfo {author}
  {\bibfnamefont {M.~F.}\ \bibnamefont {Rivet}}, \bibinfo {author}
  {\bibfnamefont {E.}~\bibnamefont {Vient}}, \bibinfo {author} {\bibfnamefont
  {E.}~\bibnamefont {Rosato}}, \bibinfo {author} {\bibfnamefont
  {G.}~\bibnamefont {Spadaccini}}, \bibinfo {author} {\bibfnamefont
  {M.}~\bibnamefont {Vigilante}}, \bibinfo {author} {\bibfnamefont
  {M.}~\bibnamefont {Bruno}}, \bibinfo {author} {\bibfnamefont
  {T.}~\bibnamefont {Marchi}}, \bibinfo {author} {\bibfnamefont
  {L.}~\bibnamefont {Morelli}}, \bibinfo {author} {\bibfnamefont
  {M.}~\bibnamefont {Cinausero}}, \bibinfo {author} {\bibfnamefont
  {M.}~\bibnamefont {Degerlier}}, \bibinfo {author} {\bibfnamefont
  {F.}~\bibnamefont {Gramegna}}, \bibinfo {author} {\bibfnamefont
  {T.}~\bibnamefont {Kozik}}, \bibinfo {author} {\bibfnamefont
  {T.}~\bibnamefont {Twar\'og}}, \bibinfo {author} {\bibfnamefont
  {R.}~\bibnamefont {Alba}}, \bibinfo {author} {\bibfnamefont {C.}~\bibnamefont
  {Maiolino}},\ and\ \bibinfo {author} {\bibfnamefont {D.}~\bibnamefont
  {Santonocito}} (\bibinfo {collaboration} {FAZIA Collaboration}),\ }\href
  {https://doi.org/10.1103/PhysRevC.88.064607} {\bibfield  {journal} {\bibinfo
  {journal} {Phys. Rev. C}\ }\textbf {\bibinfo {volume} {88}},\ \bibinfo
  {pages} {064607} (\bibinfo {year} {2013})}\BibitemShut {NoStop}%
\bibitem [{\citenamefont {Piantelli}\ \emph {et~al.}(2019)\citenamefont
  {Piantelli}, \citenamefont {Olmi}, \citenamefont {Maurenzig}, \citenamefont
  {Ono}, \citenamefont {Bini}, \citenamefont {Casini}, \citenamefont
  {Pasquali}, \citenamefont {Mangiarotti}, \citenamefont {Poggi}, \citenamefont
  {Stefanini}, \citenamefont {Barlini}, \citenamefont {Camaiani}, \citenamefont
  {Ciampi}, \citenamefont {Frosin}, \citenamefont {Ottanelli},\ and\
  \citenamefont {Valdr\'e}}]{Piantelli2019}%
  \BibitemOpen
  \bibfield  {author} {\bibinfo {author} {\bibfnamefont {S.}~\bibnamefont
  {Piantelli}}, \bibinfo {author} {\bibfnamefont {A.}~\bibnamefont {Olmi}},
  \bibinfo {author} {\bibfnamefont {P.~R.}\ \bibnamefont {Maurenzig}}, \bibinfo
  {author} {\bibfnamefont {A.}~\bibnamefont {Ono}}, \bibinfo {author}
  {\bibfnamefont {M.}~\bibnamefont {Bini}}, \bibinfo {author} {\bibfnamefont
  {G.}~\bibnamefont {Casini}}, \bibinfo {author} {\bibfnamefont
  {G.}~\bibnamefont {Pasquali}}, \bibinfo {author} {\bibfnamefont
  {A.}~\bibnamefont {Mangiarotti}}, \bibinfo {author} {\bibfnamefont
  {G.}~\bibnamefont {Poggi}}, \bibinfo {author} {\bibfnamefont {A.~A.}\
  \bibnamefont {Stefanini}}, \bibinfo {author} {\bibfnamefont {S.}~\bibnamefont
  {Barlini}}, \bibinfo {author} {\bibfnamefont {A.}~\bibnamefont {Camaiani}},
  \bibinfo {author} {\bibfnamefont {C.}~\bibnamefont {Ciampi}}, \bibinfo
  {author} {\bibfnamefont {C.}~\bibnamefont {Frosin}}, \bibinfo {author}
  {\bibfnamefont {P.}~\bibnamefont {Ottanelli}},\ and\ \bibinfo {author}
  {\bibfnamefont {S.}~\bibnamefont {Valdr\'e}},\ }\href
  {https://doi.org/10.1103/PhysRevC.99.064616} {\bibfield  {journal} {\bibinfo
  {journal} {Phys. Rev. C}\ }\textbf {\bibinfo {volume} {99}},\ \bibinfo
  {pages} {064616} (\bibinfo {year} {2019})}\BibitemShut {NoStop}%
\bibitem [{\citenamefont {Ciampi}(2021)}]{Ciampi2021}%
  \BibitemOpen
  \bibfield  {author} {\bibinfo {author} {\bibfnamefont {C.}~\bibnamefont
  {Ciampi}},\ }\href@noop {} {\bibinfo {type} {{Ph.D.} thesis}},\ \bibinfo
  {school} {{Università degli Studi di Firenze}} (\bibinfo {year}
  {2021})\BibitemShut {NoStop}%
\bibitem [{\citenamefont {Galichet}\ \emph
  {et~al.}(2009{\natexlab{b}})\citenamefont {Galichet}, \citenamefont
  {Colonna}, \citenamefont {Borderie},\ and\ \citenamefont
  {Rivet}}]{Galichet2009b}%
  \BibitemOpen
  \bibfield  {author} {\bibinfo {author} {\bibfnamefont {E.}~\bibnamefont
  {Galichet}}, \bibinfo {author} {\bibfnamefont {M.}~\bibnamefont {Colonna}},
  \bibinfo {author} {\bibfnamefont {B.}~\bibnamefont {Borderie}},\ and\
  \bibinfo {author} {\bibfnamefont {M.~F.}\ \bibnamefont {Rivet}},\ }\href
  {https://doi.org/10.1103/PhysRevC.79.064615} {\bibfield  {journal} {\bibinfo
  {journal} {Phys. Rev. C}\ }\textbf {\bibinfo {volume} {79}},\ \bibinfo
  {pages} {064615} (\bibinfo {year} {2009}{\natexlab{b}})}\BibitemShut
  {NoStop}%
\end{thebibliography}%
 \bibliographystyle{apsrev4-2.bst}
 
\end{document}